arXiv:2407.13628v1 [quant-ph] 18 Jul 2024

# UNRUH-DEWITT QUANTUM COMPUTING:
# REALIZING QUANTUM SHANNON THEORY WITH QUANTUM FIELDS

BY

ERIC W. ASPLING

BS, Binghamton University, 2018

DISSERTATION

Submitted in partial fulfillment of the requirements for
the degree of Doctor of Philosophy in Physics
in the Graduate School of
Binghamton University
State University of New York
2024

Accepted in partial fulfillment of the requirements for the degree of
Doctor of Philosophy in Physics
in the Graduate School of
Binghamton University
State University of New York
2024

03/29/2024

Stephen Levy, Chair
Department of Physics and Astronomy, Binghamton University

Michael Lawler, Faculty Advisor
Department of Physics and Astronomy, Binghamton University

Masatsugu Sei Suzuki, Member
Department of Physics and Astronomy, Binghamton University

Pegor Aynajian, Member
Department of Physics and Astronomy, Binghamton University

Mengen Wang, External Examiner
Department of Electrical and Computer Engineering, Binghamton University

Thomas Hartman, External Member
Department of Physics, Cornell University

# Abstract

Qubit-field quantum transduction provides numerous advantages to quantum computing, such as device-specific error-correcting codes, efficient scalability, and effective entanglement generation. An all-to-all connected bus of qubits implanted around the outside of a topological insulator, allowed to interact with the edge state, is a promising arena for transduction with flying fermionic qubits. Unruh–DeWitt detectors have allowed quantum information scientists to model entanglement properties of qubit-field interactions in many settings in a field known as Relativistic Quantum Information (RQI). Unruh–DeWitt detectors are useful tools to realize quantum Shannon theory, a subset of the theory of quantum communication, in condensed matter systems, aptly named Unruh–DeWitt quantum computers. These systems will provide quantitative measurements of communication in quantum materials that utilize coherent states for bosonic and fermionic fields. In this thesis, emphasis is placed on the well-studied theory of Tomonaga-Luttinger liquids, as the bosonization of a helical Luttinger liquid provides a pedagogical arena to construct RQI channels of fermionic systems. Multiple experimentally realizable systems are proposed, and design constraints are constructed to ensure maximum channel capacity. Furthermore, we elucidate the strength of these quantum channels using measurements from quantum Shannon theory such as coherent information, dephasing formalism, diamond distance and universality of Unruh–DeWitt quantum logic gates.

I dedicate this work to my loving wife Amanda, who sacrificed and endured many years to help me achieve this dream.

# Acknowledgements

I am forever grateful for the opportunities I have been given and Binghamton University. I want to thank my mentors and advisors Michael Lawler, Bruce White, and Charles Nelson. All of whom gave me room to explore the curiosity that inspired me to go back to school. Without your mentorship this would not be possible.

I would also like to thank my classmates and peers, Dennis Dempsey, Chris Renskers, Mabrur Ahmed, Saba Kharabadze, and Justin Kulp and many more from my communities at Binghamton and Cornell. The endless proof reading, conversations, and encouragement was more than I could have asked for. Without your friendship and collaboration, I couldn't have finished this degree.

Lastly, I'd like to thank my family. My beautiful and loving wife Amanda, my amazing children Ava, Xander, Emery, and Lillia, my supportive and encouraging parents, Scott and Cathy, and my brother Steev, who always encouraged my dreaming. I was lucky to have such a strong support system to help me complete this degree. I cannot begin to name the ways I received support for they are far to numerous. With all my heart, Thank you.

# Contents

# List of Figures

# Chapter 1

# Introduction

## 1.1 Motivation

Among the resources that measure the effectiveness of a good quantum computer is the ability to generate and propagate entanglement. Quantum coherent information and quantum capacity have become standard methodologies for exploring this value [2, 4, 5]. However applications of quantum communication in quantum materials will come in many forms, some of which require different interaction Hamiltonians. Thus, a prescription method for numerically simulating the parameters that determine the ability of the physical system to carry out quantum communication is necessary, if these systems are to be realized in a laboratory setting.

The past few decades have led to fruitful collaborations in Quantum Field Theory (QFT) and Condensed Matter (CM) systems, so naturally, one might expect to find this overlap in quantum computing as well [6–10]. However, with the difficulties of decoherence present in quantum wires [11], there has been little to no progress in advancing technologies that would allow such a collaboration.

The seemingly obvious application of QFT interactions for Quantum Informa-tion Science (QIS) purposes, is in the arena of photonics, where there has been a large effort in 1-D quantum state transport through waveguides [12–17]. On the other hand, lesser-known fermionic application provides a means to probe quantum communication in quantum materials [18] utilizing QFT applications of condensed matter systems. Further extensions to these models include QIS in Valleytronics [19], Moìre pattern materials [20], and doped Transition Metal Dichalcogenides [21–23], all of which qualify as models that could be used to couple spin-qubits to quantum fields for transduction purposes, but introduce new technologies for quantum wires.

Regardless, It remains a challenge to theoretically evaluate Quantum Information (QI) rigorously in quantum materials [24]. Topologically protected edge states generated by the Anomalous Quantum Hall Effect (AQHE) give rise to phenomenon such as the edge states of LLs, transport along graphene nanoribbons, and transport on the edges of Moiré patter materials [18]. These materials offer promising arenas for carrying out quantum computing [24,25]. However, to benchmark these computers theoretical tests of entanglement processing is necessary.

When information is processed through a classical computer, every step of the way, we can measure and conclude how this information has changed. Quantum mechanics forbids this knowledge during quantum computing processes. Once the classical input has entered the quantum circuit and has transformed into quantum information, any measurement of that system will cause entanglement decoherence of the measured qubits and subsequently some, if not all, of the quantum information will be lost [26]. The act of measuring will revert our quantum information back into classical information and without knowledge of the full Quantum State

Tomography of the system one cannot hope to know how the information has been changed [5].

It is well understood, in quantum computing, that encoding information (such as spin) via entanglement of particles, and measurements of the entanglement as it propagates through the quantum circuit happen stochastically through interactions with the environment [4]. Quantum error correcting codes have shown that highly entangled states provide better information propagation [6, 27–29]. Therefore, a key aspect to building quantum computing systems resides in the ability to generate and maintain entanglement between qubits.

Furthermore, transduction of this information via quantum fields can be utilized similarly to classical information, for long range communication, with the restriction of 1-D motion. Field-mediated quantum gates play a natural role in the theory of quantum computing and quantum networking, specifically in the regime of ground-to-satellite communication [30–36]. Long-range communication of photons in non-inertial frames of reference, such as the gravitational well of the earth, indicate the need for relativistic theories to explore relativistic effects [37, 38]. It is easy to bog oneself down in the details of these relativistic theories. Formulating and presenting comparisons of channels utilized in RQI and canonical channels allows practitioners to forego the details of these theories.

## 1.2 Quantum Troubles With Infinities

Quantum Mechanics has the unusual way of working out beautifully in some infinite limits, and completely breaking down in others. For example, to achieve unitarity in measurement theory, one must sample infinite measurements, a com-plication that does not exist in the Everettian interpretation of quantum mechanics [39]. However, in a laboratory setting, infinite measurements do not exist, yet generalizations of quantum state tomography, such as the theory of classical shadows, are able to approximate unknown states with calculable errors [40–42]. So long as limits of infinity are not able to be evaluated, the true quantum state tomography of a system can only ever be approximated with these errors. A brief overview with applications that show the strength of classical shadows are presented in Appx. A.

Another source of complications with quantum mechanics and infinity, is in regards to a relativistic setting. QFT introduced a perturbative methodology, that recovered (special) relativistic invariance, where accurate dynamics were modeled through summing over infinite interactions between continuous fields on Lorentz invariant backgrounds. These infinite interactions yielded infinite energies that require renormalization to match experimental observations [43–46]. Despite these complications, QFT proves to be one of the most accurate descriptions of reality and connecting the formalism of QFT to quantum computing has been a high-priority task for the last few decades through simulations [7] or transduction.

Throughout this project we identify multiple ways that infinite limits work well to realize the theory, but are not realistic in a laboratory setting. We describe the ways that our systems have side-stepped these problems through effective field theory limitations and non-perturbative solutions. However, a theory of renormalization, complete with the structure of RG flow in the Kondo model, is necessary to overcome the historical issues presented within the following models. As this thesis is an introduction to QIS properties of a novel system we save this investigation for future work.

## 1.3 Unruh–DeWitt Detectors and Quantum Computers

Over the past several decades one approach to unite the fields of QI and QFT has been accomplished utilizing tools from cosmology, namely Unruh–DeWitt detectors. As a subset of the field Relativistic Quantum Information (RQI), Unruh-DeWitt (UDW) detectors model interactions between spin structures and quantum fields [2, 47–54].

Originally proposed as two-state systems coupled to quantum fields, UDW detectors are used to model the detection of radiation in accelerating/non-inertial frames of reference [55–57]. Extensions of this model have shown that spin-qubits coupled to scalar fields for encoding and decoding quantum information, can be used to formalize quantum information channels and measure quantum Shannon properties such as quantum capacity [2, 18, 48, 58].

As mentioned above the two necessary components of quantum computing are entanglement generation and propagation. UDW detector models have realized both. Firstly, the generation of entanglement has been shown through extensive work in a process called entanglement harvesting [47, 48, 52, 59–63]. More simplistically, this can be demonstrated through singular unitary logic gate operations that will be formalized in Chapter 2.

Secondly, entanglement propagation via. state transfer [58] has been demonstrated through aforementioned quantum Shannon theory measurements such as von Neumann entropy, quantum coherent information and quantum capacity [2]. This process is understood to happen through applications of multiple controlled unitaries [47] and therefore can draw some analogy to the canonical state transfer channel which is constructed using two CNOT gates [5,64]. Moreoever, we demonstrate other quantum logic gates that can be constructed using UDW unitary gate operations such as the SWAP gate. Extensions of this kind can lead to other insights in regards to quantum communication, such as the diamond distance norm (also known as the diamond norm) [4, 5, 65].

## 1.4 Unruh–DeWitt Quantum Communication Theory

The diamond norm is a comparison measure of distinguishability between two quantum channels and becomes especially useful when constructing novel quantum channels that may have unique benefits, and relating them to well known quantum channels that may prove to be cumbersome [4, 66, 67]. A diamond norm of zero indicates that the channels in question result in the same output density matrix with zero distinguishability. A product of having UDW quantum channels written in a language analogous to canonical quantum computing, is that the diamond norm can measure the "idealness" of a UDW quantum channel. The UDW channels are set up to specifically model qubit-field interactions, but as we will show, in limits of strong coupling this diamond distance proves to demonstrate that the quantum state transfer channel, associated with UDW detectors is identical to that of the canonical state transfer.

Formalizing UDW quantum gates as analogous quantum computing gates has other benefits in quantum Shannon theory. One of the accomplishments of this thesis, is the recognition that the UDW state transfer channel can be constructed in a bosonic dephasing channel perspective. Sociologically, dephasing channels are reserved to describe the decoherence in multipartite systems through the systematic elimination of non-zero off-diagonals in the density matrix. However, we will show that the dephasing of the off-diagonals in the coherent state denstiy matrix, the computational basis of a quantum field, are exactly the intention of a functional UDWQC. For this reason we use terms such as "dephasing like" to avoid any sociological confusing.

RQI channels have a reputation for being tedious and lengthy in calculation. With correlation functions (vacuum expectation values) of eight or more vertex operators in some of the most simplistic situations. To make matters worse, if one were to consider noisy interactions in RQI quantum channels this number of vertex operators grows in size. These lengthy correlators can be not only operationally complicated, there are nontrivial computational considerations. However, as we will show in Chapter 5, in bosonic dephasing formalism, noisy calculations can be simplified by adjusting the dephasing parameter, a functional parameter in terms of three parameters; smearing, switching, and coupling between field and qubits. This simplification avoids all extensions to the complicated correlation functions, but the noise-free correlators will still need to be calculated regardless.

## 1.5 Physical Realizations

There are extensions to the model proposed in this thesis via two potential systems for realizing the study of quantum information flow in quantum materials: CdTe-HgTe-CdTe heterostructures in the quantum spin Hall phase [68, 69] and doped Transition Metal Dichalcogenides (TMDs) and their applications in either the anomalous or spin quantum-Hall regime [22, 23, 70, 71]. These systems contain spin qubits coupled to topologically protected edge states and allow a theoretical analysis combining the formalism of UDW detectors with the bosonization of fermionic systems. This theory is restricted by the ability to engineer these systems for exploring quantum information flow through quantum fields. Building on the previous work of Simidzija *et al* [2, 47, 48], we show that a non-zero channel capacities are possible in our model. Furthermore, we provide pedagogical approach to evaluate quantum Shannon theoretic values with numerical simulation. Through this pedagogy, one can extend our model to evaluate these properties of the above experimental realizations.

### 1.5.1 Case One: Luttinger Liquids and Quantum Dots

The first extension of the UDWQC is through coupling quantum dots (Qdots) of CdTe to a helical LL of HgTe [72]. A cartoon representation of this, given by Fig. 1.1 depicts how the Qdots, isolated by a potential well, can be forced to interact with the topologically protected edge states of the LL by lowering the potential barrier that separates the two. Modelling the Qdots as Kondo impurities, we recreate the two-state system common in the UDW detector formalism.

Developing the technology necessary to achieve this a formidable problem.

Luttinger liquids have only been realized experimentally in recent years in HgTe and developing a system that maintains stable edge states that reliably interact with impurities will be very challenging. Furthermore, in this thesis we focus on impurity scattering of the fermionic free theory. This is not a practical interaction block for realizing this theory physically, though extensions to other interaction blocks of the Tomonaga-Luttinger liquid are straightforward as demonstrated in Chapter 3. Instead, we aim to show the effectiveness of quantum Shannon theory on our model, which is system independent.

### 1.5.2 Case Two: Doped TMDs

Doped Monolayer TMDs are a far more stable and allow for first principle investigations into the aforementioned interests of the CdTe-HgTe-CdTe model. Dopants, such as vanadium atoms, display signs of mid-gapped, optically excitable [23] states, which makes them good candidates for qubits. Furthermore, doped monolayer TMDs such as Vanadium Tungsten Disulfide (V-$WS_2$), display room temperature ferromagnetic moments at 4% doping [73]. We can infer the qubit nature of the dopants by examining the ferromagnetic properties using experimental techniques such as cantilever magnetometry.

The stability of doped atoms as qubits would allow for bilayer stacking and interlayer communication between dopants as a first approach to examine the spin-qubit communication in a UDWQC setting. In the future, replacing one of the layers with a material such as MoTe, a material that experiences a topologically protected edge state, will allow for investigations of qubit-field transduction for quantum communication over distances in these fermionic systems.

With these two experimental setups, near-term investigations into UDWQCs provide insights into fermionic flying qubits, thermalization of quantum materials, and general transduction properties of qubit-field systems.

## 1.6 Outline of Results.

In Chapter 2, we outline the idealized model we aim to study throughout this program: Kondo-like impurities coupled to the topological edge states of the Luttinger liquid model. While these models are well-understood, our approach introduces important subtleties such as rapid switching functions, as seen in Eq. 2.9, and Gaussian-like smearing. These terms are necessary to utilize the current UDW program for RQI purposes and may potentially alter the physics of the standard system.

In Sec. 2.3.2, we provide an overview of the RQI program outlined in Ref. [2]. This involves systematically developing UDW quantum gates designed to achieve the Quantum State Transfer (QST) operation, as discussed in Sec. 2.3.3. The QST channel facilitates the unidirectional transfer of quantum information from one state to another. Chapter 4 offers detailed insights into the QI aspects of this process, including scenarios involving entanglement-breaking channels with unitaries similar to those described in Eq. 2.19.

In Chapter 3, we explore the process of bosonization as it applies to Luttinger liquids and introduce the entire Tomonaga-Luttinger liquid model. We highlight how different interaction blocks apply to various physical systems of interest beyond our idealized model. Furthermore, we illustrate the bosonization procedure using point-splitting methods in Sec. B and extend this procedure to other blocks of the Hamiltonian, as outlined in Table 3.1. If one intends to apply the results of this thesis to physical systems, the utilization of this bosonized block-Hamiltonian will likely be necessary.

In Chapter 5, we present our first major result of the thesis: Our idealized model of a Kondo-like impurity coupled to a helical Luttinger liquid can be bosonized to generate the same unitaries as the QST channel. With this formalism in mind we begin to explore our UDW quantum gates and the applications to quantum Shannon theory. We revisit the pedagogy explained in Chapter 2, but this time with the intent to show the similarities it possesses to that of bosonic dephasing channels. In fact, one can demonstrate, as we do in Eqs. 5.18 and 5.19, that the parameters of the UDW model can be wrapped up into a dephasing parameter which accomplishes the usual goal of minimizing off-diagonals in the coherent state density matrix.

Three key findings emerge through the lens of the UDW gate model resembling dephasing processes. Firstly, in Sec. 5.6.1 we show that interacting with the environment following the usual dephasing prescription introduces no additional noise to the system so long as we remain working with field-mediated channels. Secondly, in this dephasing-like perspective, introducing additional detectors in a form of cross-talk interactions do not necessarily contribute field-observables to the already cumbersome correlator in Eq. 2.25. Instead one can utilize the dephasing parameter to account for the effects of additional detectors, significantly reducing computational resources. Lastly, evaluating these cross-talk channels have provided insights into unwanted additional noisy interactions. One surprising result is that it is possible to craft such a cross-talk interaction that increases the lower bound of the coherent information as demonstrated in Fig. 5.3. These results can be found in detail in Ref. [58].

In Chapter 6, we elucidate the projector form of UDW quantum gate formalism and compare it to that of canonical logic gates. Under this prescription, we are able to generate UDW quantum logic gates such as CNOT, Hadamard $H$, $S$, and $T$-gates (See Sec. 6). This specific set of gates has the well-known property of being universal. In other words, one can utilize these quantum gates to generate any operation on a quantum computer. To prove the effectiveness of our new UDW logic gates, we turn once again to the tools of quantum Shannon theory, namely the diamond distance.

The diamond distance is a measure of distinguishability between quantum information channels. A diamond distance of one indicates that the channels are completely distinguishable, whereas a value of zero indicates indistinguishability [4]. Our first step to show effectiveness is to compare this value directly to the coherent information of our QST channel as these results were previously shown. As one can see in Figs 6.6 and 6.7, the diamond norm captures this indistinguishability quality of our UDW field-mediated QST channel when compared to the qubit-qubit QST channel. We further utilize the diamond distance metric to verify our field-mediated CNOT gates in several arrangements in Figs. 6.9 and 6.10. The primary results of this chapter can be found in Ref. [1].

A surprising consequence of the culmination of work thus far is that with the introduction of universal quantum computing gates, we have theoretically achieved all seven of the DiVincenzo criteria of quantum computing [74]. These criteria, laid out in Chapter 7, are well-known requirements of a system to carry out quantum computations. With these criteria in place, we describe future work to realize options for first-principle investigations.

We close this thesis by describing in detail current and future projects that aim to tackle several of the main open problems introduced throughout this work. Furthermore, we demonstrate the wide range of impact this work seemingly influences, including topics like faster-than-light signaling and black hole evaporation.

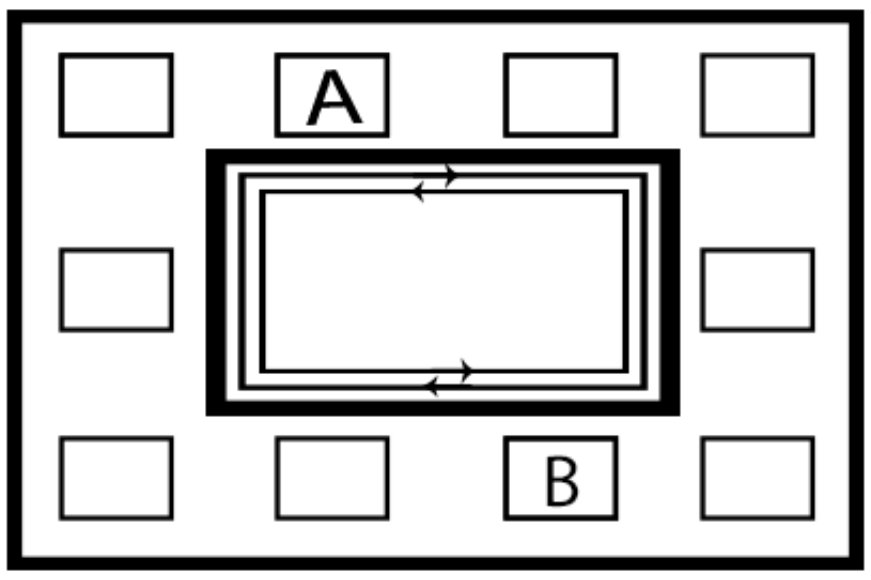

(a) Qubit A can access all qubits on the system *via* left- and right-movers.

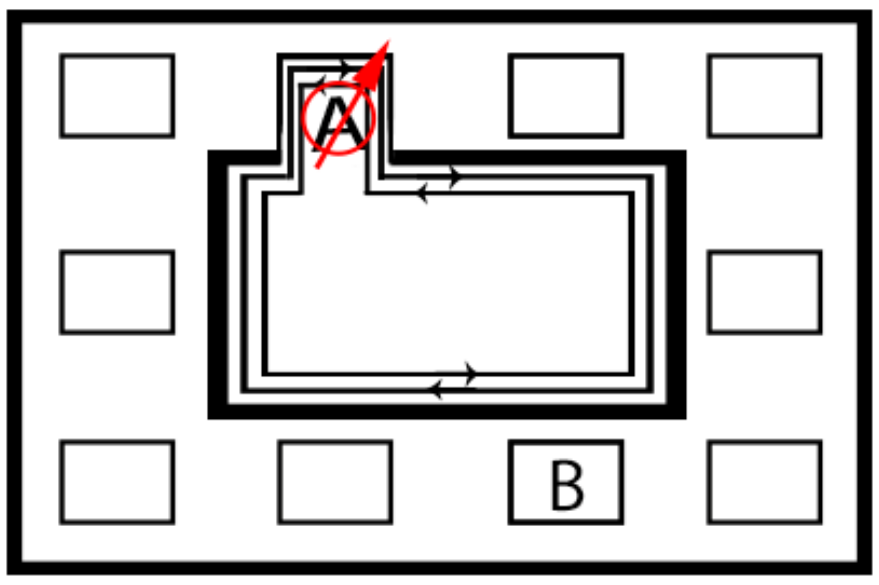

(b) Opening up qubits will cause interactions with the left- and right-movers.

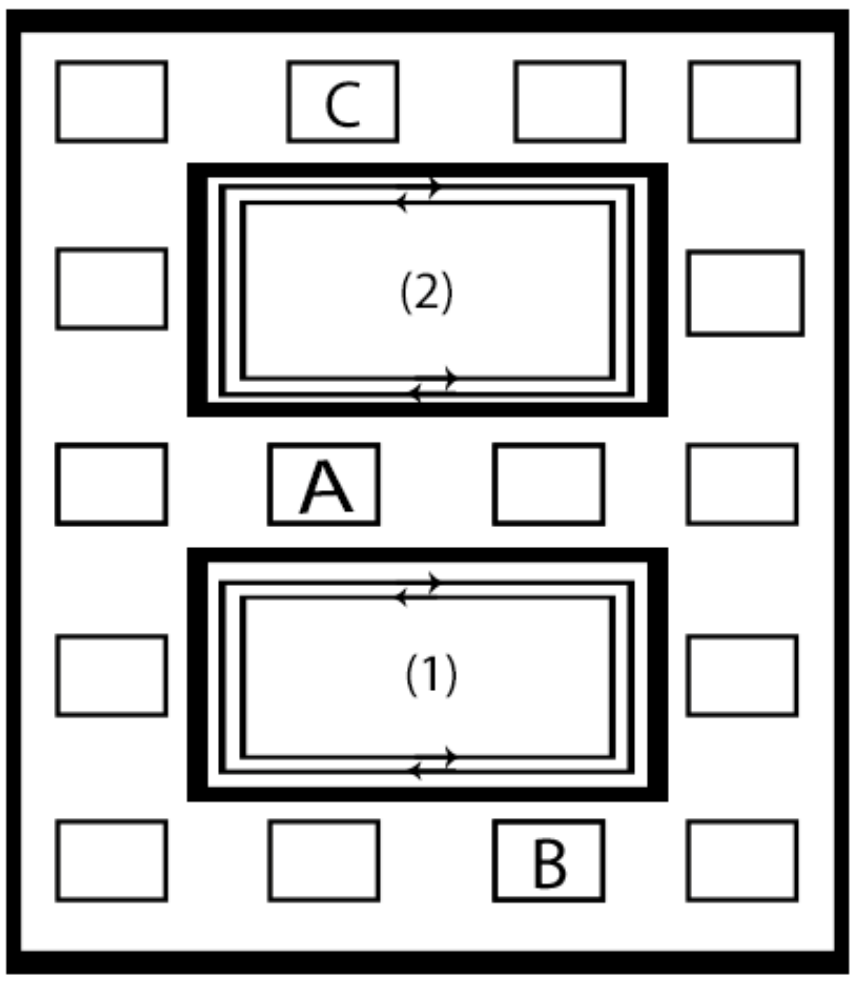

(c) Adding another block of qubits restricts of our left- and right-movers to paths (1) or (2).

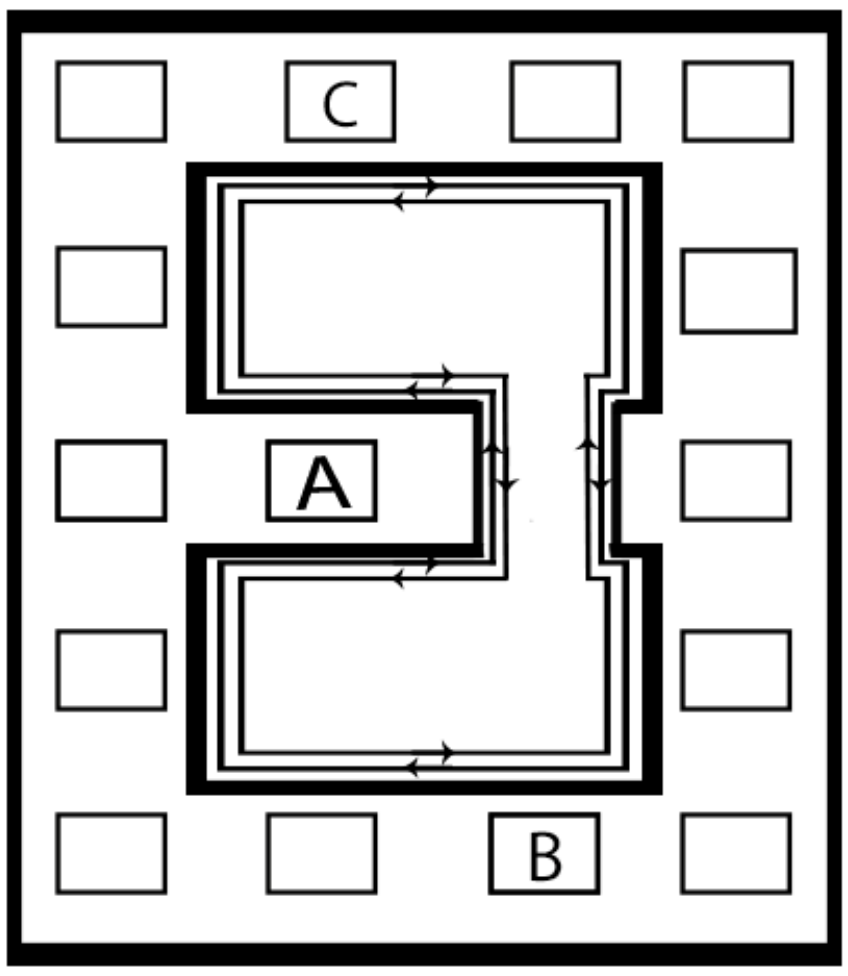

(d) Opening the bulk between the blocks gives Qubits B and C direct access to the entirety of the qubits.

Figure 1.1: Figures (a-d) show a simplistic view of our quantum bus. Qubits (such as qubits A and B) are placed around the edges of a Luttinger liquid. The potential wall is lowered, as in Fig. (b) and the qubits are able to interact with the topological edge states. Figures (c) and (d) indicate how scaling up this system can be done easily by adjusting where the bulk of our fields lives through raising and lowering the potential barrier.

# Chapter 2

# The Unruh-DeWitt Detector: The Quantum Computing Brand

## 2.1 History and Definitions

Simulations in quantum field theory (QFT) have long been expected to be a practice of quantum computers (QC) [6–10]. Sociologically, this means carrying out lattice QFT calculations using standard qubits. We hope to address a different style of QC+QFT. Namely, field-mediated QC with UDW detectors as two level systems for readouts. However, a physically realizable experiment has many difficulties to get through. Including faster-than-light signaling and information loss via entanglement decoherence [2, 50, 75, 76]. Massless Dirac fermions evaluated in UDW detector models have been a promising endeavor for relativistic quantum information processes [77–79]. Right- and left-movers of Helical Luttinger Liquids (HLL) (which have been experimentally realized on 1+1 dimensional cylindrical space-times [68, 69, 71]), provide the right flavor of flying qubit to carry out these computations, but offer a difficult arena for calculations as we'll show in Chapter 3.

Unruh–DeWitt detector models were originally designed to detect radiation,

apparent only in an accelerating frame of reference [55, 56]. However, the same framework accurately represents stationary qubits coupled to quantum fields. Through the smearing and switching functions of the UDW detector model, quantum information can be propagated and generated through the fields and collected by another detector in a process known as entanglement harvesting [80]. The implementation of UDW models into quantum computing has led to insights in information transformation in quantum computing. To understand the relevance of detectors in quantum computing we start from the ground up. We'll need to establish a Hamiltonian in a form that is friendly to quantum computing. To do this we will spend a fair deal of this chapter dealing with definitions, complications, and resolutions that make this system possible.

## 2.2 Field Observables and Monopole Moments

The process of encoding and decoding information onto fields in a quantum circuit, like that of figure 2.1, requires understanding the system and its constraints. In this thesis, we are restricted to 1+1 dimensional studies to avoid complications such as the effects of Huygens' principle which is reserved for higher dimensional spacetimes and results from the radial motion of quantum information through processes of thermalisation [2]. Furthermore, considering topological edge states of helical luttinger liquids behave similar to photonic wave guides for the right- and left-movers, our treatment can be extended, without loss of generality to photonics. This fact further exemplified in Chapter. 3 where we demonstrate that at the level of field-mediated quantum channels, we can treat fermions and bosons as equivalent through the process of bosonization [81].

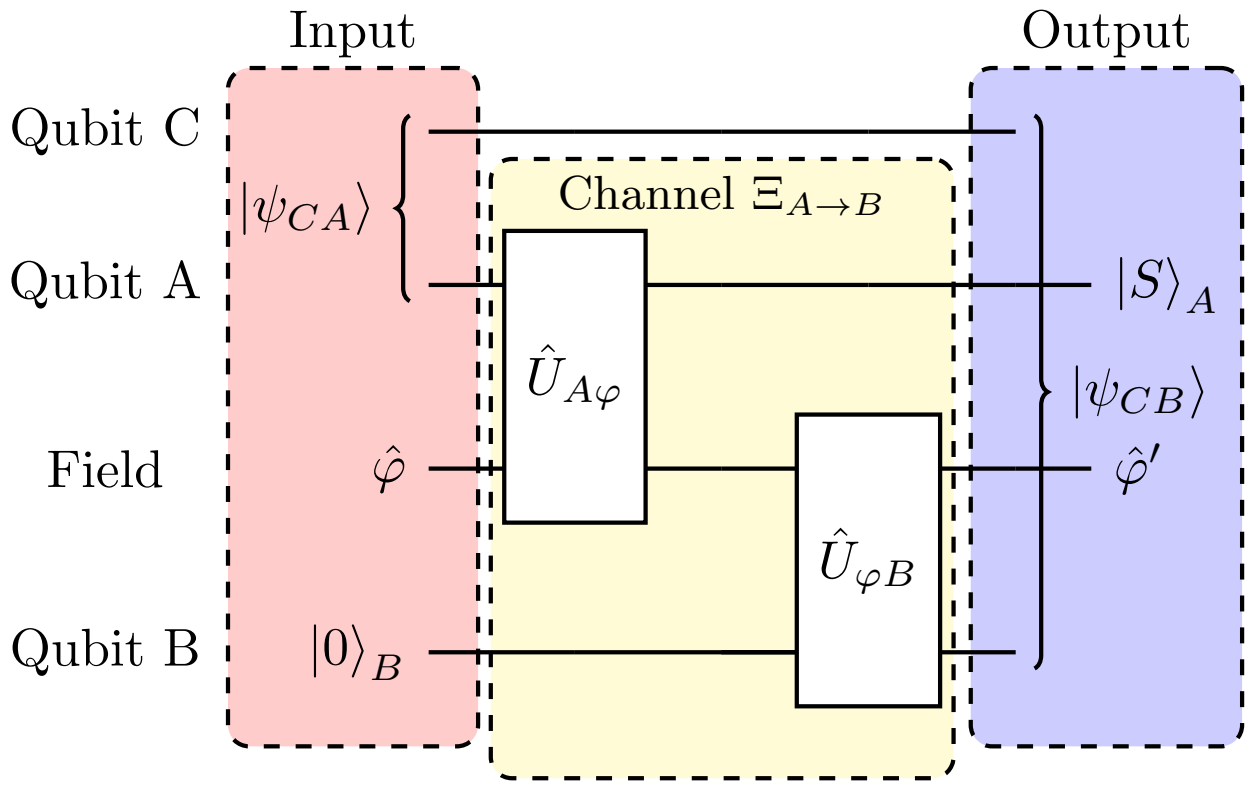

Figure 2.1: Operators $\hat{U}_{A\varphi}$ and $\hat{U}_{\varphi B}$ are employed to encode and decode quantum information onto and off of the field $\hat{\varphi}$ thus transferring the quantum information from qubit A to qubit B.

### 2.2.1 The Field and Conjugate Momentum

Let's begin by establishing some definitions and procedures. Starting with the scalar field and the conjugate momentum in the interaction picture,

$$\hat{\varphi}(x,t) = \int \frac{dk}{\sqrt{4\pi k}}[\hat{a}(k)e^{i(\omega t-kx)} + \hat{a}^\dagger(k)e^{-i(\omega t-kx)}] \tag{2.1}$$

$$\hat{\pi}(x,t) = \int \frac{dk}{\sqrt{4\pi k}}[-ik\hat{a}(k)e^{i(\omega t-kx)} + ik\hat{a}^\dagger(k)e^{-i(\omega t-kx)}] \tag{2.2}$$

where $\hat{a}^\dagger(\hat{a})$ are the usual creation (annihilation) operators that follow the standard bosonic commutation relations $[\hat{a}_i, \hat{a}_j] = [\hat{a}_i^\dagger, \hat{a}_j^\dagger] = 0$ and $[\hat{a}_i, \hat{a}_j^\dagger] = \delta_{i,j}$. When working with qubit-field interactions, it is common procedure to work with smeared out observables $\hat{\mathcal{O}}$ with form

$$\hat{\mathcal{O}} = \int dx(F_1(x\hat{\varphi}(x,t) + F_2(x)\hat{\Pi}(x,t)) \tag{2.3}$$

with smearing functions $F_1$ and $F_2$. Moving forward we will pick up the common notation that

$$\hat{\mathcal{G}}(F) = \int dx F(x)\hat{\mathcal{G}}(x,t). \tag{2.4}$$

For the 1+1 HLL, the smearing fuctions $F_1$ and $F_2$ are assumed to be equal. This not the case generally.

Sticking with the system of HLL coupled to Kondo impurities we can write down a naive detector interaction Hamiltonian defined by

$$\hat{H}_{\text{int}}(t) \;=\; \chi(t)\int_{\mathbb{R}} dy \;\; p(x(t),y)J_{\alpha,z}\hat{\mu}_{\alpha}(t) \otimes (\hat{\psi}_{+}^{\dagger}(y)\hat{\psi}_{+}(y) \;-\; \hat{\psi}_{-}^{\dagger}(y)\hat{\psi}_{-}(y)). \quad (2.5)$$

Here $\hat{\psi}_{-}$ ($\hat{\psi}_{+}$) denote our spin-down, left-moving (spin-up, right-moving) fermions (We discuss the structure of these fields in Chapter 3). This interaction term assumes the Qdot limit where factors of $e^{\pm 2ik_F y}$ have been suppressed as we restrict $\sigma >> \frac{1}{k_F}$ [82]. Notice that Eq. 2.5 has many familiar properties, but the introduction of the smearing and switching functions are common in UDW formalism [83]. Since this model includes a Kondo impurity, some may recognize our two-level system, which we treat as the qubit as,

$$J_{\alpha,z}\hat{\mu}_{\alpha}(t) = J_{xz}\hat{S}_x(t) + J_{yz}\hat{S}_y(t) + J_{zz}\hat{S}_z(t) \quad (2.6)$$

$$J_{xz}e^{i\hat{H}_0' t}\hat{S}_x e^{-iH_0' t} = \frac{J}{2}(\hat{S}_{+}e^{-i\Omega t} + \hat{S}_{-}e^{+i\Omega t}) \equiv J\hat{\mu}(t) \quad (2.7)$$

where $\hat{S}_i$ is the projector onto some spin state and $\Omega$ is some real valued number that indicates a non-trivial energy difference between spin states. The second equality follows by choosing $J_{\alpha,z} = JX_\alpha$ to point along a new $\hat{X}$ direction and time dependence generated by Hamiltonian $\hat{H'}_0 = -g\hat{\mu}_B\vec{B}\cdot\vec{S} \equiv \Omega\hat{S}_Z$ with $\hat{Z}$ perpendicular to $\hat{X}$. We've added the magnetic field to show that this system is a two-level system given by,

$$\hat{\mu}(t) = |1\rangle\langle 0|e^{-i\Omega t} + |0\rangle\langle 1|e^{i\Omega t} \tag{2.8}$$

It should be noted here that coupling to both the field and the conjugate momentum separately is necessary to ensure that our channel is not entanglement breaking. We will demonstrate that the resulting channel has a zero channel capacity which is consistent with the literature. Furthermore, the energy gap $\Omega$ cannot be vanishing, or the interaction will not happen.

Our formulation takes the usual convention of a HLL and implants it into the UDW model to describe a quantum channel between distant spin-qubits communicating via Dirac fermions. It allows for an elegant promotion to a quantum circuit model by constructing the Hamiltonian in the following way

$$\hat{H}_{\text{int}}^{(\text{r},\nu)}(t) = J\chi(t)\hat{\mu}(t) \otimes \hat{\mathcal{G}}(F)^{\dagger}_{\nu r}\hat{\mathcal{G}}(F)_{\nu r}. \tag{2.9}$$

where $\nu$ specifies the interacting spin-qubit ($\nu \in \{A, B\}$ in Fig. 2.1), $r \in \{\pm\}$. The delta-like switching function $\chi(t)$, can be seen as the instance of time $t_\nu$ of interaction for qubit $\nu$ such that we can express the coupling as $J_\nu = J\chi(t_\nu)$ . We can then construct a set of unitary operators $\hat{U}_{A\mathcal{G}}$ and $\hat{U}_{\mathcal{G}B}$ that take the form

$$\hat{U}_{\nu O} = \exp\left(-iJ_\nu\hat{\mu}_\nu(t) \otimes \hat{\mathcal{G}}(F)^{\dagger}_{\nu r}\hat{\mathcal{G}}(F)_{\nu r}. \tag{2.10}$$

Since we wish to work in a strong coupling regime, our choice of field interactions that exist in the literature is far less. One that is commonly used is a scalar boson field that couples linearly to the two state moment $\mu(t_\nu)$ [48]. In chapter 3 we discuss other possibilities for exploring gates including strong coupling of quadratic interactions which is not well understood in the literature. For now we establish the common formalism of the usual scalar fields of the strongly coupled UDW detector system.

## 2.3 Coherent States Developing Constraints

### 2.3.1 Coherent States

A quick reminder of the 1-D time-independent scalar field $\hat{\varphi}(x)$ and it's associated conjugate momentum $\hat{\Pi}(x)$ are given as

$$\hat{\varphi}(x) = \int \frac{dk}{2\pi}\sqrt{\frac{v}{2\omega(k)}}[\hat{a}(k)e^{ikx} + \hat{a}^{\dagger}(k)e^{-ikx}] \tag{2.11}$$

$$\hat{\Pi}(x) = \int \frac{dk}{2\pi}\sqrt{\frac{\omega(k)}{2v}}[-i\hat{a}(k)e^{ikx} + i\hat{a}^{\dagger}(k)e^{-ikx}]. \tag{2.12}$$

In this continuous space, these field observables can be expressed as displacement operators $D[\alpha(x)]$ with a point-wise coherent amplitude

$$\alpha_{\varphi}(x) = \sqrt{\frac{v}{2\omega(k)}}e^{-ikx}. \tag{2.13}$$

For convenience we have introduced subscripts to differentiate between coherent amplitudes of the field and conjugate momentum observables. A Fourier transform following the usual prescription

$$f(k) := \frac{1}{\sqrt{(2\pi)}}\int dx f(x)e^{ikx} \tag{2.14}$$

allows us to express a continuous displacement operator of a scalar bosonic field [48] as

$$\exp\left(\pm i\hat{\varphi}\right)|0\rangle \;=\; \hat{D}[\alpha(k)]\,|0\rangle \;=\; \exp\left(\int dk\,\left[\alpha(k)\hat{a}_k^\dagger - \alpha(k)^*\hat{a}_k\right]\right)|0\rangle \;\equiv\; |\pm\alpha\rangle\,. \tag{2.15}$$

The final equivalence is a notational convenience we will adopt for the majority of this thesis when carrying out simulations. This notational convenience can be understood as single mode excitation of a continuous spectrum in a very idealized sense. For now, as we continue to introduce structure, we proceed with the more general form.

These field observables follow the equal-time canonical commutation relations $[\hat{\varphi}(x), \hat{\varphi}(x')] = 0$, $[\hat{\Pi}(x), \hat{\Pi}(x')] = 0$, and $[\hat{\varphi}(x), \hat{\Pi}(x')] = i\delta(x-x')$ and consequently produce displacement operators that obey commutation relations

$$[\hat{a}_\varphi(k'), \hat{D}(\alpha_\varphi(k))] = \alpha_\varphi(k')\hat{D}_\varphi(\alpha(k)). \tag{2.16}$$

### 2.3.2 Unruh–DeWitt Quantum Gates

A unitary operator in the form of a quantum gate can generate and/or propagate quantum information through a quantum circuit [5]. We expect this same consequence to be achievable with UDW detector models [2]. Incorporating our scalar field from Eq. 2.11 we present the well known UDW detector model as

$$\hat{U}_{UDW}(t_1, t_2) = Te^{-i\int_{t_1}^{t_2} dt\hat{\mathrm{H}}_{int}(t)} \tag{2.17}$$

with

$$\hat{\mathrm{H}}_{int}(t) = J\chi(t)\int_{\mathbb{R}} dx\ F(k)\hat{\mu}(t)\otimes\hat{\varphi}(k,t). \tag{2.18}$$

where $\chi(t)$ and $F(k)$ are the switching and smearing functions respectively and $J$ is an effective coupling constant that indicates what gate is interacting at what time and $\hat{\mu}(t)$ is a two state (qubit) detector that has the form Eq. 2.7. By using a delta-like switching function we can indicate the time ordering of our unitary and find that it simplifies to

$$\hat{U}_{UDW} = \exp\left(-iJ\hat{\mu}\otimes\hat{\varphi}\right). \tag{2.19}$$

The structure of $\hat{\mu}$ as discussed above is that of a qubit observable and for this we can rewrite the gate by introducing projectors $\hat{P}_s$ onto the eigenstates of $\hat{\mu}$ as

$$\hat{U}_{UDW} = \sum_{s\in\pm}\hat{P}_s\otimes e^{is\hat{\varphi}(F)}. \tag{2.20}$$

In this form we see that $\hat{U}_{UDW}$ is like a controlled unitary gate, it acts with one unitary on the field in the Hilbert of the qubit and another unitary in the Hilbert space of the field. This is part of the "encoding" process that imparts the quantum information from the qubit to the field. Here we have redefined the field observables to include the smearing and coupling of the UDW model

$$\hat{\varphi}(F) \coloneqq J_{\varphi}\int dxF(k)\hat{\varphi}(k,t) \tag{2.21}$$

$$\hat{\Pi}(F) \coloneqq J_{\Pi}\int dxF(k)\hat{\Pi}(k,t). \tag{2.22}$$

One of the concerns of this theory is the way in which we treat the Hilbert space of the field. For instance, we are allowing coupling between a Hilbert space of dimension two $\mathcal{H}_A$ , with an infinite dimensional Hilbert space $\mathcal{H}_\phi$ in the full space of $\mathcal{H} = \mathcal{H}_A \otimes \mathcal{H}_\phi \otimes \mathcal{H}_B$ where $\phi$ is a scalar field. In the literature, it is common to treat the Hilbert space of the field as two-dimensional coherent state Hilbert space, a computational basis of the quantum field. This is a result of infinitely large coupling strength and Gaussian-like spatial smearing. In reality, these things are non-physical. However, for significantly large coupling the extra dimensions of the Hilbert space minimally contribute in some calculable way, and can be accounted for as noise.

Equation 2.20 captures one of the two criteria discussed in the beginning of this section, namely entanglement generation. To accomplish the propagation of QI and establish metric of effectiveness of that communication we'll need to evaluate a Quantum State Transfer (QST)

### 2.3.3 UDW State Transfer Channel

#### Setting the Channel Up

For a comprehensive review of quantum information channels and the history and pedagogy involved, see chapter 4. It was shown in Ref. [2], that to write down a channel that preserves entanglement and performs a state transfer, we would need two of the controlled unitaries presented in Eq. 2.20. This will lead to the unitary

$$\hat{U}_{\nu\varphi} = \sum_{z,x\in\pm} \hat{P}_x \hat{P}_z \otimes e^{ix\hat{\Pi}(F)_\nu} e^{iz\hat{\varphi}(F)_\nu} \tag{2.23}$$

where we introduce the notation that $\hat{P}_x$, with an $x$-index, and $\hat{P}_z$, with a $z$-index, are the projection operators onto the eigenstates of Pauli matrices $\hat{\sigma}_x$ and $\hat{\sigma}_z$ respectively and $\nu$ indicates the qubit interacting with the field at a given time $t_\nu$. Figure 2.1 presents an idealized circuit diagram for this channel which has the mathematical structure

$$\Xi_{A\to B} = \mathrm{Tr}_{A\varphi}[U_{A\hat{\varphi}}U_{\varphi B}(\hat{\rho}_{A,0}\otimes\hat{\rho}_{\varphi}\otimes\hat{\rho}_{B,0})U^{\dagger}_{\varphi B}U^{\dagger}_{A\varphi}] \tag{2.24}$$

and can be expanded out as

$$\begin{aligned}\Xi_{A\to B} = \\ &\sum_{l,l',x_i,z_i}\langle 0|e^{iz_1\hat{\varphi}_A}e^{ix_1\hat{\Pi}_A}e^{ix_2\hat{\Pi}_A}e^{iz_2\hat{\varphi}_A}\\ &\times e^{iz_1\hat{\varphi}_A}e^{ix_1\hat{\Pi}_A}e^{ix_2\hat{\Pi}_A}e^{iz_2\hat{\varphi}_A}|0\rangle\\ &\times\langle l'_z|\hat{P}_{-z_1}\hat{P}_{-x_1}\hat{P}_{x_4}\hat{P}_{z_4}|l_z\rangle_C\langle l'_z|\\ &\otimes\hat{P}_{-z_3}\hat{P}_{-x_3}|+_y\rangle_B\langle +_y|\hat{P}_{x_2}\hat{P}_{z_2}.\end{aligned} \tag{2.25}$$

The correlator in Eq. 2.25, which is a result of the trace over the coherent states of the field, is a tricky value to calculate. One may recognize the correlator as a vacuum expectation value of eight vertex operators, often utilized in conformal field theory and string theory. If utilizing coherent states, we recognize that they will lead to non-orthogonal inner products $|\langle +\alpha(k)|-\alpha(k)\rangle|$ that contain unwanted states, that if we could dephase away, would improve the coherent information of the channel. Furthermore, any additional interactions on the field will only yield a larger correlator. In chapter 5 we also explore the possibility that a dephasing perspective may eliminate the necessity to increase the size of the correlator for additional UDW interactions.

**Choosing the Correct Parameters**

Within Eq. 2.18 and Eqs. 2.21 and 2.22, there are still a few free parameters remaining. Namely, the coupling constants $J_\varphi$ and $J_\Pi$ as well as the smearing function $F(k)$. We therefore utilize these parameters to remove unwanted states that contribute to the noise in the qubit-field interactions.

These unwanted states show up in two ways, firstly during the application of $e^{ix\hat{\Pi}}$, which when applied to a coherent state will generally change the value of the state, and secondly with non-zero values of the inner product $|\langle +\alpha(k)| - \alpha(k)\rangle|$. In RQI literature, it is common to redefine the coherent amplitude, given in Eq. 2.13, to include the coupling constants and the smearing function. By doing this we can utilize equation 2.16 and the following identity

$$e^{\pm i\Pi}\hat{a}_k e^{\mp i\Pi} = \hat{a}_k \mp \alpha_\Pi(k) \tag{2.26}$$

to prove the relation

$$\hat{\Pi}\,|\pm\alpha(k)\rangle = \pm\gamma\,|\pm\alpha(k)\rangle + e^{i\Pi}\Pi\,|0\rangle \tag{2.27}$$

which in the limit of $\gamma^2 \gg \langle 0|\Pi^2|0\rangle$ gives us the following constraint in (1+1) dimensions

$$\left(J_\varphi \int dk\,|\tilde{F}_\nu(k)|^2\right)^2 \gg \frac{1}{2}\int dk\,\omega(k)\,|\tilde{F}_\nu(k)|^2. \tag{2.28}$$

This constraint specifically allows for the application of $e^{ix\Pi}$ to a coherent state and results in a phase constant (see Appx. C for a demonstration of this constraint.). Furthermore, constraining this phase constant by

$$\Gamma := J_{\Pi} J_{\varphi} \int dk \, |\tilde{F}_{\nu}(k)|^2 = \frac{\pi}{4} \bmod 2\pi \tag{2.29}$$

results in a Bloch rotation of the output qubit A's states that eliminates the first batch of unwanted states.

The second group of unwanted states, and the states directly targeted for dephasing, follow from the non-orthogonal inner product discussed at the end of the previous section. The inner product has the form,

$$|\langle +\alpha(k)| - \alpha(k)\rangle| = \exp\left[-(J_{\varphi})^2 \int \frac{dk}{2\omega(k)} \, |\tilde{F}_{\nu}(k)|^2\right]. \tag{2.30}$$

Choosing a Gaussian smearing function with Gaussian width $\sigma$, and setting $J \gg \sigma$ renders these terms approximately zero. With these constraints, the coherent information of the field mediated communication channel from qubit A to qubit B in Fig. 2.1 grows toward one as we increase the strength of the coupling. Which is demonstrated in Fig. [4] of Ref. [2], as well as Figs. 5.3.

These recent results are well understood in the RQI community, but follow from the linear nature of the field observables in Eqs. 2.11 and 2.12. For certain nonlinear cases, it becomes a challenge to redefine the coherent amplitudes to include the coupling and smearing constants.

With the introduction of the QST channel, from the RQI literature, we claim that the UDWQC system in some idealized scenario, exhibits effective generation and propagation of entanglement, a minimal requirement for quantum communica-tion. We will revisit and derive this result in detail in Chapter 5. In the remainder of the thesis we aim to show that the this idealized scenario has some realistic attributes that may prove to be replicable in a laboratory setting. Furthermore, we show several theoretical advances achieved through the UDWQC program, that may prove to further our knowledge of quantum transduction between qubits and fields. Before we demonstrate these theoretical breakthroughs, we first outline the procedure of bosonization that allows us to evaluate our fermionic fields as scalar bosons, making correlators in Eq. 2.25 far more palatable.

# Chapter 3

# Abelian Bosonization of a Helical Luttinger Liquid

## 3.1 Motivations

The UDWQC program connects spin-qubits to condensed matter systems that are understood through the guise of QFT. The current literature on UDW detector model gives a perturbative prescription to some fermionic systems [79], but a non-perturbative method is still lacking. A common prescriptive method in condensed matter that connects fermions and bosons is through the process of bosonization [71,82,84,85]. With Bosonization we need only examine literature of bosonic interaction. Furthermore, for maximal quantum information generation and propagation, we are required in the realm of quantum computing, to consider only strong couplings [2,51].

A key fact of this thesis is that we are working with systems of right- and left-moving fermions. However, the ability to bosonize our fermions in the HLL allows our usage of the UDW model as there is a fair bit of literature on the UDW model specifically with scalar bosons, and we intend to continue to build on that literature. In this chapter, we will lay down the foundations and formalisms of bosonization. Along the way we will recover Eqs. 2.20 and 2.23 and show the fermion to boson equivalence at the level of the correlator [81].

When working in the interaction picture we follow the standard prescription of Wick rotating our variables as

$$\begin{aligned} z &= -i(x - vt) & \bar{z} &= i(x + vt) \\ \partial_z &= -\frac{i}{2}\left(\frac{1}{v}\partial_t - \partial_x\right) & \partial_{\bar{z}} &= -\frac{i}{2}\left(\frac{1}{v}\partial_t + \partial_x\right) \end{aligned} \tag{3.1}$$

and therefore are working in Euclidean space. There are several key interactions left out of this thesis that we will describe in Chapter 7 such as the time-evolution of the field after the information is encoded. For this section we treat the system as a free fermion field interacting with a Kondo impurity.

### 3.1.1 Kondo Model Bosonization of Fermions

Our "naive" Hamiltonian of Eq. 2.5 that was introduced in the previous chapter, may not have been as niave as we stated. With topological edge states of a HLL, our right and left movers stem from the Hamiltonian

$$\hat{H}_{\text{int}}(t) = \sum_{k,k',\sigma,\sigma'} J_{\alpha,\beta}\hat{\mu}_\alpha(t) \cdot (\hat{c}^\dagger_{k,\sigma}(k)\hat{\tau}^{\sigma,\sigma'}_\beta \hat{c}_{k',\sigma'}(k) + h.c.). \tag{3.2}$$

Where $J_{\alpha,\beta}$ is the coupling constant, $\hat{\mu}_\alpha(t)$ is some "spin-state" which will be defined in more detail later, $\hat{\tau}$ a the Pauli matrix and $\hat{c}^\dagger_\sigma(\hat{c}_\sigma)$ are fermionic creations (annihilation) operators. We can split this operator into left- and right-moving parts such that $\hat{c}_\sigma = \hat{c}_\uparrow + \hat{c}_\downarrow$ and likewise for $\hat{c}^\dagger_\sigma$. Furthermore,

$$\begin{aligned}\hat{c}_\uparrow(k) &= \hat{\psi}_{\uparrow,+}e^{ik_F x}\\ \hat{c}_\downarrow(k) &= \hat{\psi}_{\downarrow,-}e^{-ik_F x}\end{aligned} \tag{3.3}$$

and plugging the left- and right-moving fields into Eq. 3.2 we carry out the dot product and recover a Hamiltonian,

$$\hat{H}_{\text{int}}(t) = J\hat{\mu}(t)\left(\hat{\psi}^\dagger_+\hat{\psi}_+ - \hat{\psi}^\dagger_-\hat{\psi}_- + \hat{\psi}^\dagger_+\hat{\psi}_-(e^{2ik_F x}) - \hat{\psi}^\dagger_-\hat{\psi}_+(e^{-2ik_F x})\right) \tag{3.4}$$

where $\hat{\psi}_+(\hat{\psi}_-)$ is shorthand for $\hat{\psi}_+(\bar{z})(\hat{\psi}_-(z))$ which are the right- and left-moving fermion fields and we have dropped the indices $\uparrow$ and $\downarrow$ as they have become redundant.

### 3.1.2 Bosonizing the Hamiltonian

We aim to express our fermionic fields in left- and right-moving bosonic fields $\hat{\phi}$ and $\hat{\bar{\phi}}$.

$$\hat{\phi(z)} = \int_{k>0} \frac{dk}{2\pi}\frac{1}{\sqrt{2k}}[\hat{b}(k)e^{-kz} + \hat{b}^\dagger(k)e^{kz}] \tag{3.5}$$

$$\hat{\bar{\phi}}(\bar{z}) = \int_{k>0} \frac{dk}{2\pi}\frac{1}{\sqrt{2k}}[\hat{\bar{b}}(k)e^{-k\bar{z}} + \hat{\bar{b}}^\dagger(k)e^{k\bar{z}}] \tag{3.6}$$

Following the prescription in [85], we define bosonic fields $\hat{\varphi}$ and conjugate momentum $\hat{\Pi}$ defined by $\hat{\Pi} = \frac{1}{v}\partial_t\hat{\varphi}$,

$$\hat{\varphi}(x) = \int \frac{dk}{2\pi}\sqrt{\frac{v}{2\omega(k)}}[\hat{b}(k)e^{ikx} + \hat{b}^{\dagger}(k)e^{-ikx}] \tag{3.7}$$

$$\hat{\Pi}(x) = \int \frac{dk}{2\pi}\sqrt{\frac{\omega(k)}{2v}}[-i\hat{b}(k)e^{ikx} + i\hat{b}^{\dagger}(k)e^{-ikx}]. \tag{3.8}$$

as well as the dual boson $\hat{\vartheta}$ such that the following relations hold.

$$\hat{\phi} = \frac{1}{2}(\hat{\varphi} + \hat{\vartheta}), \quad \hat{\bar{\phi}} = \frac{1}{2}(\hat{\varphi} - \hat{\vartheta}) \tag{3.9}$$

from here we evaluate the left and right mode expansion of our scalar field and following commutation relations we develop the following expressions for our fermion fields.

$$\begin{aligned} \hat{\psi}_{+}(z) &= \frac{1}{\sqrt{2\pi}}e^{-i\sqrt{4\pi}\hat{\phi}(z)}, \quad \hat{\psi}_{+}^{\dagger}(z) = \frac{1}{\sqrt{2\pi}}e^{i\sqrt{4\pi}\hat{\phi}(z)}, \\ \hat{\psi}_{-}(\bar{z}) &= \frac{1}{\sqrt{2\pi}}e^{i\sqrt{4\pi}\hat{\bar{\phi}}(\bar{z})}, \quad \hat{\psi}_{-}^{\dagger}(\bar{z}) = \frac{1}{\sqrt{2\pi}}e^{-i\sqrt{4\pi}\hat{\bar{\phi}}(\bar{z})}. \end{aligned} \tag{3.10}$$

These equations are an essential part of bosonization and will allow us to transform the fermions of Eq. 3.2 into bosons.

When evaluating the left- and right-mover terms of the interaction Hamiltonian we employ the method of point-splitting to find,

$$\hat{\psi}_{+}^{\dagger}(z)\hat{\psi}_{+}(z) = \frac{i}{\sqrt{\pi}}\partial_{z}\hat{\phi}(z) \tag{3.11}$$

$$\hat{\psi}_{-}^{\dagger}(\bar{z})\hat{\psi}_{-}(\bar{z}) = -\frac{i}{\sqrt{\pi}}\partial_{\bar{z}}\hat{\bar{\phi}}(\bar{z}) \tag{3.12}$$

these derivations and the derivation of the resulting Hamiltonian can be found in Appendix B. Here we list the result

$$\hat{H}_{\text{int}}(t) = J\hat{\mu}(t)(\frac{1}{\sqrt{\pi}}(\partial_z\hat{\phi} + \partial_{\bar{z}}\hat{\bar{\phi}})) \tag{3.13}$$

$$= J\hat{\mu}(t)(\frac{1}{\sqrt{\pi}}(2\hat{\Pi})) \tag{3.14}$$

This is a significant consequence as we are able to take our quadratic fermionic interaction and turn it into a linear coupling with a scalar boson. This gives us a rank-one unitary like that of Eq. 2.20. As we'll find out in Chapter 4 a channel made with solely this unitary is inherently entanglement breaking. However, our bosonization process will give us some alternatives to consider.

### 3.1.3 A Zoo of interactions

The entirety of the Tomonaga–Luttinger liquid can be expressed as

$$\hat{H} = \hat{\psi}^\dagger(x)\left(J_1 i\nu_F\partial_x + J_2\Delta\right)\hat{\psi}(x) + J_3(\hat{\rho}+\hat{j}) + J_4(\hat{\psi}_+^\dagger\hat{\psi}_-(e^{2ik_Fx}) - \hat{\psi}_-^\dagger\hat{\psi}_+(e^{-2ik_Fx}))$$
$$+ 2J_5\hat{\rho}_+(x)\hat{\rho}_-(x) + J_6\left(\hat{\rho}_+(x)^2 + \hat{\rho}_-(x)^2\right). \tag{3.15}$$

For simplicity we introduce the shorthand $\hat{\rho}_+ = \hat{\psi}_+(x)^\dagger\hat{\psi}_+(x)$ and $\hat{\rho}_- = \hat{\psi}_-(x)^\dagger\hat{\psi}_-(x)$ so that we can express the electron and current densities as

$$\hat{\rho} = \hat{\rho}_+ + \hat{\rho}_- = \frac{1}{\sqrt{\pi}}\partial_x\hat{\varphi} \tag{3.16}$$

$$\hat{j} = \hat{\rho}_+ - \hat{\rho}_- = \frac{1}{\sqrt{\pi}}\hat{\Pi}. \tag{3.17}$$

The electron density Eq. 3.16 is not time-reversal symmetric when combined with our Kondo impurity spin structure. Thus, this interaction is not physi-cal. Nonetheless, it is clear to see that a Hamiltonian that includes both densities can be constructed as a toy model to demonstrate a LL Hamiltonian that realizes nonzero channel capacity.

| Interaction Type | Fermion Form | Boson Form | Coupling type |
|---|---|---|---|
| Electron Density + Fermion Density | $\hat{\psi}_+^\dagger\hat{\psi}_+ - \hat{\psi}_-^\dagger\hat{\psi}_- + \hat{\psi}_+^\dagger\hat{\psi}_+ + \hat{\psi}_-^\dagger\hat{\psi}_-$ | $\frac{1}{\sqrt{\pi}}\partial_x\hat{\varphi} + \frac{1}{\sqrt{\pi}}\hat{\Pi}$ | Linear |
| Free Dirac | $\hat{\psi}^\dagger(x)\,(J_1 i\nu_F\partial_x)\,\hat{\psi}(x)$ | $\hat{\Pi}^2 + (\partial_x\hat{\varphi})^2$ | Quadratic |
| Backscattering | $\hat{\psi}_+^\dagger\hat{\psi}_-(e^{2ik_Fx}) - \hat{\psi}_-^\dagger\hat{\psi}_+(e^{-2ik_Fx})$ | $\frac{1}{2\pi}\cos\sqrt{4\pi}(\hat{\varphi})$ | Nonlinear |
| Umklapp | $\sum_{si\in\pm}\hat{\psi}_{s1}^\dagger\hat{\psi}_{s2}^\dagger\hat{\psi}_{s3}\hat{\psi}_{s4}$ | Cosines and Sines | Nonlinear |

Table 3.1: Different Blocks of the Tomonaga Luttinger liquid Hamiltonian and thier respective bosonized form.

We have in the Tomonaga–Luttinger liquid plus Kondo impurity model, block interactions that are relevant to particular condensed matter systems. For an example we consider the term $J_3\rho$ we maybe be describing a Helical Luttinger Liquid such as HgTe. Therefore, evaluating the interaction blocks of the Hamiltonian will provide insights to individual systems.

Bosonization of Eq. 3.15 is presented in Table 3.1. We propose that some of these interaction blocks provide nonzero channel capacities that could be experimentally realized. Through the prescription method provided in this thesis, one can pick and choose components to realize these experiments, but not all of these blocks are made equally.

## 3.2 Charge and Spin Bosonization

The spinless bosonization treatment does not allows for the proper anticommutation relations. Namely for the terms $\hat{\psi}_+^\dagger\hat{\psi}_-(e^{2ik_Fx})$ and $\hat{\psi}_-^\dagger\hat{\psi}_+(e^{-2ik_Fx})$. In-teractions that allow backscattering and also maintain proper anticommutation relations, we introduce two spin bosons $\hat{\varphi}_\uparrow$ and $\hat{\varphi}_\downarrow$, that are related by the charged and spin bosons,

$$\hat{\varphi}_c = \frac{1}{\sqrt{2}}(\hat{\varphi}_\uparrow + \hat{\varphi}_\downarrow) \quad \hat{\varphi}_s = \frac{1}{\sqrt{2}}(\hat{\varphi}_\uparrow - \hat{\varphi}_\downarrow) \tag{3.18}$$

as well as chiral fields $\hat{\phi}_{c,s}$ and $\hat{\bar{\phi}}_{c,s}$. Furthermore, with the extra species of fermions, we utilize the Klein Factors $\eta_\mu$ which follow anticommutation relations,

$$\{\eta_\mu, \eta_\nu\} = \{\bar{\eta}_\mu, \bar{\eta}_\nu\} = 2\delta_{\mu,\nu} \quad \{\eta_\mu, \bar{\eta}_\nu\} = 0 \tag{3.19}$$

and preserve the fermionic anticommutation relations discussed above. We now have a new form of bosonized fermion fields given as,

$$\begin{aligned} \hat{\psi}_+(z) &= \frac{1}{\sqrt{2\pi}}\eta_\mu e^{-i\sqrt{4\pi}\hat{\phi}_\mu(z)}, \quad \hat{\psi}_+^\dagger(z) = \frac{1}{\sqrt{2\pi}}\eta_\mu e^{i\sqrt{4\pi}\hat{\phi}_\mu(z)}, \\ \hat{\psi}_-(\bar{z}) &= \bar{\eta}_\mu \frac{1}{\sqrt{2\pi}} e^{i\sqrt{4\pi}\hat{\bar{\phi}}_\mu(\bar{z})}, \quad \hat{\psi}_-^\dagger(\bar{z}) = \frac{1}{\sqrt{2\pi}}\bar{\eta}_\mu e^{-i\sqrt{4\pi}\hat{\bar{\phi}}_\mu(\bar{z})}. \end{aligned} \tag{3.20}$$

Using these definitions and recognizing that for this system we set $\eta_\mu\bar{\eta}_\nu = 1$ [85, 86], we can see now that our bosonized Hamiltonian can be split into two sections impurity scattering (the terms without the factors of ($e^{\pm 2ik_F x}$)) which result follows from the pointscattering methods in Appendix A, and our impurity backscattering terms. The impurity scattering bosonized Hamiltonian is given by,

$$\hat{H}^F_{\text{int}} = J\hat{\mu}(t)(\frac{i}{\sqrt{\pi}}(2\hat{\Pi}_c + 2\hat{\varpi}_s)) \tag{3.21}$$

where $\hat{\varpi} \equiv \frac{1}{v}\partial_t\hat{\vartheta}$ is the conjugate momentum of the dual field. Notice here if we suppress the spin terms (make the system spinless) we retrieve the same Hamiltonian in Eq. 3.13. If we were able to suppress the ($e^{\pm 2ik_F x}$) terms as mentioned before but consider both spin and charge we have two noncommuting observables that could be used to create our non-zero capacity quantum channel.

The back-scattering Hamiltonian follows from eqs. 3.20,

$$\hat{H}^{B}_{\text{int}} = J\hat{\mu}(t)(\frac{1}{2\pi}\cos\sqrt{2\pi(\hat{\varphi}_c + \hat{\vartheta}_s)}) \tag{3.22}$$

This could be used as well to create a different arrangement of operators necessary to test for a non-zero capacity quantum channel. As we mentioned in the previous chapter, to evaluate these systems we turn to the computational basis states of a quantum field, coherent states.

## 3.3 Coherent States and Their Problems.

One recurring theme in this work, is the need to tackle the nonorthogonality of coherent states. In Chapter 2, we showed that while you can never totally eliminate the inner product of different coherent states, you can greatly reduce it's value through parameter tuning. While this is true for all bosonized form of table 3.1, other problems begin to develop stemming from the inner product of coherent states. To demonstrate this we look at the bosonized form of the backscattering block.

Starting with the standard definitions of coherent states in field theory,

$$|\pm\alpha\rangle \equiv \exp(\pm i\varphi)|0\rangle = \hat{D}(\alpha)|0\rangle = \exp\left(\int d^n\mathbf{k}\left[\alpha(k)\hat{a}^{\dagger}_{\mathbf{K}} - \alpha(k)^{*}\hat{a}_{\mathbf{K}}\right]\right). \tag{3.23}$$

We want to be able to calculate

$$e^{i\cos\sqrt{4\pi}\varphi}\left|0\right\rangle, \tag{3.24}$$

but to do this we first evaluate

$$e^{\hat{D}(\alpha)}\left|0\right\rangle \tag{3.25}$$

in the Fock Basis. This will allow for a straightforward Taylor expansion of Eq. 3.24. Coherent states in the Fock basis are defined by

$$\left|\alpha\right\rangle \equiv e^{\frac{-|\alpha|^2}{2}}\sum_{n=0}^{\infty}\frac{\alpha^n}{\sqrt{n!}}\left|n\right\rangle. \tag{3.26}$$

Using the Eq. 3.26, Baker–Campbell–Hausdorff (BCH), and the following relation,

$$\hat{D}(\alpha)\hat{D}(\beta) = e^{(\alpha\beta^*-\alpha^*\beta)/2}\hat{D}(\alpha+\beta), \tag{3.27}$$

we can expand Eq. 3.25 to

$$e^{\hat{D}(\alpha)}\left|0\right\rangle = \left|0\right\rangle + \hat{D}(\alpha)\left|0\right\rangle + \frac{1}{2!}(\hat{D}(\alpha))^2\left|0\right\rangle + \frac{1}{3!}(\hat{D}(\alpha))^3\left|0\right\rangle ... \tag{3.28}$$

$$= \left|0\right\rangle + \left|\alpha\right\rangle + \frac{1}{2!}\left|2\alpha\right\rangle + \frac{1}{3!}\left|3\alpha\right\rangle ... \tag{3.29}$$

$$= \left|0\right\rangle + e^{\frac{-|\alpha|^2}{2}}\sum_{n=0}^{\infty}\frac{\alpha^n}{\sqrt{n!}}\left|n\right\rangle + \frac{1}{2!}e^{\frac{-|(2\alpha)|^2}{2}}\sum_{n=0}^{\infty}\frac{(2\alpha)^n}{\sqrt{n!}}\left|n\right\rangle$$

$$+ \frac{1}{3!}e^{\frac{-|3\alpha|^2}{2}}\sum_{n=0}^{\infty}\frac{(3\alpha)^n}{\sqrt{n!}}\left|n\right\rangle + ... \tag{3.30}$$

$$= \sum_{m=0}^{\infty}\frac{1}{m!}\left|m\alpha\right\rangle \tag{3.31}$$

$$= \sum_{m=0}^{\infty}\sum_{n=0}^{\infty}e^{-\frac{|m\alpha|^2}{2}}\frac{(m\alpha)^n}{m!\sqrt{n!}}\left|n\right\rangle. \tag{3.32}$$

While coherent states are not orthogonal, fock states are and therefore can be normalized using the scheme;

$$1 = \sum_n \langle\beta|n\rangle\langle n|\beta\rangle \tag{3.33}$$

$$= N^2 \sum_{m=0}^{\infty}\sum_{n'=0}^{\infty} e^{-|m\alpha|^2}\frac{(m\alpha)^{2n'}}{(m!)^2 n'!} \tag{3.34}$$

where we have set $|\beta\rangle \equiv e^{\hat{D}(\alpha)}|0\rangle$. Using the same line of reasoning above and remaining in the coherent state basis we find

$$e^{i\cos\sqrt{4\pi}\varphi}|0\rangle = e^{\hat{D}(-\alpha)}e^{\hat{D}(\alpha)}|0\rangle = N\sum_{m=0}^{\infty}\sum_{m'=0}^{\infty}\frac{1}{m!}\frac{1}{(m+m')!}(|m\alpha\rangle). \tag{3.35}$$

The first equality follows from Eulers formula and BCH and the phase factors have been absorbed into the displacement operator $\hat{D}(\pm\alpha)$. From here we can calculate the correlator,

$$\langle 0|e^{i\cos\sqrt{4\pi}\varphi}|0\rangle = \langle 0|e^{\hat{D}(-\alpha)}e^{\hat{D}(\alpha)}|0\rangle \tag{3.36}$$

$$= \sum_{m,m'=0}^{\infty}\sum_{s,s'\in\pm}\frac{1}{m!}\frac{1}{m'!}\langle s'm'\alpha|sm\alpha\rangle \tag{3.37}$$

$$= N'\sum_{m=0}^{\infty}\sum_{m'=0}^{\infty}\frac{1}{m!}\frac{1}{m'!}\left(e^{-\frac{1}{2}|\alpha|^2(m-m')^2} + e^{-\frac{1}{2}|\alpha|^2(m+m')^2}\right) \tag{3.38}$$

where the last equality follows from the identity

$$\langle\beta|\alpha\rangle = e^{-\frac{1}{2}(|\beta|^2+|\alpha|^2-2\beta^*\alpha)}. \tag{3.39}$$

As one can see, calculating these correlators is highly nontrivial. Numerically it appears these series converge, but without analytic proofs, it's challenging to pin down the normalization factors. Moreover, implementing these series into the numerical simulations of coherent information for the electron density and current blocks of the LL Hamiltonian is extremely taxing computationally.

In the next chapter, we introduce the formalism of quantum Shannon theory which focuses on evaluating quantum information channels. To understand the advances we have made to RQI using quantum Shannon theory, we demonstrate the background of quantum channels from a first principles formalism. Readers who are not familiar with quantum Shannon theory will hopefully find this chapter very readable.

# Chapter 4

# Quantum Information and Quantum Computing

## 4.1 Introduction to Information Channels

Those who study information sciences, have developed a series of mathematical tools to help understand the inter-workings behind how information transfers from inputs to outputs. These transfers, are aptly named *information channels* [4, 87].

Claude Shannon's remarkable works in the 1940s [88] paved the way for a deeper understanding of how to treat classical information. Shannon theory has since been developed and blossomed into a vibrant field to study for mathematicians and scientists alike. The community of QI scientists have subsequently coined quantum communication theory as "Quantum Shannon Theory", as an homage to Shannon and the works that he produced [4].

Just as with other fields of physics, classical and quantum information channels follow separate rules. The quantum revolution of the 1920s brought with it a confusing perspective of how we are to view the governing mechanics of the micro-world. Up until that moment, nature could be described by a set of deterministic rules. Only limited by the complexity of each system and the ignorance of the observer. The quantum founding fathers showed that at the smallest of scales, nature is fundamentally determined by probabilities [26, 89].

The probabilistic nature of quantum mechanics would thus play a lofty role in how we understand information. In this chapter, we introduce a guide to quantum information channels and do this through the guise of an "interpretation free" version of quantum mechanics. Measurement theory is a popular topic in quantum communication theory, as the physical interpretation of the measurement has consequences regarding the information that is being treated. Despite this, physical interpretations of the evolution of QI do not impact the results experienced in a laboratory. For this reason we provide a simple and clear, interpretation free, definition to quantum information.

**Definition 1.** *Quantum Information is the information contained in the entanglement between states and therefore cannot exist in systems whose states are "separable".*

This definition offers a straight-forward test of the "quantum-ness" of a channel. Namely, at any given point in the processing of the information, do the states become separable. At the end of this chapter we outline the mathematics that determines the separability of states and demonstrate how one can distinguish quantum information from classical information channels.

Despite the differences in the types of information, certain properties of these information channels will be analogous to each other. Therefore, we dive first into a more familiar domain of classical information channels while preparing to understand the more complicated but coinciding quantum information channels. The aim of this chapter, is to introduce some of the tools necessary to understand the properties of information channels. While the mathematics is deep and complex, I do my best to keep it self-contained as much as possible, with the intent to offer a broader audience the ideas behind information theory.

## 4.2 Classical Information Channels

As mentioned previously, complexity of a classical system can shroud its deterministic properties. Probabilities are often deployed as a simplification to these complex systems. While there is nothing quantum about flipping a fair coin, the fact that quantum theory is probabilistic, allows us to draw similar allusions by following such a system. To unpack the information hidden in this random process of coin flipping, we need a theory to measure information as it changes through these classical processes. For that, we turn to Shannon's information theory.

Shannon's theory is best understood with the usage of *bits.* A bit is a single piece of information that describes the state of an object. Consider the above example of a fair coin. The state of the coin can be described by assigning the value of zero to heads and one to tails. Therefore, reading out the bit is equivalent to measuring the state of the coin. To measure the information of a system[1], we first consider the Shannon entropy $H(X)$ given by,

$$H(X) = \sum_{x} -p_X(x) \log p_X(x) \tag{4.1}$$

with probability distribution $p_X(x)$. $H(x)$ will measure the *randomness* or *surprise* in the output of the system. If we consider our above example of the fair coin we expect a Shannon Entropy of one, which is the maximum amount of uncertainty for this system. This can easily be seen by replacing the fair coin with a weighted coin. A weighted coin with two-thirds probability of heads and one-third probability of tails, has a Shannon Entropy of 0.918. A weighted coin with nine-tenths probability of heads and one-tenth tails will yields 0.469. As the coin gets more weighted, the uncertainty in the result of the coin flip goes down, and subsequently the measure of Shannon Entropy does as well. Generally, we can say if $x$ is the probability of obtaining heads, the probability of obtaining tails is $1 - x$ and Eq. 4.1 becomes,

$$H(X) = x \log x + (1 - x) \log(1 - x). \tag{4.2}$$

[1]We will be using $\log_2$ for the entirety of this chapter for consistency.

We can see in figure 4.1, that the information content of $X$ is maximized when both the events are equally likely. This makes sense because having equal probability outcomes, makes the task of guessing the hardest. So in essence, the entropy is closely related to the contents of the information channel.

Shannon entropy specifically measures information content of the random variable $X$. If we refer to this "surprise" or "randomness" of an outcome $x$ as the information content $i(x)$, and define it as $i(x) = \log(\frac{1}{p(x)}) = -\log(p(x))$. This measurement has the property that it is high for low probability outcomes and low for high probability outcomes. Moreover, this also has another desirable property of "additivity" i.e. the information content of two outcomes $x$ and $y$ get added since the joint probability $p(x, y) = p(x).p(y) \implies i(x, y) = -\log(p(x)p(y)) = -\log(p(x)) - \log(p(y)) = i(x) + i(y)$. Thus, the total information content of the random variable $X$ can be written as the weighted average of the information of possible outcomes.

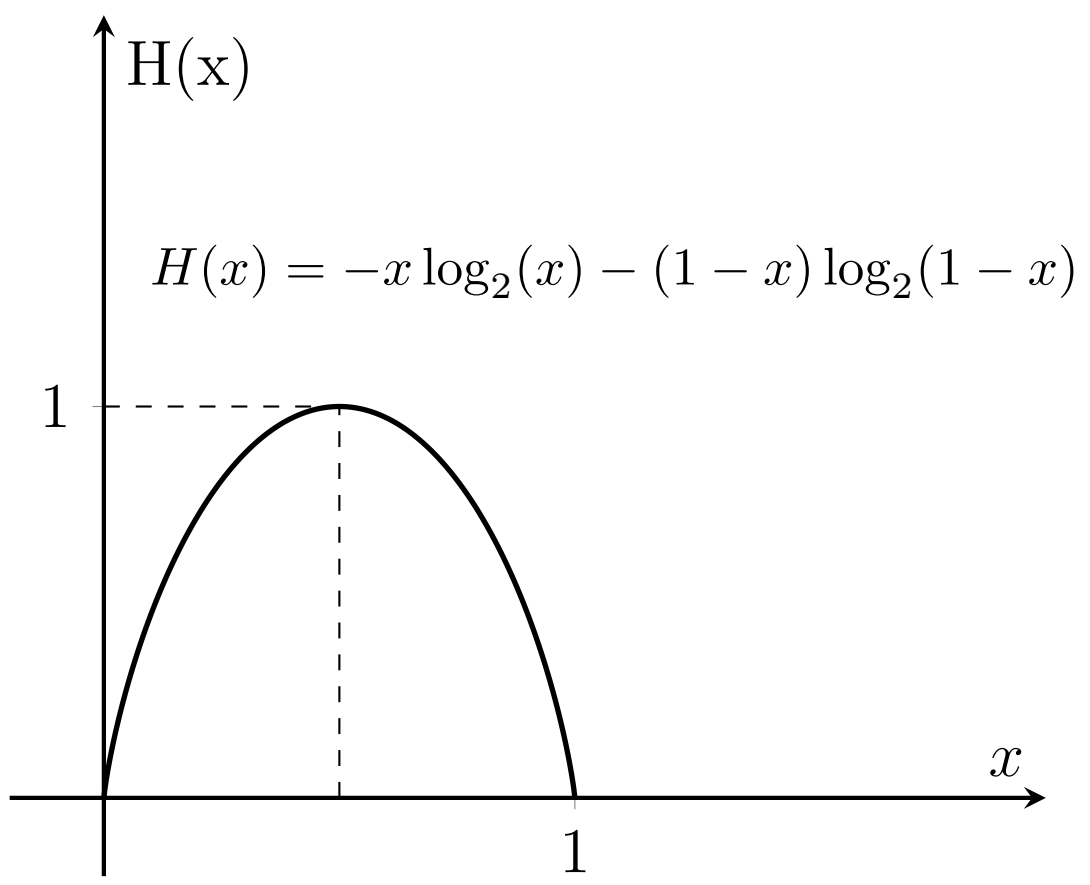

Figure 4.1: Graphical representation of H(x) for our fair-coin flipping system. Notice the maximum entropy occurs when our probability $x = 1/2$.

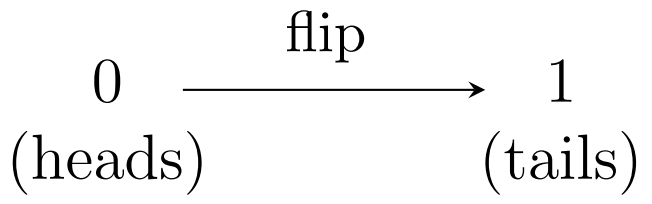

Figure 4.2: A classical information channel that shows a flip operation acting on a coin. The coin is initially in state of heads but the flip operation changes this to tails.

For both the classical and the quantum case of information channels, we will model a very simple operation that has purpose both in classical and quantum computing. For the classical system we will use a *flip* operation. The flip operation will take our input bit and flip the value of it. For instance, our fair coin modeled with heads as zero and tails as one, will do just as it says. Flip the side of the coin from its initial configuration to the opposite. Figure 4.2 gives a pictorial representation of this process.

A useful calculation in information theory is *Channel Capacity*, which is a measure of information loss (or conversely propagated) through our circuit. Information lost is often do to the noisiness of a channel. Noise is attributed to unwanted interactions and will affect the input states by altering them. For instance, a sender Alice and a receiver Bob may be attempting to communicate classically. Choosing a noisy channel to save resources could alter the inputs enough that Bob would be unable to decipher the original message Alice has sent. If instead of a fair coin, suppose Alice has access to a set of random variables $X$. She encodes a message onto the elements of $X$ and sends it via some noisy channel to Bob who receives a set of random variables $Y$ to decipher. Our aim now must be to mathematically evaluate the channel that led from $X$ to $Y$. This will show if the channel reliably transferred Alice's message. Pictorially this is shown in figure 4.3.

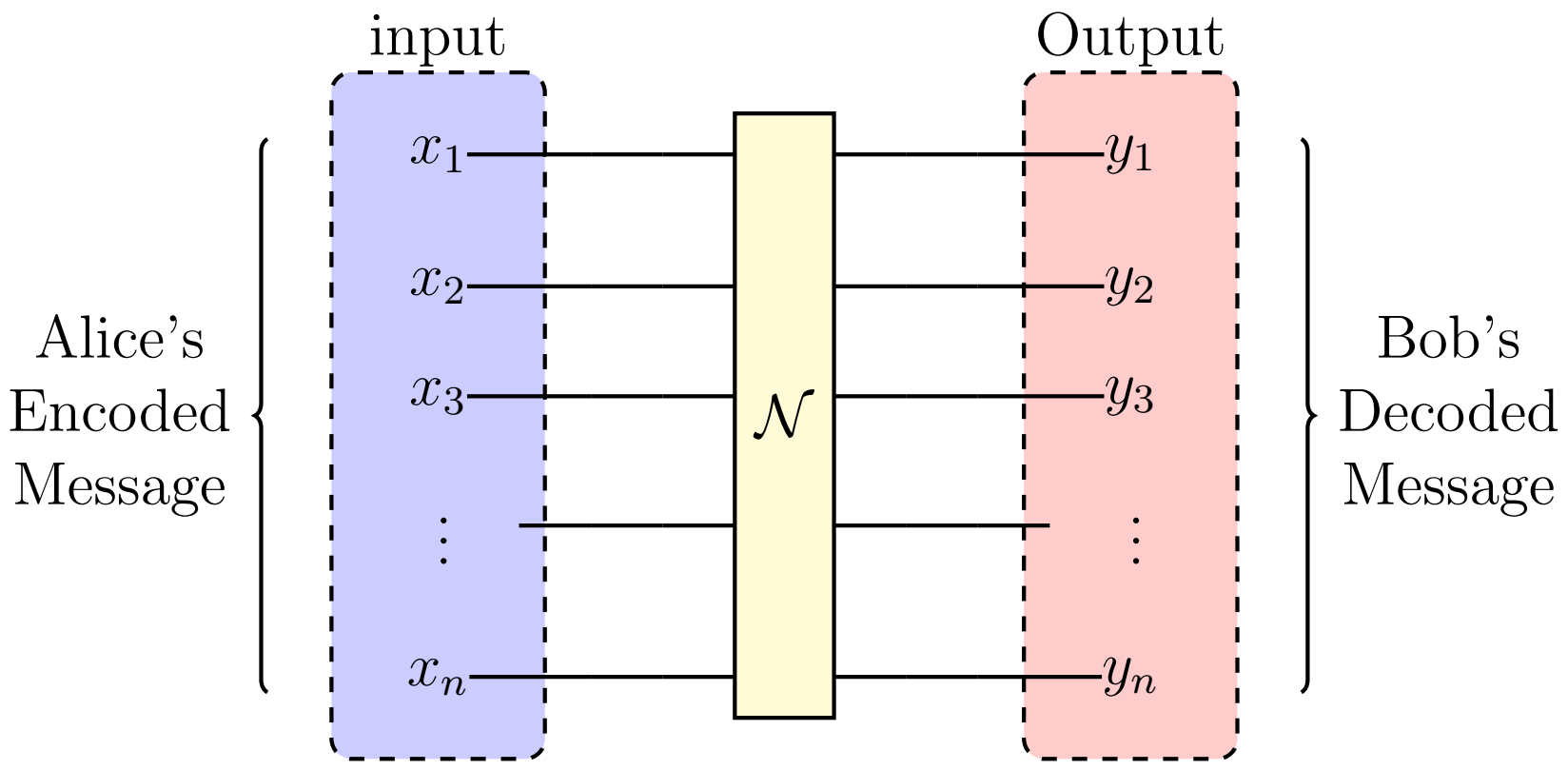

Figure 4.3: Alice's message is encoded into the elements $x_i$ where it is passed through some noisy channel $\mathcal{N}$ and later deciphered by Bob.

After the information transfer, Bob has access to $Y$, and the amount of his randomness/surprise about $X$ is quantified by the *Conditional Entropy.* Mathematically this has the shape,

$$H(Y|X) = -\sum_{x,y} p_{Y|X}(y|x) \log\left(\frac{p_{Y|X}(y|x)}{p_Y(y)}\right) \tag{4.3}$$

where $p_{X,Y}(x,y)$ is an expression of the multiplicity of probability namely, $p_{X,Y}(x,y) = p_X(x)p_Y(y)$. Notice that $H(X) \geq H(X|Y)$, and subsequently $I(X;Y) \geq 0$. Here these probabilities are a result of the interactions within the noisy quantum chan-nels. The measure of how much information Bob recovers, can be expressed as the *Mutual Information* $I(X;Y)$ where,

$$I(X;Y) = H(Y) - H(Y|X). \tag{4.4}$$

The mutual information in essence quantifies the similarity or correlations between the random variables of Alice and Bob. This is because $H(Y|X)$ quantifies the amount of error due to noise. In a perfect channel this value is zero because $Y$ is fully aware of how $X$ will change.

Up until this point, we have very purposefully not discussed the contents of the noisy information channels. We want to have a means to describe an information channel either classically or quantum mechanically, but the contents of the channels may be messy and hard to understand. This is where channel capacity comes into play. Channel capacity uses information to quantifiably measure the reliability of an information channel $\mathcal{N}$. **Note**: We have reserved the notation for $\mathcal{N}$ to describe non-relativistic information channels and $\Xi$ we have utilized for quantum information channels that utilize qubit-field interactions. The means as to which the information passes through that channel is equivalent to knowledge of the information given and received. After all, the channel is a probabilistic mapping of the set $X$ onto the set $Y$,

$$\mathcal{N} : p_{Y|X}(y|x). \tag{4.5}$$

Under this construction, we define the channel capacity as,

$$\mathcal{C}(\mathcal{N}) = \max_{p_X(x)} I(X;Y) \tag{4.6}$$

which has a physical interpretation such that, if Alice can arrange her information in a way that allows a maximal amount of correlation with Bob's, than this is the upper bound to the capacity of the channel. Using what we learned about information, we can quantify the capacity of a channel which will help identify if these channels are efficient uses of resources.

## 4.3 Quantum Information Channels

A quantum channel is a linear, completely positive, and trace preserving mapping corresponding to a physical evolution [4, 90]. That is to say, that it too, is a probabilistic mapping from initial states to final states. Yet something deeper is going on here. The probabilities involved the information transfer is not necessarily due to our inability to evaluate a complex system. Instead, the mechanics are fundamentally probabilistic. This leads to a relationship deeper than correlation, entanglement.

Entanglement, while mysterious in many ways, is a fundamental ingredient in quantum mechanics and is necessary for successful quantum computing. Similar to our classical bit, we introduce a quantum bit or *qubit*, that plays the same role as arbiter of information, but for quantum systems. If we assign the spin degrees of freedom of an electron to a value of our qubit we can create a similar arrangement to the classical coin flip.

For our purposes, lets assign spin up to our qubit $|0\rangle$, and spin-down to $|1\rangle$

and construct the state of our entangled pair (a Bell State),

$$|\psi\rangle = \frac{|0\rangle\,|0\rangle + |1\rangle\,|1\rangle}{\sqrt{2}} \tag{4.7}$$

where the coefficient of $\frac{1}{\sqrt{2}}$ guarantees that 50% of the time we will measure both spin-up and 50% of the time we will measure both spin down.

Now that our descriptive states are set up as qubits, we can begin to evaluate the information in our states. Lets, first establish a parallel to our Shannon Entropy. It may surprise you that quantum mechanical entropy came first, this was accomplished by John von Neumann in 1932. Nevertheless, our *von Neumann entropy* has a similar form to the Shannon Entropy,

$$S(\hat{\rho}) = -\operatorname{Tr}(\hat{\rho}\log\hat{\rho}) \tag{4.8}$$

where $\hat{\rho}$ is a density matrix whose purpose is to describe, in entirety, the probabilities and states of our system. If these states are eigenstates (eigenstates are states we can measure eg. spin-up and spin-down) then our density matrix takes the appearance of $\hat{\rho} = \sum_j n_j\,|j\rangle\,\langle j|$ with $|j\rangle$ denoting an eigenstate, and we can simplify equation 4.8 to,

$$S(\hat{\rho}) = -\sum_j n_j \log n_j \tag{4.9}$$

where $n_j$ is the probability of measuring an eigenstate. But what does this quantify? In essence, this quantifies the *purity* of the state. While purity has some analogous meaning to classical information, we should press on with our development of quantum information and maybe a more intuitive understanding will present itself. By now you have probably noticed that equations 4.1 and 4.9 are mostly identical. That changes quickly when we introduce *Quantum Mutual Information.*

To understand Quantum Mutual Information, lets give Alice one qubit and Bob another. We can entangle them such that our input state matches that of equation 4.7. We can then model the mutual information between Alice's and Bob's qubits as,

$$I(A:B) = S(\hat{\rho}_A) + S(\hat{\rho}_B) - S(\hat{\rho}_{AB}) \tag{4.10}$$

where $\hat{\rho}_A = \mathrm{Tr}_B(\hat{\rho}_{AB})$ and $\rho_A$ is the density matrix describing the state of the Alice's qubit. Tracing over part of the system, in essence ignores that part of the system. When we trace over Bob's information we are focused solely on the information of Alice. The physical interpretation of mutual information is thus a quantifiable measurement of the entanglement shared between Alice and Bob. For a highly entangled state such as our bell state, we find a value of two which indicates a maximum entanglement between the two qubits. While quantum mutual information is insightful for understanding entanglement between states, these states remain as initial states. We have made a modification to our system between the classical and quantum cases. The modification is the usage of Bob. In our classical noisy channel arrangement, Bob received some data $Y$ from Alice whose input was $X$. The classical mutual information was a correlation between input and output. Therefore, We need another tool that is analogous to classical mutual information, that will evaluate how a quantum channel could alter this entanglement. Luckily, we'll find a familiar analogy to classical mutual

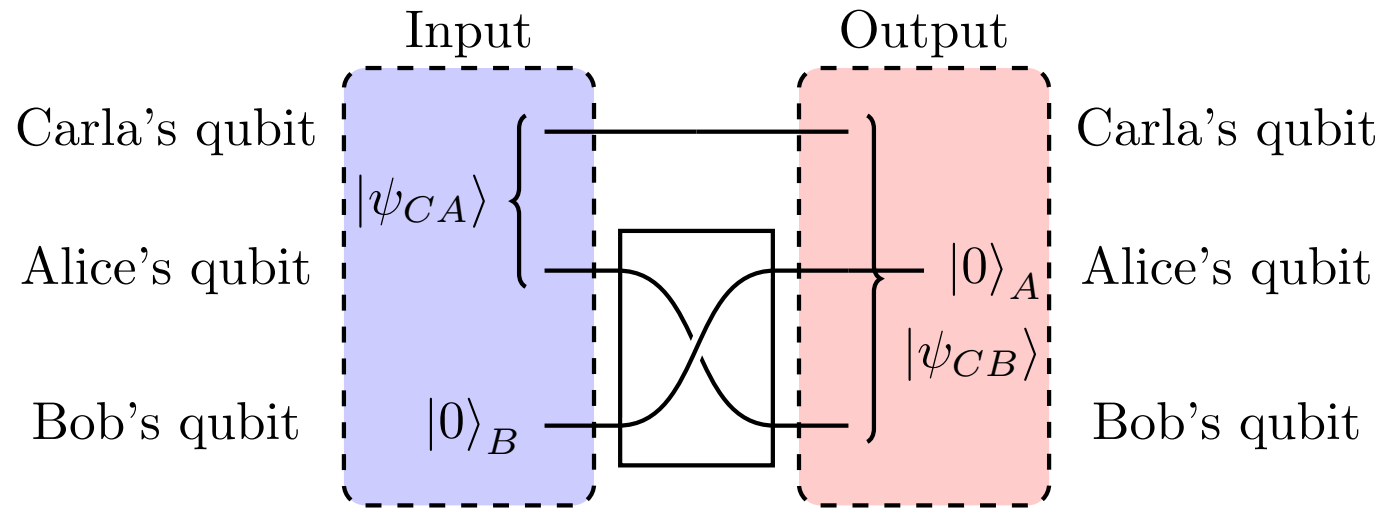

Figure 4.4: SWAP channels exchange quantum information between qubits as demonstrated with a reference qubit (Carla).

information.

The flow of quantum information is quantifiable by how much entanglement passes from the initial state to the final state. *Quantum Coherent information* is a quantitative measurement of this transfer. To understand it better, lets bring in a third qubit belonging to Carla. Let's suppose Alice and Carla entangle their qubits together and create a Bell Pair like that of Eq. 4.7. Bob's qubit is initially in a $|0\rangle$ (spin-up) state with 100% probability. We now define a quantum channel from Alice to Bob where we aim to pass the entanglement. A popular quantum channel that is well known to preserve entanglement is done using a SWAP gate. The SWAP gate will do just as it says; swap the values of Alice's and Bob's qubits and with it the information associated. We will explore the mathematical structure of the SWAP gate in Chapter 6. Figure 4.4 shows this quantum information channel, outlining the SWAP of entanglement from Alice's qubit to Bob's.

We can now define our quantum coherent information as,

$$I_C(\hat{\rho}_{in,AB}, \mathcal{N}) = S(\hat{\rho}_{out,B}) - S(\hat{\rho}_{out,BC}). \tag{4.11}$$

where, $\hat{\rho}_{in,AB}$ is the input to the channel and $\mathcal{N}$ is the channel itself. One should note here, that the Channel $\mathcal{N}$ is equivalent to $\hat{\rho}_{out,B}$. The difference in notations is that $\mathcal{N}$ expresses the operators that make up the channel, and their acting on the initial states. Whereas, $\hat{\rho}_{out,B}$ is the output state which is a result of this channel. This is because $\mathcal{N}$, mathematically is a mapping from initial states to final states. The analogy to the classical mutual information (Eq. 4.4) should be more evident now, and as we did in the classical case, we elevate this to the Quantum Channel Capacity by taking,

$$Q(\mathcal{N}) = \max_{\hat{\rho}_{in,AB}} I_C(\hat{\rho}_{in,AB}, \mathcal{N}). \tag{4.12}$$

Just as it was in classical information theory, the channel capacity is a measurement of reliability in transferring quantum information (ie. entanglement). Entanglement is one of the best metrics of a quantum computer, measuring the propagation of this entanglement, remains to be an important calculation as we further progress quantum computers.

We have spoken in detail about two channels of information, yet there remains two more. Channels that change classical information to quantum information (the process that creates entanglement) and vice versa (measurement). One thing that is a consequence of our Def. 1, is that when states are separable, the entanglement is gone and what is left is a tensor product of states. The procedure for verifying this is well established but will be fruitful for us to see written out.

## 4.4 Determining if a Quantum Channel is Entanglement Breaking

### 4.4.1 The Quantum Channel

Up until this point we have spoken about the quantum channel at surface level. For the remainder of the chapter we unpack some of the mathematics and show how these aforementioned ideas are shaped into understanding quantum information in quantum circuit.

Suppose there exists a quantum system A and an environment system $\phi$. We define a quantum channel $\Xi_{A\phi}$ as a linear, completely positive, and trace preserving mapping corresponding physical evolution from a state $\hat{\rho}_A \in \mathcal{H}_A$ to an environment (in our case the field $\phi$) $\hat{\rho}_\phi \in \mathcal{H}_\phi$ which is arranged as a pure state (eg. $|0\rangle\langle 0|_\phi$). The initial state of the whole system is $\hat{\rho}_{A\phi} = \hat{\rho}_A \otimes |0\rangle\langle 0|_\phi$ where $\hat{\rho}_{A\phi} \in (\mathcal{H}_A \otimes \mathcal{H}_\phi)$ and there exists some interaction between the systems with Unitary Operator $\hat{U}_{A\phi}$ such that after the interaction, we can find a state $\Xi_{A\phi}(\hat{\rho}_A)$ from the channel

$$\Xi_{A\phi}(\hat{\rho}_A) = \mathrm{Tr}_A\{\hat{U}_{A\phi}(\hat{\rho}_A \otimes |0\rangle_\phi \langle 0|)\hat{U}^\dagger_{A\phi}\} \tag{4.13}$$

it is helpful here to rewrite our initial state as

$$\hat{\rho}_{A\phi} = \hat{\rho}_A \otimes |0\rangle_\phi \langle 0| = (\mathbb{1}_A \otimes |0\rangle_\phi)(\hat{\rho}_A)(\mathbb{1}_A \otimes \langle 0|_\phi) \tag{4.14}$$

as well as understand a partial trace to be defined as

$$\mathrm{Tr}_B\{X_{AB}\} \equiv \sum_l \langle l|_B X_{AB} |l\rangle_B \tag{4.15}$$

plugging these two identities into Eq. 4.13 we find,

$$\Xi_{A\phi}(\hat{\rho}_A) = \sum_l (\mathbb{1}_A \otimes \langle l|_\phi)\hat{U}_{A\phi}(\mathbb{1}_A \otimes |0\rangle_\phi)(\hat{\rho}_A)(\mathbb{1}_A \otimes |0\rangle_\phi)\hat{U}^\dagger_{A\phi}(\mathbb{1}_A \otimes |1\rangle_\phi) \tag{4.16}$$

$$= \sum_l \hat{M}_l \hat{\rho}_A \hat{M}_l^\dagger \tag{4.17}$$

where we have defined $\hat{M}_l \equiv (\mathbb{1}_A \otimes \langle l|_\phi)\hat{U}_{A\phi}(\mathbb{1}_A \otimes |0\rangle_\phi)$. This operator will become important when finding our if our channel is entanglement breaking.

### 4.4.2 Separable States

Before, we get into entanglement breaking we need to identify separability of states for our system. To do this we describe a joint ensemble of our systems A and $\phi$

$$\{p_l(l), |A\rangle_l \otimes |\phi\rangle_l\} \tag{4.18}$$

where $p_l(l)$ is some probability distribution and the set $\{|A_l\rangle\}$ and $\{|\phi_l\rangle\}$ are sets of pure states on their respective system. We define the density operators of our system A and environment $\phi$ separately (ignoring the other exists) to be,

$$\hat{\rho}_A = \sum_l p_l(l)|A_l\rangle\langle A_l| \quad \hat{\rho}_\phi = \sum_l p_l(l)|\phi_l\rangle\langle\phi_l| \tag{4.19}$$

so the density operator of this ensemble is given as,

$$\hat{\rho}_{A\phi} = \sum_{l} p_l(l)(|A_l\rangle\langle A_l| \otimes |\phi_l\rangle\langle\phi_l|) \tag{4.20}$$

which classifies the density operator $\hat{\rho}_{A\phi}$ as "separable" by definition. It can be noted here that a density operator that is not separable is classified as "entangled".

### 4.4.3 Breaking Entanglement

For a deeper dive into this material, please see Refs. [2, 4]. Now, we revisit our operators $\hat{M}_l$ in Eq. 4.17 and $\hat{U}_{A\phi}$ that exists in $\hat{M}_l$. We wish to show that if $\hat{U}_{A\phi}$ is a Schmidt rank-one unitary like that of Eq. 2.20, than the density operator $\hat{\rho}_{A\phi}$ is separable[2]. Let us suppose we entangle our qubit In system A with some ancillary qubit in another system C. We define the bipartite density operator $\hat{\rho}_{AC} \in \mathcal{H}_A \otimes \mathcal{H}_C$ and since $\hat{\rho}_{AC}$ is entangled than it is not separable.

We then define a Channel from A and C to the environment $\phi$

$$(\mathbb{1}_C \otimes \Xi_{A\phi})(\hat{\rho}_{CA}) = \mathrm{Tr}_A\{\hat{U}_{A\phi}(\hat{\rho}_{AC} \otimes |0\rangle_\phi \langle 0|)\hat{U}^\dagger_{A\phi}\} \tag{4.21}$$

The observable defined by our two-state qubit can be rewritten as a projection onto spin up or spin down,

$$\hat{\mu}_A = \sum_{s\in\{\pm\}} s\hat{P}_s \tag{4.22}$$

[2]we show this for a single qubit in system A but it remains true for any number of qubits in A.

With this decomposition of $\hat{\mu}_A$ we can express our single rank one unitary $\hat{U}_{A\phi}$ as,

$$\hat{U}_{A\phi} = \sum_{s\in\{\pm\}} \hat{P}_s \otimes \hat{U}_s \tag{4.23}$$

Where our operator $\hat{U}_s$ follows from Eq. 2.20, and is therefore $\hat{U}_s = \exp(is\lambda\hat{\mathcal{O}})$. We can then rewrite Eq. 4.21 as,

$$(\mathbb{1}_C \otimes \Xi_{A\phi})(\hat{\rho}_{CA}) = \sum_s \mathrm{Tr}_A\{\hat{P}_s(\hat{\rho}_{AC} \otimes \hat{U}_s \left|0\right\rangle_\phi \left\langle 0\right| \hat{U}_s^\dagger). \tag{4.24}$$

We then recognize the following;

$$\begin{aligned} p_s(s) &= \mathrm{Tr}\{\hat{P}_s\hat{\rho}_{AC}\} \\ \mathrm{Tr}_A\{\hat{P}_s\hat{\rho}_{AC}\} &= \sum_s \left\langle s\right|_A \hat{P}_s\hat{\rho}_{AC} \left|s\right\rangle_A = p_s(s)\hat{\rho}_{AC}(s) \\ \hat{\rho}_\phi(s) &= \hat{U}_s \left|0\right\rangle_\phi \left\langle 0\right| \hat{U}_s^\dagger \end{aligned} \tag{4.25}$$

plugging these quantities into Eq. 4.24 we get our final form of the channel,

$$(\mathbb{1}_C \otimes \Xi_{A\phi})(\hat{\rho}_{CA}) = \sum_s p(s)\hat{\rho}_{AC}(s) \otimes \hat{\rho}_\phi(s). \tag{4.26}$$

which we recognize to be separable.

# Chapter 5

# Applications of Shannon Theory to UDWQCs

## 5.1 The Skeleton of a Quantum Computer

So far we have introduced the UDW detector formalism as well as scalar fields $\hat{\varphi}$ and $\hat{\Pi}$ and have seen some overlap between the formalism of UDW detectors and HLLs. In Chapter 4, quantum communication theory was introduced as the means to find quantum channels that can process and transmit quantum information. In this chapter, we demonstrate the quantum Shannon properties that we can develop using the UDWQC model.

One might now understand the figure 1.1 better. In our model, if we implant a two-state Qdot (Alice) into a HLL, we can detect if information has been embedded into the field[1]. The information will then be attached to right- and left-moving fields which can be collected at another Qdot (Bob). Entanglement of an ancilliary qubit with Alice can be evaluated to understand the effectiveness of the entanglement propagation of the channel, otherwise known as the quantum

[1] This is the fundamental underpinning detector models. See [91]

channel's capacity.

## 5.2 A Brief Review of the System

We have shown that our "naive" Hamiltonian Eq. 2.5 was constructed specifically to demonstrate the connection of UDW detectors to a CFT (namely HLLs) to realize qubit-field transduction. Given that we wish to take our system and extend it into general quantum Shannon theory, we redefine our Eq. 2.5 as

$$\hat{H}_{\text{int}}(t) = \lambda\chi(t)\int_{\mathbb{R}} dy\ p(x(t),y)\hat{\mu}(t)(\hat{\psi}_+^\dagger\hat{\psi}_+ - \hat{\psi}_-^\dagger\hat{\psi}_-) \tag{5.1}$$

so that we can begin to implement constraints and simulate numerically, quantum Shannon theory properties. Notice in Eq. 5.1 we have used a more generic form of coupling $\lambda$ which is common in QIS. We will remain with this coupling notation for this chapter. As we did in Chapter 2, we have suppressed the terms containing the $e^{\pm 2ik_F x}$ through a Gaussian smearing function.

An interpretation of this quantum computer thus far, may be considered as follows through the methodology of UDW detectors: Consider the spin-state of our Qdot Alice at some initial time $t_A$. At this time we allow an interaction between Alice and the HLL. The information is then smeared onto the left- and right-moving components of the field through the smearing function $p(x(t_A), y)$[2]. It is then collected by Qdot Bob at some time $t_B$. Given the time evolution of the field one could expect that the the function required for decoding the information will be different than that of the encoding function. At least at the level of the correlator that is surely true and a calculable quantity. At most, for field-mediated processes, this extra value will end up being an overall phase. For now, we presume that in one dimension, it remains to be the same shape as the smearing function at Alice.

[2]Smearing and collecting information will also be referred to as encoding and decoding the information respectively.

Without strong coupling, our channel from Alice to Bob has a perturbatively limited quantum capacity. Whereas, the goal of quantum computing is to reach maximum quantum capacity of one. With a perturbative approach we can find an array of literature describing the interaction Hamiltonians to analyze. However, the literature on strong coupling is severely more limited. We are subject to study a case that is well understood with the aim to uncover more systems in future work. What is understood in earnest are interactions between our two state Qdot system and a scalar bosonic field. Implementing what was shown in Chapter 3 we can express Eq. 5.1 as,

$$\hat{H}_{\text{int}}(t) = \lambda\chi(t)\int_{\mathbb{R}} dy\ p(x(t),y)\hat{\mu}(t)(\frac{1}{\sqrt{\pi}}(\partial_z\hat{\phi} + \partial_{\bar{z}}\hat{\bar{\phi}})) \tag{5.2}$$

$$= \lambda\chi(t)\int_{\mathbb{R}} dy\ p(x(t),y)\hat{\mu}(t)(\frac{1}{\sqrt{\pi}}(2\hat{\Pi})) \tag{5.3}$$

These two equations provide us with our first to versions of quantum gates to explore numerically. Unfortunately, Eq. 5.3 constructs a rank-one unitary operator which behaves as a controlled unitary operator for qubit-field interactions. Controlled-NOT (CNOT) gates, also controlled unitaries, are known to be destructive to QI and as was shown at the end of Chapter 4, so are the channels made up of this controlled unitary. However two CNOT gates, in series, is known as a quantum state transfer and acts as the ideal comparison for our UDWQC when considering the propagation of QI. This comparison will be demonstrated in Chapter 6 via the diamond norm.

## 5.3 UDWQC Quantum Channels

Recall, an entanglement breaking channel has the necessity that the information at some point throughout the process becomes classical. This is an indication that whether some observer recognizes it or not, a measurement takes place throughout the evolution processes. Initializing our field in a pure state $|0\rangle\langle 0|_\phi$, we define our mapping $\Xi_{A\to\phi}$ as the mapping from initial states of system A and field $\hat{\phi}$ to final states of $\hat{\phi}$ which is given by the equation,

$$\Xi_{A\to\phi}(\hat{\rho}_A) = \mathrm{Tr}_A\{\hat{U}_{A\phi}(\rho_A \otimes |0\rangle_\phi\langle 0|)\hat{U}^\dagger_{A\phi}\} \tag{5.4}$$

where $\Xi_{A\to\phi}(\hat{\rho}_A)$ is some state of the field $\hat{\phi}$ after the interaction between $A$ and $\hat{\phi}$ and $\hat{U}_{A\phi}$ is a Unitary operator. We will discuss in Chapter 7, the relevance that this system is relativistic. While we remain in euclidean space (through Wick rotation) through this thesis, we can see that these circuits fit naturally onto any given manifold.

Now suppose we entangle our system A with some external system C. Our new quantum channel takes the shape,

$$(\mathbb{1}_C \otimes \Xi_{A\phi})(\hat{\rho}_{CA}) = \mathrm{Tr}_A\{\hat{U}_{A\phi}(\hat{\rho}_{AC} \otimes |0\rangle_\phi\langle 0|)\hat{U}^\dagger_{A\phi}\} \tag{5.5}$$

where our $\hat{\rho}_{AC}$ is an entangled bipartite state between System $A$ and $C$. If at some point we find that the system C and the field states in $\hat{\rho}_{A\phi}$ are separable ie. $\hat{\rho}_{AC} \otimes \hat{\rho}_{\phi}$ than our entanglement has been broken. As shown in Chapter 4, and consistent with the literature [2, 47, 92] this channel is broken as it can be written in the form,

$$(\mathbb{1}_C \otimes \Xi_{A\to\phi})(\sigma_{CA}) = \sum_s p(s)\rho_{AC}(s) \otimes \rho_{\phi}(s). \tag{5.6}$$

The entire quantum channel in question here, is from Alice to Bob (our two Qdots). This will be represented by a longer quantum channel,

$$\Xi_{A\to B}(\hat{\rho}_A) = \mathrm{Tr}_{A\phi}\{\hat{U}_{\phi B}\hat{U}_{A\phi}(\hat{\rho}_{A\phi}\,|0\rangle_{\phi}\langle 0|\,\hat{\rho}_{\phi B})\hat{U}^{\dagger}_{A\phi}\hat{U}^{\dagger}_{\phi B}\} \tag{5.7}$$

and visualized in Fig. 5.1. We now aim to address the types of unitaries that preserve entanglement and propagate it through this channel from Alice to Bob.

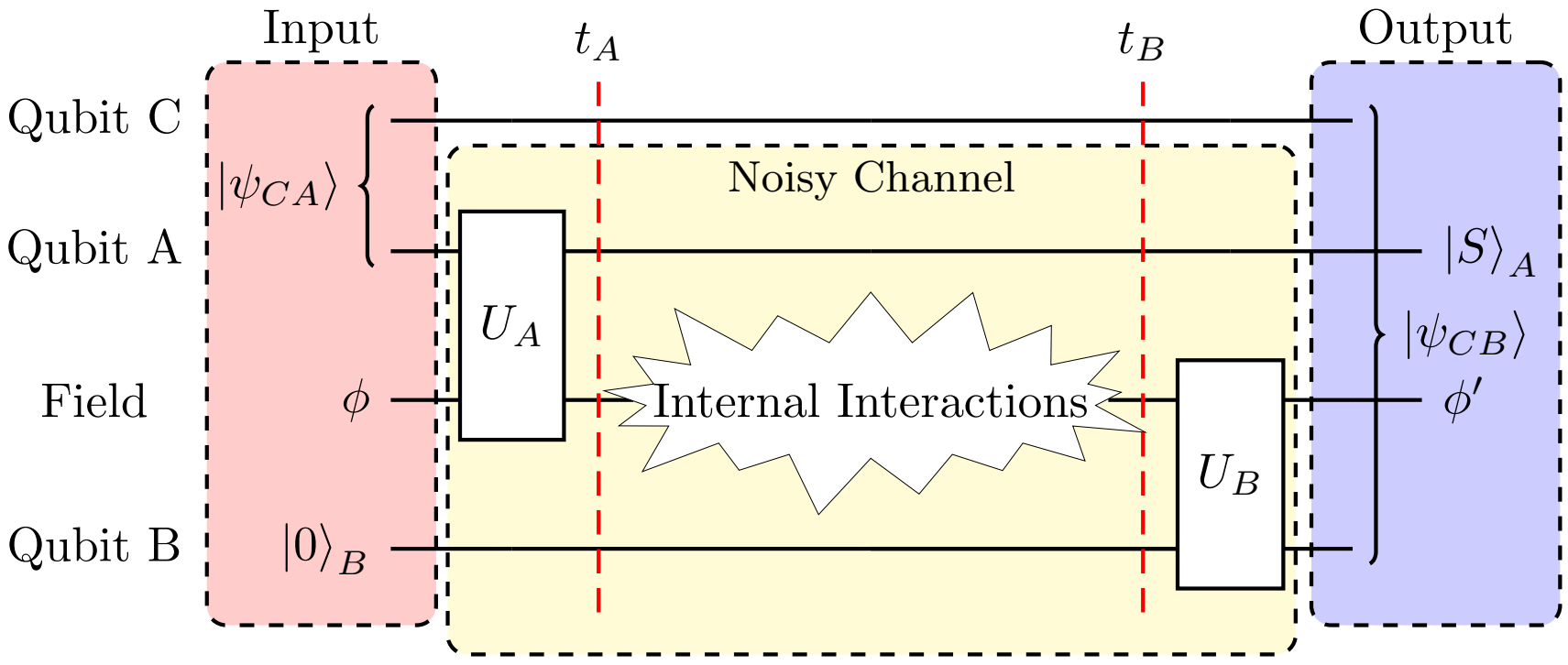

Figure 5.1: Encoding and Decoding quantum information onto and off of the field $\hat{\varphi}$.

## 5.4 Possible Gates For Our Quantum Computer

### 5.4.1 Gate Formations of Interest

We have found that our system (UDWQC) is adequate for transferring information onto and off of our fields that have the same form as the fields of Eq. 2.1. As discussed in Chapter 2, it is known in the literature that the minimum necessity for an entanglement safe channel is two Schmidt rank-one unitaries that have the shape,

$$\hat{U}_\nu = \exp\left(i\lambda_{\nu 2}\hat{\mu}_{\nu 2} \otimes \hat{\mathcal{O}}_{\nu 2}\right)\exp\left(i\lambda_{\nu 1}\hat{\mu}_{\nu 1} \otimes \hat{\mathcal{O}}_{\nu 1}\right). \tag{5.8}$$

We now explore regimes that have potential to realize this type of unitary in our LL scheme.

### 5.4.2 Left- and Right-Moving Gates

If we rewrite Eq. 5.1 as the following,

$$\hat{H}_{\text{int}}(t) = \chi(t)\int_{\mathbb{R}} dy\ p(x(t),y)\hat{\mu}(t)(\lambda_\alpha \hat{\psi}^\dagger_{\alpha,+}\hat{\psi}_{\alpha,+} - \lambda_\beta \hat{\psi}^\dagger_{\beta,-}\hat{\psi}_{\beta,-}) \tag{5.9}$$

where the coupling constants are now defined with separate coupling to left- and right-movers. Following the formalism in Chapter 3 we can find a bosonized form of Eq. 5.2

$$\hat{H}_{\text{int}}(t) = \chi(t)\int_{\mathbb{R}} dy\ p(x(t),y)\hat{\mu}(t)\left(\frac{1}{\sqrt{\pi}}(\lambda_\alpha\partial_z\hat{\phi} + \lambda_\beta\partial_{\bar{z}}\hat{\bar{\phi}})\right). \tag{5.10}$$

Decomposing this Hamiltonian into $\hat{H}_{\text{int}}(t) = \hat{H}_\alpha(t) + \hat{H}_\beta(t)$ we can evaluate a single left- or right-moving interaction Hamiltonian and we express the $\partial_z\hat{\phi}$ terms as field and momentum observables such that,

$$\hat{H}_\alpha(t) = \int_{\mathbb{R}} dy\ p(x(t), y)\hat{\mu}(t)(\frac{1}{\sqrt{\pi}}(\lambda_{\alpha\phi}\partial_x\hat{\phi}(t,x) - \lambda_{\alpha\pi}\hat{\Pi}(t,x))). \tag{5.11}$$

We also include the switching function into the coupling as the expectation here is that we time order the couplings with delta-like functions. Doing this, we outline two observables that are able to satisfy Eq. 5.8. Namely,

$$\hat{U}_\nu = \exp{(i\lambda_{\alpha\phi}\hat{\mu}_\phi \otimes \partial_x\hat{\phi})}\exp{(i\lambda_{\alpha\pi}\hat{\mu}_\pi \otimes \hat{\Pi})}. \tag{5.12}$$

The change in this system is that we only work with left- and right-movers separately. This systems exists as a Chiral Luttinger Liquid (CLL) [93, 94] and as such, could offer a more complex quantum computer set up.

### 5.4.3 Dirac-type Current Hamiltonians

Another gate that will be examined thoroughly throughout this project is a dirac style Hamiltonian that considers the current $\hat{\psi}^\dagger_\pm\partial_x\hat{\psi}_\pm$. This hamiltonian applied to our UDW+HLL system takes the following shape

$$\hat{H}^D_{\text{int}}(t) = \lambda\chi(t)\int_{\mathbb{R}} dy\ p(x(t), y)\hat{\mu}(t)(\hat{\psi}^\dagger_+\partial_x\hat{\psi}_+ - \hat{\psi}^\dagger_-\partial_x\hat{\psi}_-) \tag{5.13}$$

we can follow the point-splitting procedure from Appendix B to find a different form of this Dirac Hamiltonian,

$$\hat{H}^D_{\text{int}}(t) = \int_{\mathbb{R}} dy\ p(x(t), y)\hat{\mu}(t)(\lambda_\Pi\hat{\Pi}^2 + \lambda_\varphi(\partial_x\hat{\varphi})^2) \tag{5.14}$$

which brings back our original fields from Eq. 2.1. This however is a quadratic coupling to our Qdot and has a Unitary gate of the form,

$$\hat{U}_\nu = \exp\left(i\lambda_{\nu 2}\hat{\mu}_{\nu 2} \otimes \hat{O}^2_{\nu 2}\right) \exp\left(i\lambda_{\nu 1}\hat{\mu}_{\nu 1} \otimes \hat{O}^2_{\nu 1}\right). \tag{5.15}$$

This regime of strong coupling in the UDW detector literature has not been explored but perturbatively it has been [79]. Further investigations into strong coupling of this interaction would provide unique quantum channels to explore numerically.

Presented here are two versions of our LL model that can provide quantum channels to numerically simulate quantum Shannon properties. There are of course other blocks of the Tomonaga-Luttinger liquid Hamiltonian to explore as well, but as demonstrated in the end of Chapter 3, there are other problems with the coherent states of those unique field operators to tackle first. With the concrete (and not-so-concrete) channels in place we move into the numerical simulations and introduce new formalism that decreases our computational workload.

## 5.5 Unruh–DeWitt Bosonic Dephasing-like Channels

### 5.5.1 The UDW Channel as a Bosonic Dephasing Channel

In quantum Shannon theory it is often useful to discuss entanglement in multipartite channels that experience decoherence as dephasing channels, usually under the guise of interactions with the environment [95, 96]. To that regard, we may see a similarly behaving channel in the Fock basis as

$$\mathcal{N}_\gamma(\hat{\rho}) = \sum_{m,n=0}^{\infty} e^{-\frac{\gamma}{2}(n-m)}(c_n c_m^*)|n\rangle\langle m| \tag{5.16}$$

where $\gamma$ is the factor that determines the rate of decoherence. For example as $\gamma \to \infty$ the diagonal terms vanish. We revisit the concept of dephasing due to the environment in Sec. 5.6.1.

In the case of the standard UDW channel, we aim to show that the dephasing is a necessity following the non-orthogonality of the coherent states. Generating the off-diagonals of the coherent state density matrix, scrambles the encoded information from $\hat{U}_{A\varphi}$ as the controlled structure of our unitaries encodes (decodes) QI onto (off of) the diagonals of the coherent state density matrix. Utilizing the dephasing perspective, along with our free parameters to send these states to zero, can accomplish near perfect channel capacity.

To show the relation between the current formalism and the proposed dephasing perspective, we can reformulate the state transfer channel of Ref. [2] to

$$\begin{aligned}\hat{U}_{\nu\varphi} = \sum_{z,x\in\pm} \hat{P}_x \hat{P}_z \otimes \exp\left[ix \int dk\, \sqrt{\gamma_\Pi(k)}\hat{\Pi}_\nu(k)\right] \\ \times \exp\left[iz \int dk\, \sqrt{\gamma_\varphi(k)}\hat{\varphi}_\nu(k)\right]\end{aligned} \tag{5.17}$$

where we have set

$$\hat{\varphi}(\tilde{\gamma})_\nu := \lambda_\varphi \int dk \tilde{F}_\nu(k)\hat{\varphi}(k,t_\nu) = \int dk \sqrt{\gamma_\varphi(k)}\hat{\varphi}(k,t_\nu) \tag{5.18}$$

$$\hat{\Pi}(\tilde{\gamma})_\nu := \lambda_\Pi \int dk \tilde{F}_\nu(k)\hat{\Pi}(k,t_\nu) = \int dk \sqrt{\gamma_\Pi(k)}\hat{\Pi}(k,t_\nu). \tag{5.19}$$

Instead of redefining the coherent amplitudes to include our free parameters as we did in Sec. 2.3.3, we have wrapped the free parameters up in some function $\gamma_O(k)$ and leave the coherent amplitudes as they exist in the field observables $\hat{O}$. This simplification allows for general treatments of the channel regardless of the linearity of the ladder operators in the field observables.

Now carrying out the inner product of Eq. 2.30 yields

$$|\langle+\sqrt{\gamma_\varphi(k)}\alpha(k)|-\sqrt{\gamma_\varphi(k)}\alpha(k)\rangle| = e^{-2\gamma_\varphi(k)|\alpha(k)|^2} \tag{5.20}$$

where we have utilized the non-orthogonality of coherent states identity

$$\langle\beta(k)|\alpha(k)\rangle = \exp\left(-\frac{1}{2}|\alpha(k)|^2 - \frac{1}{2}|\beta(k)|^2 + \beta^*(k)\alpha(k)\right). \tag{5.21}$$

It is obvious from Eq. 5.20 that the coherent states are part of a bosonic dephasing channel $\Xi_{\varphi\to\varphi'}$ with dephasing constant $\sqrt{\gamma_O(k)}$. However, unlike canonical dephasing channels the coupling, smearing, and switching are the mechanisms of dephasing. Sending our dephasing function $\gamma_\varphi(k) \to \infty$ for any value $k$, removes off diagonal elements from this channel. This should be expected following the constraints in Sec. 2.3.3.

## 5.6 Applications of a UDW Channel From a Bosonic Dephasing perspective.

The main result of the previous section is that in the RQI channel outlined in Fig. 2.1, we can model the coupling constant and smearing function as a de-phasing function (similar to the dephasing parameter). For the remainder of this chapter, we demonstrate the results of this dephasing-like model using numerical simulations. For this purpose, we utilize a Gaussian smearing function and subsequently treat the dephasing function as a constant $\gamma_{\hat{O}}$ with a single mode $k$. Since we aim to take advantage of the new perspective, let's first consider the canonical dephasing channel.

### 5.6.1 Canonical Bosonic Dephasing Channel

The canonical dephasing channel is one where the system interacts with the environment, and the QI signal depreciates. Therefore, given the paradoxical result in our UDW system, outlining such a dephasing channel (one where the field is allowed to interact with the environment), may provide such a boost in QI through our channel. However, given that dephasing channels acts on the diagonal components of the density matrix we may find that the unwanted states remain undeterred.

To evaluate this channel, we begin with a standard unitary that takes advantage of the number operator

$$\hat{U}_{\mathrm{E}} = e^{-i\sqrt{\gamma_E}\hat{a}^\dagger\hat{a}(\hat{b}+\hat{b}^\dagger)} \tag{5.22}$$

where $\gamma_E$ is the dephasing constant associated with the interaction between the field and environment and $\hat{b}^\dagger(\hat{b})$ are the creation (annihilation) operators of the environment. We can model the channel's effect on the field by writing out the

composite channel

$$\Xi_{\varphi\to\varphi',E} = \mathrm{Tr}_E[\hat{U}_\mathrm{E}(\hat{\rho}_{1,\varphi}\otimes|0\rangle_E\langle 0|)\hat{U}_\mathrm{E}^\dagger]. \tag{5.23}$$

Implanting this in the usual prescription of bosonic dephasing channels, similar to that of Eq. 5.30, we get

$$\Xi_{\varphi\to\varphi',E}(\hat{\rho}_\varphi) = \int_{-\infty}^{\infty} d\phi\, p(\phi)\, e^{-i\hat{a}^\dagger\hat{a}\phi}\hat{\rho}_{1,\varphi}e^{i\hat{a}^\dagger\hat{a}\phi}, \tag{5.24}$$

where $\Xi_{\varphi\to\varphi',E}(\hat{\rho}_\varphi)$ represents the dephasing channel and the effects on our field $\hat{\varphi}$, and $p(\phi)$ is a Gaussian probability density given by

$$p(\phi) = \sqrt{\frac{1}{2\pi\gamma_E}}e^{-\frac{1}{2}\frac{\phi^2}{\gamma_E}}. \tag{5.25}$$

To evaluate this channel we convert to the Fock basis and continue by substituting in Eq. 5.32 and simplifying Eq. 5.24 which then becomes

$$\Xi_{\varphi\to\varphi',E}(\hat{\rho}_\varphi) = \sum_{s,s'\in\pm}\int_{-\infty}^{\infty} d\phi\, p(\phi)\times e^{-i\hat{a}^\dagger\hat{a}\phi}|s\sqrt{\gamma_\varphi}\alpha\rangle\langle s'\sqrt{\gamma_\varphi}\alpha|e^{i\hat{a}^\dagger\hat{a}\phi}$$
$$= \sum_{s,s'\in\pm}\sum_{m,n}\int_{-\infty}^{\infty} d\phi\, p(\phi)e^{-\frac{1}{2}(\gamma_\phi|\alpha|^2(s^2+s'^2))}\times\frac{(s\sqrt{\gamma_\varphi}\alpha)^n(s'\sqrt{\gamma_\varphi}\alpha^*)^m}{\sqrt{n!}\sqrt{m!}}e^{-i\hat{a}^\dagger\hat{a}\phi}|n\rangle\langle m|e^{i\hat{a}^\dagger\hat{a}\phi}. \tag{5.26}$$

Evaluating the operators we obtain

$$\Xi_{\varphi\to\varphi',E}(\hat{\rho}_\varphi) = \sum_{s,s'\in\pm}\sum_{m,n}\int_{-\infty}^{\infty} d\phi\, p(\phi) e^{-i\phi(m-n)} \times e^{-\frac{1}{2}(\gamma_\phi|\alpha|^2(s^2+s'^2))} \frac{(s\sqrt{\gamma_\varphi}\alpha)^n(s'\sqrt{\gamma_\varphi}\alpha^*)^m}{\sqrt{n!}\sqrt{m!}}|n\rangle\langle m|. \quad (5.27)$$

Carrying out the integral yields

$$\Xi_{\varphi\to\varphi',E}(\hat{\rho}_\varphi) = \sum_{s,s'\in\pm}\sum_{m,n} e^{-\frac{1}{2}(m-n)^2\sqrt{\gamma_E}} \times e^{-\frac{1}{2}(\gamma_\varphi|\alpha|^2(s^2+s'^2))} \frac{(s\sqrt{\gamma_\varphi}\alpha)^n(s'\sqrt{\gamma_\varphi}\alpha^*)^m}{\sqrt{n!}\sqrt{m!}}|n\rangle\langle m|. \quad (5.28)$$

It is evident that when $m = n$, we get no effect from the dephasing channel. Moreover, when considering the change that this dephasing makes to our original correlator in Eq. 2.25, we trace over the final state of the field and assess the inner product of our new coherent states. What we find is that non-zero results require $m = n$. Therefore, acting on the field with the number operator results in an inner product identical to Eq. 5.20 and subsequently will not change the coherent information of the UDW channel. This result indicates that in this idealized dephasing channel, the environment will have no effect on the quantum information through our channel. Coincidentally, the condensed matter systems where these channels have been proposed, take advantage of topologically protected edge states to transmit quantum information [18]. Therefore, interactions with the environment were expected to be minimal regardless.

### 5.6.2 Cross-Talk Noise

#### Setting up a UDW Noise Channel

One possible generation of unwanted disturbances of quantum information in proposed condensed matter approaches to UDW channels, is unwanted interactions with detectors, a type of Cross-Talk (CT) noise (see Fig. 5.2). Traditional calculations of these additional interactions would increase the amount of vertex operators in the correlator of Eq. 2.25, making computational times significantly longer as well as more involved. However, considering the dephasing perspective, we instead can incorporate additional detector effects into the dephasing parameter.

Let's assume that the new gate is defined as

$$\hat{U}_{\mathrm{CT}} = \sum_{z\in\pm} \hat{P}_z \otimes \exp\left[iz\sqrt{\gamma_N}\hat{\varphi}_\nu\right], \tag{5.29}$$

similar to that of Eq. 2.20. Following the prescription in Sec. 5.6.1 we model the channel as

$$\Xi_{\varphi\to\varphi',\mathrm{N}}(\hat{\rho}_\varphi) = \int_{-\infty}^{\infty} d\phi\, p(\phi)\, e^{i\hat{\varphi}\phi}\hat{\rho}_{1,\varphi}e^{-i\hat{\varphi}\phi} \tag{5.30}$$

which expresses the channel $\Xi_{\varphi\to\varphi',\mathrm{N}}(\hat{\rho}_\varphi)$, a noisy composite channel describing the field evolution throughout the channel $\Xi_{A\to B,\mathrm{N}}$. The channel indicates the probability (which follows a random probably distribution $p(x)$ [95, 96]) that the CT detector produces a different coherent state. For this calculation we will utilize the probability distribution of Eq. 5.25 with the substitution of the dephasing parameter $\gamma_N$. $\hat{\rho}_{1,\varphi}$ is the state after the interaction between the field and qubit

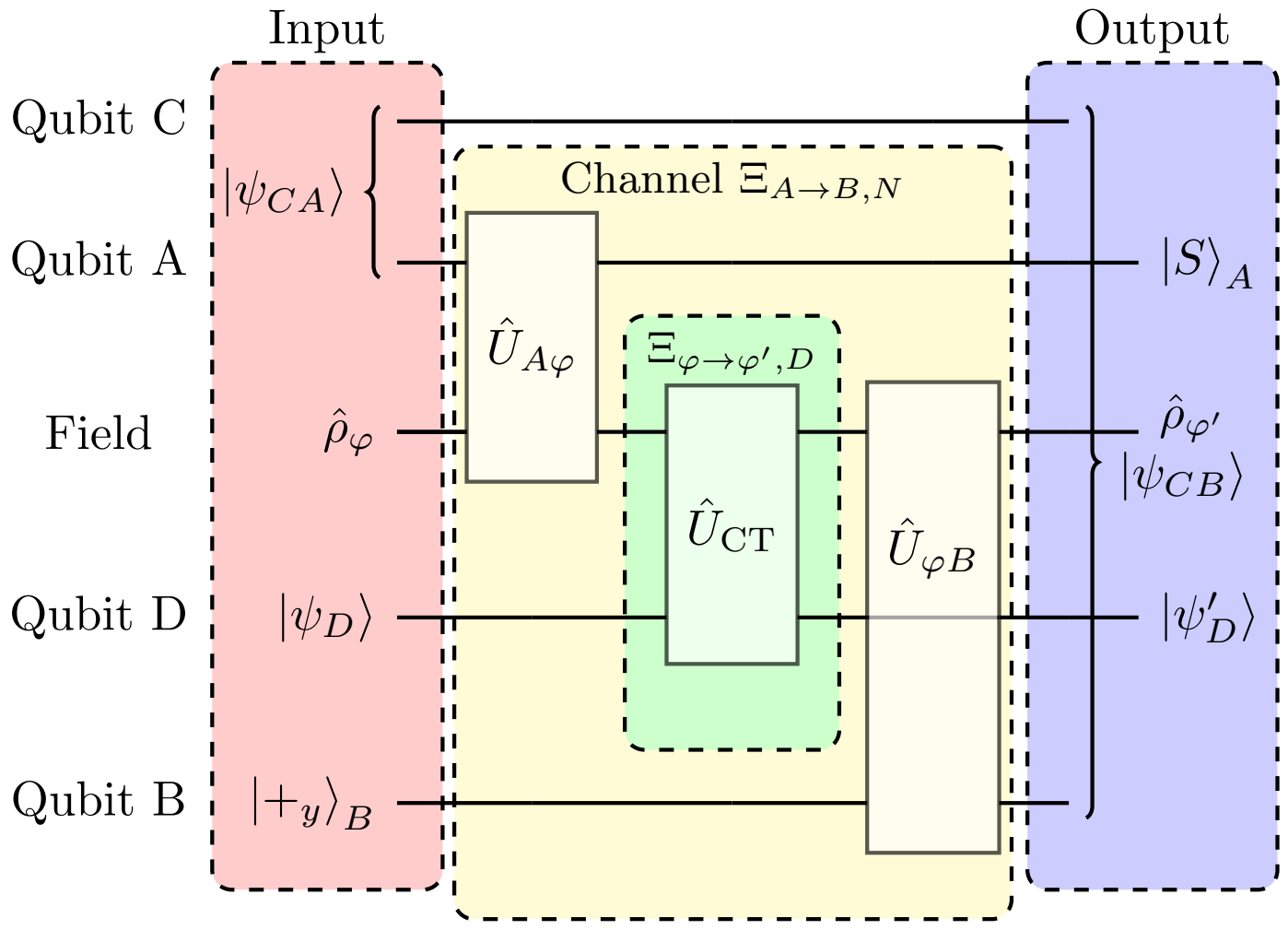

Figure 5.2: Introducing another UDW gate acts as additional information entering the channel and subsequently will introduce cross-talk noise to the original channel in Eq. 2.24.

A and has the explicit density form

$$\hat{\rho}_{1,\varphi} = \sum_{s,s'\in\pm} |s\sqrt{\gamma_\varphi}\alpha\rangle\langle s'\sqrt{\gamma_\varphi}\alpha| \tag{5.31}$$

$$= \sum_{s,s'\in\pm} e^{s\sqrt{\gamma_\varphi}\hat{\varphi}}|0\rangle\langle 0|e^{-s'\sqrt{\gamma_\varphi}\hat{\varphi}}. \tag{5.32}$$

Substituting Eq. 5.32 into Eq. 5.30 we get

$$\Xi_{\varphi\to\varphi',\mathrm{N}}(\hat{\rho}_\varphi) = \sum_{s,s'\in\pm} \int_{-\infty}^{\infty} d\phi\, p(\phi)$$

$$\times\, e^{i(\phi+\sqrt{\gamma}_\varphi)\hat{\varphi}}\hat{\rho}_\varphi e^{i(\phi+\sqrt{\gamma}_\varphi)\hat{\varphi}} \tag{5.33}$$

$$= \sum_{s,s'\in\pm} \int_{-\infty}^{\infty} d\phi\, p(\phi)$$

$$\times\, |s(\phi+\sqrt{\gamma}_\varphi)\alpha\rangle\langle s'(\phi+\sqrt{\gamma}_\varphi)\alpha|. \tag{5.34}$$

As was shown in Sec. 2.3.3, the unwanted states result from the correlator formed by the partial trace over the field and the non-orthogonality of coher-ent states. Tracing over $\Xi_{\varphi\to\varphi',\mathrm{N}}(\hat{\rho}_\varphi)$ to reproduce the effect of the correlator in Eq. 2.25 will demonstrate the changes made to the inner product of the coherent states which we have denoted by $\langle\Xi_{\varphi\to\varphi',\mathrm{N}}(\hat{\rho}_\varphi)\rangle$ and defined as

$$\mathrm{Tr}\,\Xi_{\varphi\to\varphi',\mathrm{N}}(\hat{\rho}_\varphi) \equiv \langle\Xi_{\varphi\to\varphi',\mathrm{N}}(\hat{\rho}_\varphi)\rangle = \sum_{s,s'\in\pm}\int_{-\infty}^{\infty} d\phi\, p(\phi)\times\langle s'(\phi+\sqrt{\gamma_\varphi})\alpha|s(\phi+\sqrt{\gamma_\varphi})\alpha\rangle$$
$$= \sum_{s,s'\in\pm}\int_{-\infty}^{\infty} d\phi\, p(\phi)\,\exp\Big(-\frac{1}{2}s'^2(\phi+\sqrt{\gamma_\varphi})^2|\alpha|^2-\frac{1}{2}s^2(\phi+\sqrt{\gamma_\varphi})^2|\alpha|^2+s's(\phi+\sqrt{\gamma_\varphi})^2|\alpha|^2\Big)$$
$$= \sum_{s,s'\in\pm}\int_{-\infty}^{\infty} d\phi\, p(\phi)\,\exp\Big(-\frac{1}{2}(\phi+\sqrt{\gamma_\varphi})^2|\alpha|^2(s'-s)^2\Big) \tag{5.35}$$

where we have utilized Eq. 5.21 as we did in Sec. 5.5.1. Notice Eq. 5.35 has the same structure as Eq. 5.20 but allows new dephasing construction of the parameters. As expected when $s'=s$ the expectation value is trivially one and will not effect the outcome of channel Eq. 2.24. However, when $s'\neq s$ we now have two parameters, $\gamma_\varphi$ and $\phi$, to enforce dephasing of unwanted states.

Plugging the above probability distribution into Eq. 5.35 and evaluating the integral for $s'\neq s$ we get

$$\langle\Xi_{\varphi\to\varphi',\mathrm{N}}(\hat{\rho}_\varphi)\rangle = \frac{\exp\Big(-\frac{2\gamma_\varphi|\alpha|^2}{1+4|\alpha|^2b^2\gamma_\varphi}\Big)}{\sqrt{1+4|\alpha|^2b^2\gamma_\varphi}} \tag{5.36}$$

where we have redefined our CT dephasing parameter as a multiple of the original $\gamma_N = b\gamma_\varphi$. Eq. 5.36 is directly comparable to Eq. 5.20. It is straightforward to see in Fig. 5.3a as $b\to 0$ the interaction with the noise is turned off, at $b=1$ the channel is the noisiest, and as $b\to\infty$ $\gamma_N$ acts as the primary dephasing factor for small values of $\gamma_\varphi$ which is demonstrated in Fig. 5.3b.

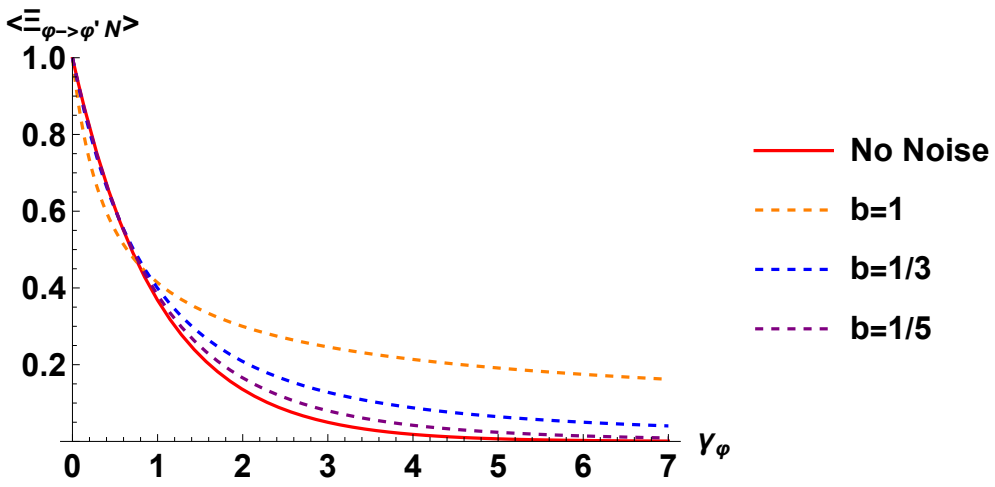

(a) As $b \to 0$ the coupling as well as the noise is turning off.

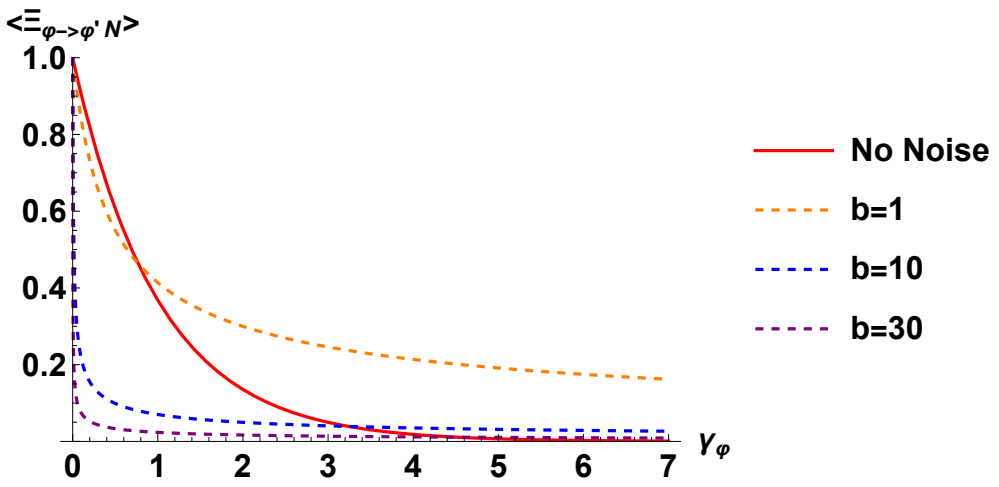

(b) As $b \to \infty$ it takes over the dephasing process for smaller values of $\gamma_\varphi$

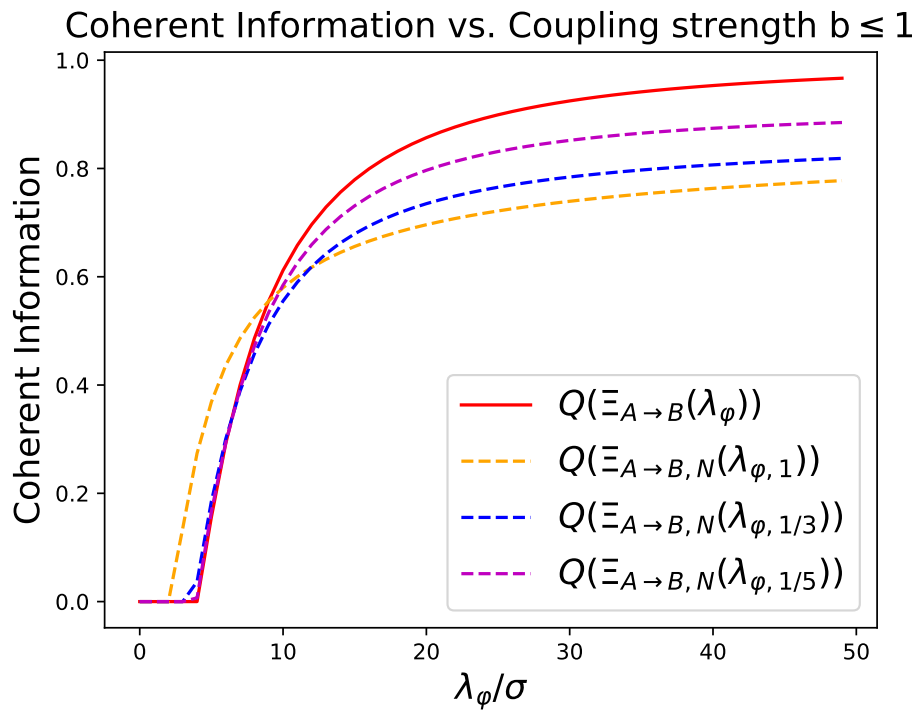

(c) As the noise is turning off the quantum capacity is increasing.

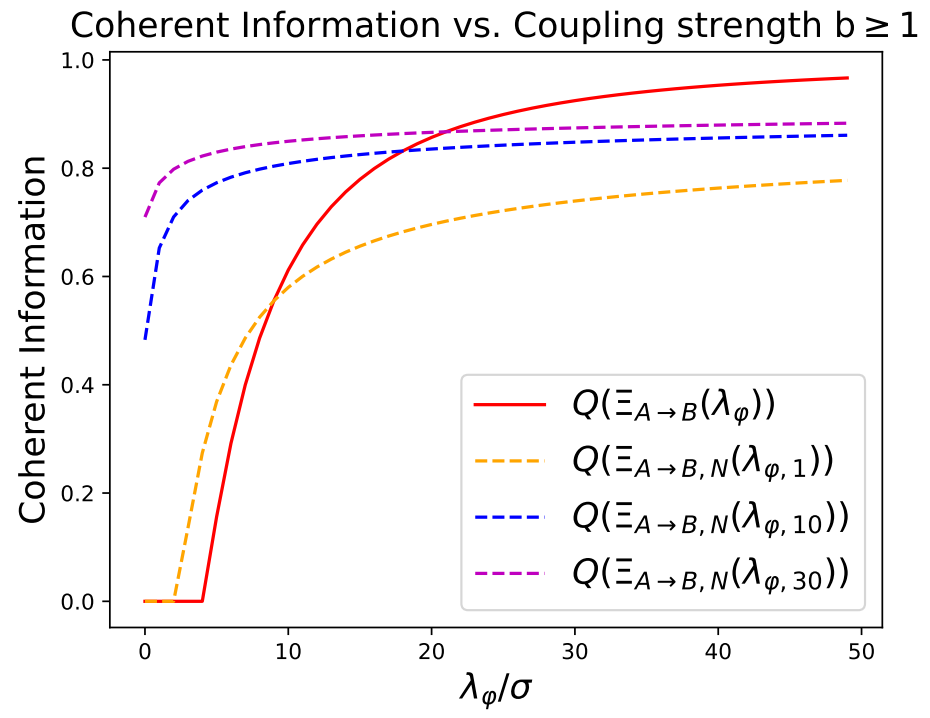

(d) For large $b$ and small values of $\lambda_\varphi$ the additional interaction can result in better quantum capacity.

Figure 5.3: Comparing the coherent information of channels with and without noise, it is clear to see that the noisy channel $\Xi_{A\to B,N}(\lambda_{\varphi,b})$ reaches a channel capacity of one at a slower rate. At $b = 1$ the noisy signal is the highest, this is due to a signal boost in the unwanted states. Regardless, we can see in (d) an increase in the lower bound of coherent information for small values of $\gamma_\varphi$. Code for simulations can be found at Ref. [1].

### Noisy UDW Effects on Coherent Information

To understand the difference these new parameters make to the coherent information we can look at the definition of $\gamma_\varphi$ in Eq. 5.18. Since $\gamma_\varphi$ is a function of $\lambda_\varphi$ we can set up the new inner product in terms of a new coupling constant $\lambda_{\varphi,b}$ given by the relation

$$\frac{\exp\left(-\frac{2\gamma_\varphi|\alpha|^2}{1+4b^2\gamma_\varphi|\alpha|^2}\right)}{\sqrt{1+4b^2\gamma_\varphi|\alpha|^2}} = \exp\left[-2(\lambda_{\varphi,b})^2 \times \int dk\, |\tilde{F}_\nu(k)|^2|\alpha|^2\right]. \quad (5.37)$$

We then solve for $\lambda_{\varphi,b}$ in terms of $\lambda_\varphi$ and ascertain

$$\lambda_{\varphi,b} = \left[\frac{\lambda_\varphi^2}{1+\frac{4b^2|\alpha|^2\lambda_\varphi^2}{\sqrt{(2\pi)^3}\sigma}} - \frac{\sqrt{(2\pi)^3}\sigma}{2|\alpha|^2} \times \ln\left(1+\frac{4b^2|\alpha|^2\lambda_\varphi^2}{\sqrt{(2\pi)^3}\sigma}\right)\right]^{\frac{1}{2}} \quad (5.38)$$

where the term $\sqrt{(2\pi)^3}\sigma$ is a consequence of Gaussian smearing with width $\sigma$. Calculating channel capacity with this new value $\lambda_{\varphi,b}$, while keeping the remaining parameters intact, still reaches near perfect channel capacity but reaches it slower, when $b \leq 1$ demonstrated by Fig. 5.3c. Furthermore, for high values of $b$ and low values of $\lambda_\varphi$ one would expect $b$ to act as the primary dephasing factor as shown in Fig. 5.3d.

In this chapter, We have shown that the formalism of quantum channels produced by UDW detectors provides a bosonic dephasing channel perspective. With this perspective, we have demonstrated that the purpose of the dephasing is to remove unwanted states that lead to the scrambling of the QI encoded on the field. We have shown that the dephasing constant can be written in terms of the strength of coupling between the qubit and field.

We aimed to demonstrate several applications of the dephasing perspective that provide interpretations of unwanted interactions. Firstly, we applied the canonical form of a bosonic dephasing channel, allowing the system to interact with the environment. In this idealized dephasing channel, the prior constraints on the off-diagonals were the only effects that remained when tracing over the field. Subsequently, this indicated no additional noise generated from interacting with the environment under this prescription.

Secondly, how does coherent information change given some noise due to ad-ditional UDW detectors? To evaluate this numerically, we presented a unitary operation that introduces CT noise to the system and calculated how that noise affects the strength of coupling between the qubits and fields. Figure 5.3 demonstrates how the additional noise can affect the propagation of quantum information through the channel given in Eq. 2.24. These values make intuitive sense as one may expect an additional signal boost to non-orthogonal off-diagonal elements of the density matrix in Eq. 5.34 which act to scramble the QI.

An open problem remaining is the possibility of writing down a unitary dephasing channel that increases coherent information overall. One might notice that non-unitary interactions that increase the strength of the diagonal elements of the coherent state density matrix while decreasing the off-diagonal elements are possible. However, given the nature of dephasing channels, accomplishing this with unitary operations requires much care.

# Chapter 6

# Universal UDWQC with Analogous Quantum Logic Gates

Elementary quantum logic gates, from which universal quantum computing is established [87,97], are a vital part of quantum information theory but are absent in the developing theory of Relativistic Quantum Information (RQI). Unruh–DeWitt (UDW) detectors have given RQI a formalism that allows for applications in quantum Shannon theory, such as mutual information, coherent information, and quantum capacity in field-mediated qubit interactions [2]. Experimental realizations of UDW-style qubits have been proposed during the last few years [18], and a set of quantum logic gates will provide insights into quantum communication and computational power in two-dimensional quantum materials [24].

The introduction of elementary quantum logic gates, designed for UDW qubits to carry out quantum processing tasks, allows for cross-discipline applications for qubit-field quantum computing. To illustrate this, we introduce elementary UDW quantum logic gates, a field-mediated equivalent to canonical qubit logic gates, to demonstrate the diamond distance, a measure of similarity, between a UDW state transfer channel and the canonical state transfer channel. Measurements like diamond distance allow for a rigorous comparison between field-mediated transduction through UDW detectors and local quantum mechanical operations.

## 6.1 Canonical Quantum Logic Gates

### 6.1.1 Projectors and Quantum Logic Gates

Behind every quantum logic gate is a series of Pauli and projection operators in computational basis states. A common example of this is the quantum "NOT" gate. Commonly denoted as $\hat{X}$, the quantum NOT gate, also known as the Pauli-X gate, has the mathematical forms

$$\hat{X} = \hat{\sigma}^x = \sum_{\mu \in \pm} \mu \hat{P}_X^\mu = |+\rangle\langle+| - |-\rangle\langle-| = |1\rangle\langle 0| + |0\rangle\langle 1| = \begin{pmatrix} 0 & 1 \\ 1 & 0 \end{pmatrix}. \tag{6.1}$$

Here we have defined the projector $\hat{P}_X^\mu \equiv |\mu_x\rangle \langle\mu_x|$ with $|\mu_x\rangle$ being the basis states in the x-basis. Similar projectors can be constructed for the z- and y-basis.

A truth table is a stylistic representation of input and output states that displays the changes that are undergone through a given process. We can generate a truth table for the NOT gate that has the following form.

NOT Gate

| Input | Output |
|---|---|
| $\|0\rangle$ | $\|1\rangle$ |
| $\|1\rangle$ | $\|0\rangle$ |

Table 6.1: A truth table is a simplified look at operating on an input state with a computational, or in our case a quantum computational operation. In this table we are acting on the input state with the NOT gate to achieve the Output state.

The other Pauli matrices have a similar formalism

Pauli-Z Gate

| Input | Output |
|---|---|
| $\|0\rangle$ | $\|0\rangle$ |
| $\|1\rangle$ | $-\|1\rangle$ |

Table 6.2: Pauli-Z truth table negates the state $|1\rangle$ while leaving the $|0\rangle$ state untouched.

Pauli-Y Gate

| Input | Output |
|---|---|
| $\|0\rangle$ | $-i\|1\rangle$ |
| $\|1\rangle$ | $i\|0\rangle$ |

Table 6.3: The Pauli-Y truth table demonstrates the rotation onto the complex plane.

$$\mathrm{Z} = \sigma^z = \sum_{z\in\pm} P_z^Z = |0\rangle\langle 0| - |1\rangle\langle 1| = \begin{pmatrix} 1 & 0 \\ 0 & -1 \end{pmatrix} \tag{6.2}$$

$$\mathrm{Y} = \sigma^y = \sum_{y\in\pm} P_y^Y = -i(|1\rangle\langle 0| - |0\rangle\langle 1|) = \begin{pmatrix} 0 & -i \\ i & 0 \end{pmatrix}, \tag{6.3}$$

with corresponding truth tables.

With these Pauli and projection operators we can now examine controlled unitary gates. Controlled gates act on multiple qubits, where some qubits are the "controls" and the other qubits act as the "targets". For reasons to be made clear, we will utilize the example of the Controlled NOT (CNOT) gate with projector form

$$\mathrm{CNOT}(1,2) = \sum_{\mu\in\pm} \hat{P}_z^\mu \otimes \hat{X}^{\frac{-\mu+1}{2}} = |0\rangle\langle 0| \otimes \mathbb{1} + |1\rangle\langle 1| \otimes \hat{X}. \tag{6.4}$$

To construct a reverse CNOT gate we can interchange the tensored values such as

$$\mathrm{CNOT}(2,1) = \sum_{\mu\in\pm} X^{\frac{-\mu+1}{2}} \otimes P_z^\mu \tag{6.5}$$

but the CNOT gate also has the alternative form

$$\text{CNOT}(2,1) = \sum_{\mu\in\pm} \hat{P}_x^{\mu} \otimes \hat{Z}^{\frac{-\mu+1}{2}} = |+\rangle\langle+| \otimes \mathbb{1} + |-\rangle\langle-| \otimes \hat{Z}. \tag{6.6}$$

This alternative form will be helpful when considering field-mediated logic gates.

These two CNOT gates are demonstrated with the following truth tables;

CNOT(1,2)

| Input | | Output | |
|---|---|---|---|
| Control | Target | Control | Target |
| $\|0\rangle$ | $\|0\rangle$ | $\|0\rangle$ | $\|0\rangle$ |
| $\|0\rangle$ | $\|1\rangle$ | $\|0\rangle$ | $\|1\rangle$ |
| $\|1\rangle$ | $\|0\rangle$ | $\|1\rangle$ | $\|1\rangle$ |
| $\|1\rangle$ | $\|1\rangle$ | $\|1\rangle$ | $\|0\rangle$ |

Table 6.4: A CNOT(1,2) will flip the entry of the target gate when the control gate is the $|1\rangle$ state.

CNOT(2,1)

| Input | | Output | |
|---|---|---|---|
| Target | Control | Target | Control |
| $\|0\rangle$ | $\|0\rangle$ | $\|0\rangle$ | $\|0\rangle$ |
| $\|0\rangle$ | $\|1\rangle$ | $\|1\rangle$ | $\|1\rangle$ |
| $\|1\rangle$ | $\|0\rangle$ | $\|1\rangle$ | $\|0\rangle$ |
| $\|1\rangle$ | $\|1\rangle$ | $\|0\rangle$ | $\|1\rangle$ |

Table 6.5: The CNOT(2,1) gate switched the roles of the qubits in the system.

### 6.1.2 Quantum Computing Operations from Logic Gates

#### Quantum State Transfer

It is well-known that some fundamental quantum processes can be achieved through the combinations of single and multiple qubit quantum logic gates. One such example of this process is the QST channel. A canonical QST passes quantum information uni-directionally through a quantum circuit. In generic quantum computing terms, QST is accomplished with two CNOT gates and has the projector form of

$$\text{QST} = \text{CNOT}(2,1)\text{CNOT}(1,2) = \sum_{\mu,\mu'\in\pm} P_z^{\mu} P_x^{\mu'} \otimes X^{\frac{-\mu+1}{2}} Z^{\frac{-\mu'+1}{2}} \tag{6.7}$$

and demonstrated in the circuit diagram of Fig. 6.1.

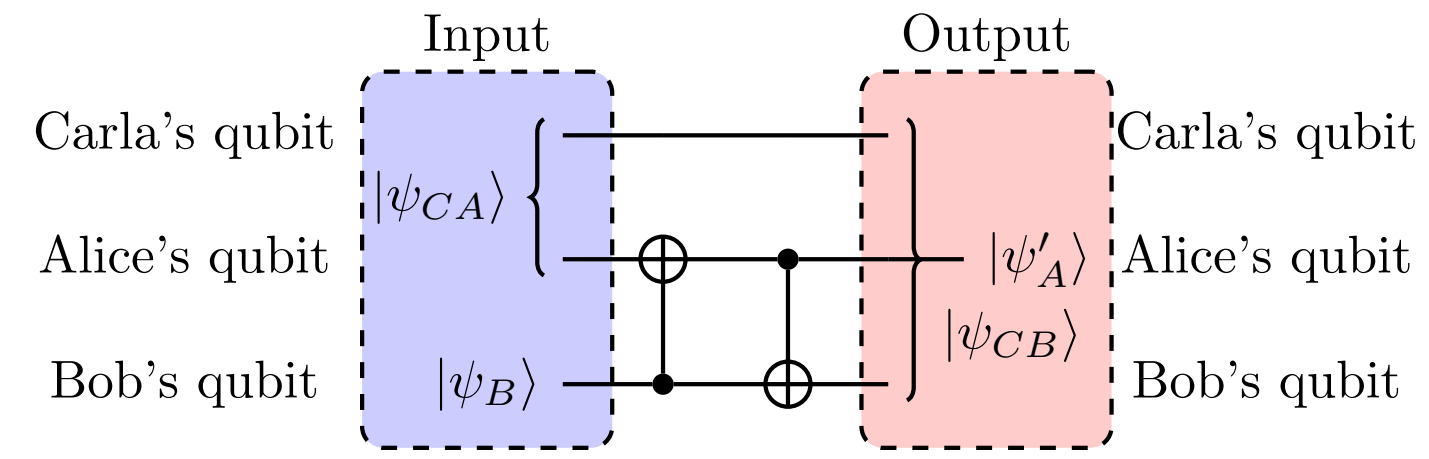

Figure 6.1: Two CNOT gates form a State Transfer Gate

In matrix form this operation between Alice and Bob can be expressed as,

$$\text{QST} \equiv \begin{pmatrix} 1 & 0 & 0 & 0 \\ 0 & 0 & 1 & 0 \\ 0 & 0 & 0 & 1 \\ 0 & 1 & 0 & 0 \end{pmatrix} \tag{6.8}$$

and as shown in Fig. 6.1 it will take the state $|\psi_{CA}\rangle \otimes |\psi_B\rangle$ to $|\psi'_A\rangle \otimes |\psi_{CB}\rangle$. Notice that the output state $|\psi'_A\rangle$ may not have the same form as $|\psi_B\rangle$. This result can be made more evident by adding a reference state $|\psi_D\rangle$ to Bob's qubit as in Fig. 6.3.

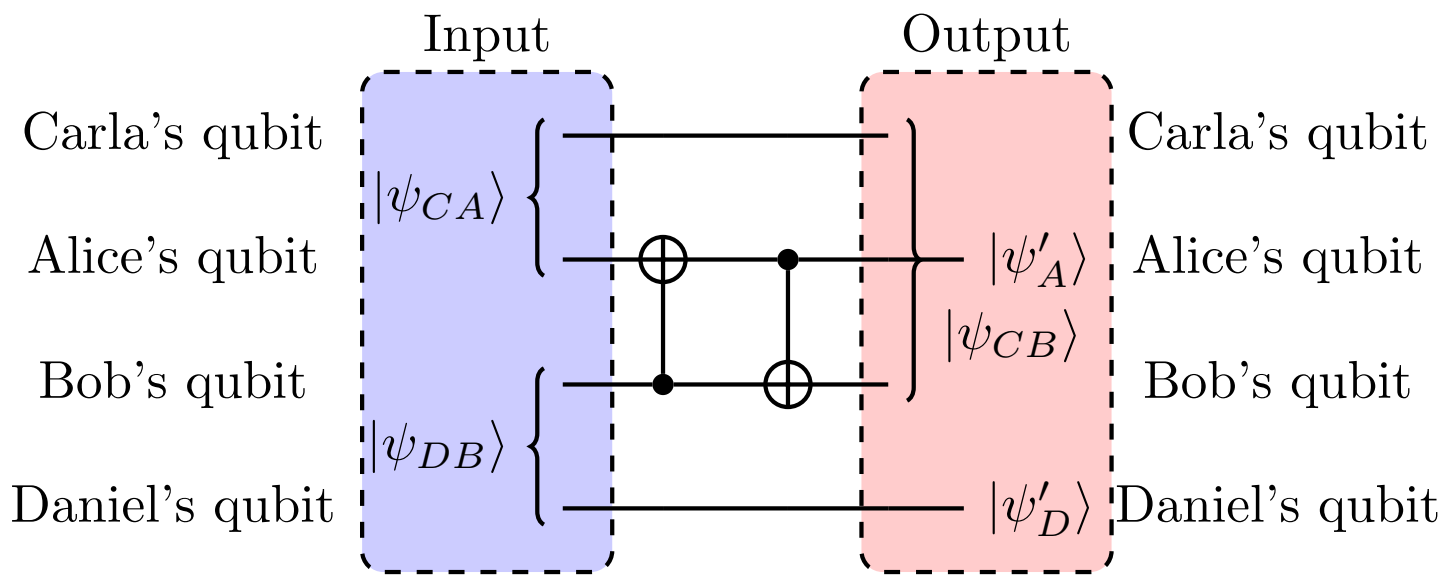

Figure 6.2: Two CNOT gates form a State Transfer Gate as shown with reference qubits.

**SWAP gate**

To create a proper SWAP a third CNOT gate is needed.

A canonical form of the SWAP gate can be seen using three CNOT gates in the following forms

$$
\begin{aligned}
\mathrm{SWAP} &= \mathrm{CNOT}(2,1)\mathrm{CNOT}(1,2)\mathrm{CNOT}(2,1) \\
&= \sum_{\mu,\mu',\mu''\in\pm} P_x^{\mu''} P_z^{\mu} P_x^{\mu'} \otimes Z^{\frac{-\mu''+1}{2}} X^{\frac{-\mu+1}{2}} Z^{\frac{-\mu'+1}{2}} \\
&= (|+\rangle\langle+| \otimes \mathbb{1} + |-\rangle\langle-| \otimes \mathrm{Z})(|0\rangle\langle0| \otimes \mathbb{1} + |1\rangle\langle1| \otimes \mathrm{X})(|+\rangle\langle+| \otimes \mathbb{1} + |-\rangle\langle-| \otimes \mathrm{Z})
\end{aligned}
\tag{6.9}
$$

Figure 6.3 demonstrates the entanglement transfer between Alice and Bob's qubits. Which unlike the exchange in Fig. 6.2, the SWAP gate exchanges the quantum information between these two qubits bidirectionally.

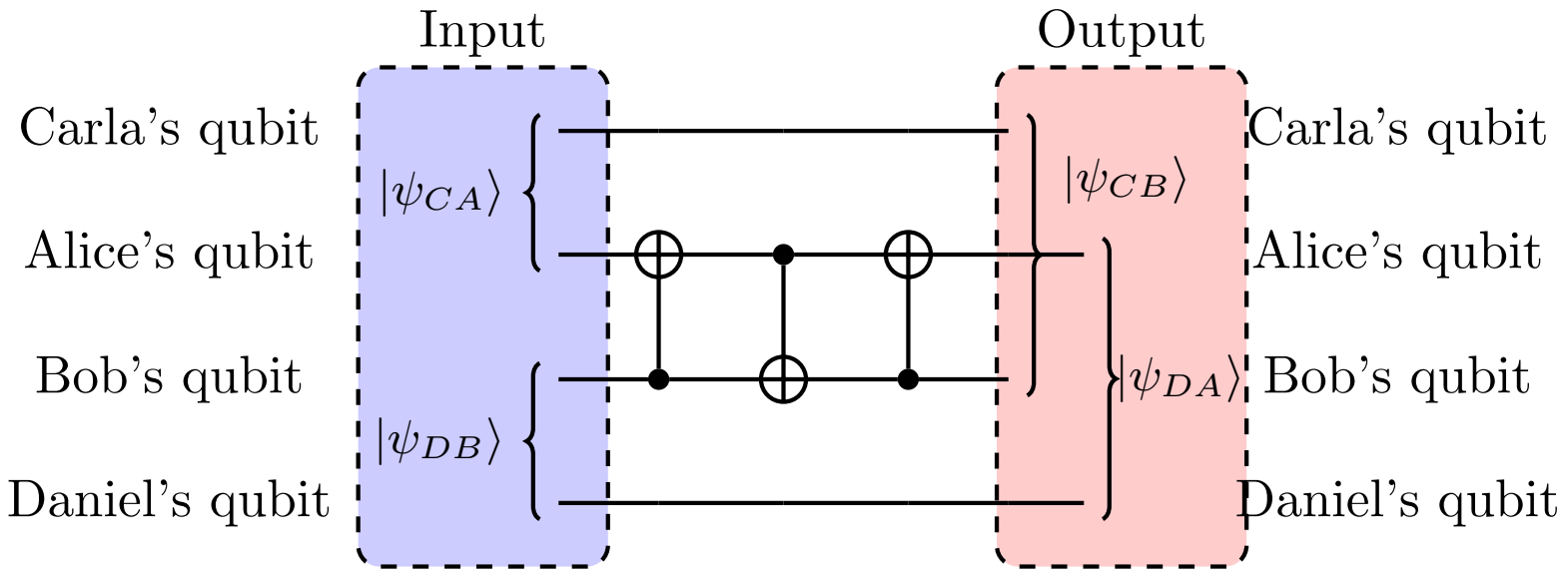

Figure 6.3: Proper SWAP circuit between qubits.

### 6.1.3 Intermediate Qubit

One primary goal in this thesis, is to outline the necessary formalism to generalize RQI using Unruh–DeWitt detectors, to that of standard quantum information theory with spin-qubits. As a comparison to the field-mediated channels, we first demonstrate what these processes look like as qubit mediated processes. The circuit diagram in Fig. 6.4 demonstrates a canonical SWAP interaction between two qubits in terms of CNOT gates. It is through this quantum channel $\mathcal{N}$ we exchange entanglement freely between Alice and Bob. To construct this channel we can assign our SWAP gates to unitary operators. Let's call $\hat{U}_{AF}$ the unitary operator that swaps between Alice and Frank and $\hat{U}_{FB}$ for Frank and Bob. Our channel from Alice to Bob is therefore

$$\mathcal{N}_{A\to B} = \mathrm{Tr}_{AF}[\hat{U}_{AF}\hat{U}_{FB}\hat{U}_{AF}(\hat{\rho}_{A,0} \otimes |\psi_F\rangle\langle\psi_F| \otimes \hat{\rho}_{B,0})\hat{U}^\dagger_{AF}\hat{U}^\dagger_{FB}\hat{U}^\dagger_{AF}]. \tag{6.10}$$

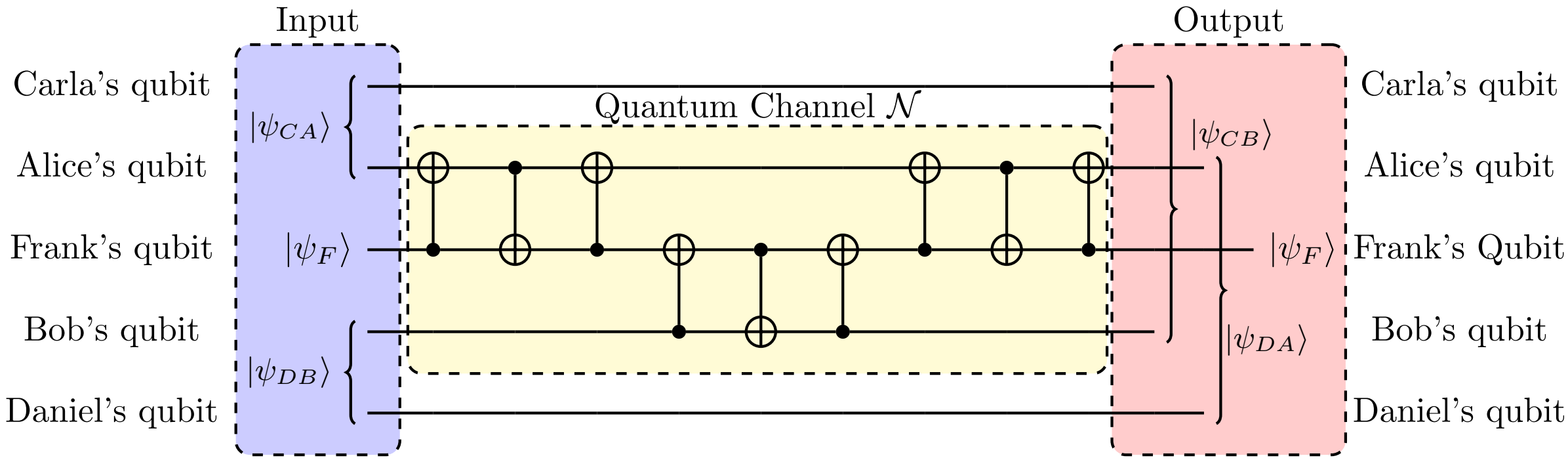

Figure 6.4: Information is swapped between Alice and Bob through Frank.

The channel from Bob to Alice, $\mathcal{N}_{B\to A}$ can be seen by tracing over Bob and Frank.

A quantitative measure of the ability for Bob and Alice to communicate with quantum information can then be seen through calculating the channel capacity,

$$Q(\mathcal{N}_{A\to B}) = \lim_{n\to\infty} \frac{1}{n} I_{max}(\mathcal{N}^{\otimes n}_{A\to B}) \tag{6.11}$$

a measure of maximum coherent information, where $I_{max}(\mathcal{N}_{A\to B}) = \mathrm{supp}(I_c(\mathbb{1}_C \otimes \mathcal{N}_{A\to B}, \mathcal{N}_{A\to B}))$. For our particular case, and throughout this thesis we only refer to bipartite states, and this allows us to set $n = 1$ in Eq. 6.11 as well as $I_{max}(\mathcal{N}_{A\to B}) = I_c((\mathbb{1}_C \otimes \mathcal{N}_{A\to B}), \mathcal{N}_{A\to B})$. Furthermore, we assume the maximally entangled state between our system qubits and reference qubits, results in the highest coherent information through our system. These two simplifications allow us to look at the coherent information of our system as a measure of channel capacity.

## 6.2 Introducing Fields into the Channels

### 6.2.1 Coherent Basis States of the Field

We now upgrade Frank from a two-state qubit to an entire quantum field. To make progress on this complex setting, we will work with the set of coherent states that spans the infinite Hilbert space $\mathcal{H}_\varphi$ of the quantum field

$$|\{\beta(\mathbf{k})\}\rangle = \exp\left(\int d^n\mathbf{k}\left[\beta(\mathbf{k})\hat{a}^\dagger_{\mathbf{k}} - \beta(\mathbf{k})^*\hat{a}_{\mathbf{k}}\right]\right)|0\rangle \tag{6.12}$$

where $|0\rangle$ is the ground state of the field satisfying $\hat{a}(\mathbf{k})|0\rangle = 0\,\forall\,\mathbf{k}$. Within this Hilbert space, there exists a *particular* two-dimensional subspace spanned by $|\pm\alpha\rangle$ of $\mathcal{H}_\varphi$ that we can utilize for encoding and decoding quantum information

$$|\pm\alpha\rangle \equiv \hat{D}(\alpha(k))\,|0\rangle = \exp\left(\int d^n\mathbf{k}\left[\alpha(k)\hat{a}^\dagger_{\mathbf{K}} - \alpha(k)^*\hat{a}_{\mathbf{K}}\right]\right)|0\rangle \tag{6.13}$$

where $\hat{D}(\alpha(k))$ is the unitary displacement operator, and subsequently sets the normalization as $|\langle\pm\alpha|\pm\alpha\rangle| = 1$. The function $\alpha(\mathbf{k})$ that defines our special two-dimensional subspace is a function of parameters that are chosen such that $|\pm\alpha\rangle$ is nearly orthogonal, and therefore *sets up a qubit encoding.* We describe these parameters in detail in the following section.

To realize the displacement function $\hat{D}(\alpha(k))$, we utilize the 1-D scalar field

$\hat{\varphi}(x)$ and it's associated conjugate momentum $\hat{\Pi}(x)$ which are traditionally given in the Heisenberg picture as

$$\hat{\varphi}(x) = \int \frac{dk}{2\pi}\sqrt{\frac{v}{2\omega(k)}}[\hat{a}(k)e^{ikx} + \hat{a}^{\dagger}(k)e^{-ikx}] \tag{6.14}$$

$$\hat{\Pi}(x) = \int \frac{dk}{2\pi}\sqrt{\frac{\omega(k)}{2v}}[-i\hat{a}(k)e^{ikx} + i\hat{a}^{\dagger}(k)e^{-ikx}]. \tag{6.15}$$

These field operators allow us to generate coherent states with amplitude

$$\alpha_{\varphi}(x) = \sqrt{\frac{v}{2\omega(k)}}e^{-ikx}. \tag{6.16}$$

For convenience we have introduced subscripts to differentiate between coherent amplitudes of the field and conjugate momentum observables. A Fourier transform following the usual prescription

$$f(k) := \frac{1}{\sqrt{(2\pi)}}\int dx f(x)e^{ikx} \tag{6.17}$$

allows us to recover our function $\alpha(k)$ (in one-dimension) but to make these states near-orthogonal we turn to the UDW formalism detailed in Ref. [2].

### 6.2.2 Encoding and Decoding and the UDW QST

Since quantum fields acting on the vacuum generate coherent states, our computational basis of quantum fields, we can utilize the field operators above to generate unitary operations which will encode and decode quantum information onto and off of our fields via coherent states. As we will show, the effectiveness of this encoding and decoding relies on constraints that enforce an orthognality condition on our coherent states.

### Necessary Constraints

For our states $|\pm\alpha\rangle$ to be near orthogonal certain constraints need to be implemented. The hallmark of the UDW model is the smearing and switching functions in the standard UDW interaction Hamiltonian

$$\hat{\mathrm{H}}_{int}(t) = J_\varphi \chi(t) \int_{\mathbb{R}} dk\ \tilde{F}(k)\hat{\mu}(t) \otimes \hat{\varphi}(k,t) \tag{6.18}$$

with coupling $J_\varphi$, smearing function $\tilde{F}(k)$, switching function $\chi(t)$, and two-state magnetic moment $\hat{\mu}(t)$. For simplicity we only evaluate the interaction block Eq. (6.18). However in Chapter 7, we discuss how other blocks of the Hamiltonian can be used to calculate noise with this prescription.

A delta-like switching function allows for the promotion of our time-dependent Hamiltonian to unitary gates [2, 58]. Following the standard UDW prescription we absorb the coupling and smearing function into smeared-out field observables

$$\hat{\varphi}(F) \coloneqq \lambda_\varphi \int dk F(k)\hat{\varphi}(k,t) \tag{6.19}$$

$$\hat{\Pi}(F) \coloneqq \lambda_\Pi \int dk F(k)\hat{\Pi}(k,t). \tag{6.20}$$

For the sake of brevity, we will use these smeared versions of the field observables when written out in unitary gate form, and therefore suppress the function $F$ from our notation via $\hat{\varphi}(F) \equiv \hat{\varphi}$.

We summarize the constraints given in Ref. [2] and Chapter 5 here; in the regime of strong coupling we find $|\langle +\alpha| -\alpha\rangle| \approx 0$, an atypical result of the

nonorthogonal coherent state basis. This result is made evident by constraining properties of the UDW model through the inner product of these coherent states given by

$$|\langle +\alpha | -\alpha \rangle| = \exp\left(-(J_\varphi)^2 \int \frac{dk}{2\omega_k} |\tilde{F}(k)|^2\right) \tag{6.21}$$

which follows from the identity

$$\langle \beta | \alpha \rangle = \exp\left(-\frac{1}{2}|\alpha|^2 - \frac{1}{2}|\beta|^2 + \beta^* \alpha\right). \tag{6.22}$$

It is clear from Eq. (6.21) that as the coupling $J_\varphi$ increases, the states become increasingly orthogonal.

### UDW QST

Elevating the Hamiltonian in Eq. (6.18) to include both the field and scalar momentum

$$\hat{\mathrm{H}}_{\mathrm{int}}^{QST}(t) = J_\varphi \chi(t) \int_{\mathbb{R}} dx\ F(x) \hat{\mu}(t) \otimes (\hat{\varphi}(x,t) + \hat{\Pi}(x,t)). \tag{6.23}$$

and using a delta like $\chi(t)$, we can use Eq. (6.23) to formulate unitaries that carry out the encoding and decoding demonstrated in the circuit diagram in Fig. 5.1

$$\hat{U}_{\mathrm{QF}}^{QST} = \sum_{\mu,\mu' \in \pm} \hat{P}_x^{\mu} \hat{P}_z^{\mu'} \otimes e^{i\mu\hat{\Pi}} e^{i\mu'\hat{\varphi}} = (\hat{P}_x^- \otimes e^{-i\hat{\Pi}} + \hat{P}_x^+ \otimes e^{i\hat{\Pi}})(\hat{P}_z^+ \otimes e^{i\hat{\varphi}} + \hat{P}_z^- \otimes e^{-i\hat{\varphi}}) \tag{6.24}$$

which looks remarkably similar to Eq. (6.7). This unitary behaves like two controlled unitaries, as we will see at the end of this section.

In order to remain in the two-dimensional subspace of $\mathcal{H}_\varphi$ we follow the pre-scription of Ref. [2] and enable the constraint

$$\hat{\Pi}\,|\pm\alpha\rangle \approx \pm\gamma\,|\pm\alpha\rangle \tag{6.25}$$

where the value of $\gamma$ is set by

$$\gamma := \lambda_\Pi \lambda_\varphi \int dk\,|\tilde{F}_\nu(k)|^2 = \frac{\pi}{4} \bmod 2\pi \tag{6.26}$$

a consequence of the restriction

$$\left(\lambda_\varphi \int dk\,|\tilde{F}_\nu(k)|^2\right)^2 \gg \frac{1}{2}\int dk\,\omega(k)\,|\tilde{F}_\nu(k)|^2. \tag{6.27}$$

Constraining the parameters in this way allows the operator $e^{i\hat{\Pi}}$ to act on the state and introduce a phase

$$e^{i\hat{\Pi}}\,|\mu\alpha\rangle = e^{i\mu\gamma}\,|\mu\alpha\rangle\,. \tag{6.28}$$

Practically, we can see this through the following example. If we initialize our state as $|\psi_{CA}\rangle \otimes |\psi_\varphi\rangle = |000\rangle + |110\rangle$ and apply our unitary from Eq. (6.24) then our output state is given by

$$e^{\gamma}(|0\rangle_C\,|+\rangle_A\,|\alpha\rangle_\varphi - |1\rangle_C\,|-\rangle_A\,|-\alpha\rangle_\varphi) + e^{-\gamma}(|0\rangle_C\,|-\rangle_A\,|\alpha\rangle_\varphi + |1\rangle_C\,|+\rangle_A\,|-\alpha\rangle_\varphi). \tag{6.29}$$

From this output state we can see that the qubit-field QST operation is different than the qubit-qubit QST. It is not a simple quantum state transfer from qubit to field. While qubit to field QST may be attainable through this prescription, for this work we maintain focus on field-mediated qubit communications. Despite the difference in the way these QST operations treat quantum information, we will find in Sec. 6.3 that there is an equivalence in the field-mediated QST channel and the qubit-mediated counterpart.

### 6.2.3 SWAP gate with fields

We can define projection operators of coherent states without the field observable form and carry out standard quantum logic gates. Let's define these projectors as

$$\hat{\mathrm{P}}_{+\alpha} = |+\alpha\rangle\langle+\alpha| \qquad \hat{\mathrm{P}}_{-\alpha} = |-\alpha\rangle\langle-\alpha| \tag{6.30}$$

$$\hat{\mathrm{P}}^{Z}_{\alpha} = |+\alpha\rangle\langle+\alpha| - |-\alpha\rangle\langle-\alpha| \qquad \hat{\mathrm{P}}^{X}_{\alpha} = |-\alpha\rangle\langle+\alpha| + |+\alpha\rangle\langle-\alpha| \tag{6.31}$$

$$\hat{\mathrm{P}}^{\frac{\pi}{2}}_{+\alpha} = |+\alpha\rangle\langle+\alpha| + |-\alpha\rangle\langle+\alpha| + |+\alpha\rangle\langle-\alpha| + |-\alpha\rangle\langle-\alpha| \tag{6.32}$$

$$\hat{\mathrm{P}}^{\frac{\pi}{2}}_{-\alpha} = |+\alpha\rangle\langle+\alpha| - |-\alpha\rangle\langle+\alpha| - |+\alpha\rangle\langle-\alpha| + |-\alpha\rangle\langle-\alpha|. \tag{6.33}$$

However, to utilize these projection operators, there are operational costs depending on the field initialization. Commonly, the field is initialized in the ground state, which results in an extra phase factor of $e^{-\frac{1}{2}|\alpha|^2}$ as a result of the non-orthogonality identity of coherent states

$$\langle\beta|\alpha\rangle = \exp\left(-\frac{1}{2}|\alpha|^2 - \frac{1}{2}|\beta|^2 + \beta^*\alpha\right). \tag{6.34}$$

Furthermore, even if the state is initialized in either $|\pm\alpha\rangle$, given Eq. 6.34 we expect some non-zero value where there shouldn't be. As mentioned in Chapter 2, we can place constraints on this model to deal with the non-orthogonality of the coherent states. With the CNOT program in place we can revisit the canonical SWAP gate outlined in Sec. 6.1.3. As emphasized before, the field mediated action should be as close to equivalent as possible with the canonical SWAP gate between two qubits. Ignoring physical implementation for a moment we can utilize the projectors defined in Eqs. 6.30 which will allow for a straightforward SWAP gate of the form

$$\hat{U}_{\mathrm{QF}}^{\mathrm{SWAP}} = (|0\rangle\langle 0|\otimes\mathbb{1}+|1\rangle\langle 1|\otimes\hat{\mathrm{P}}_{\alpha}^{X})(|+\rangle\langle +|\otimes\mathbb{1}+|-\rangle\langle -|\otimes\hat{\mathrm{P}}_{\alpha}^{Z})(|0\rangle\langle 0|\otimes e^{-i\hat{\varphi}_{\nu}}+|1\rangle\langle 1|\otimes e^{i\hat{\varphi}_{\nu}}). \tag{6.35}$$

Alternatively we can initialize our field in the state $|+\alpha\rangle$ and we get the familiar

$$\hat{U}_{\mathrm{QF}}^{\mathrm{SWAP}} = (|0\rangle\langle 0|\otimes\mathbb{1}+|1\rangle\langle 1|\otimes\hat{\mathrm{P}}_{\alpha}^{X})(|+\rangle\langle +|\otimes\mathbb{1}+|-\rangle\langle -|\otimes\hat{\mathrm{P}}_{\alpha}^{Z})(|0\rangle\langle 0|\otimes\mathbb{1}+|1\rangle\langle 1|\otimes\hat{\mathrm{P}}_{\alpha}^{X}). \tag{6.36}$$

While the field-mediated operator formalism for the SWAP gate is not inherently evident, we will see in Sec. 6.4.1 that with some care, one can write down a two-qubit single CNOT analogy with our field-mediated gates. For this reason, we can minimally write down a cumbersome and computationally exhaustive SWAP gate by gluing three of these together. However, it is likely that a more simplified version of this gate exists in field operator formalism, just as Eq. 6.24 exists as a simplified version of two individual field-mediated CNOTs glues together.

## 6.3 Applications in Quantum Shannon Theory: Diamond Norm

The UDW quantum logic gates provide insight into field mediated transduction between qubits. The importance of this transduction can not be understated for future investigations into long range photonics and fermionic systems (quantum materials) mentioned in Chapter 1. Coherent information and quantum capacity has so far allowed one to examine the communication properties of UDW channels, however a straightforward comparison between the UDW logic gate channels and the canonical quantum logic gate channels provides a deeper insight to a systems ability to match processes. We therefore examine the measure of diamond distance between these channels.

The diamond norm can indicate the differences between two quantum channels [4]. It is well understood that the SWAP gate leads to ideal channels for transferring quantum information between qubits. Similarly QST, in the form of two CNOT gates, provides an ideal unidirectional information transfer.

If we compare the two QST channels discussed above, we can formulate the diamond norm as

$$\|(\mathbb{1}_C \otimes \Xi_{A\to B}) - (\mathbb{1}_C \otimes \mathcal{N}_{A\to B})\|_\diamond \tag{6.37}$$

where

$$\Xi_{A\to B} = \mathrm{Tr}_{A\varphi}[\hat{U}^{QST}_{\varphi B}\hat{U}^{QST}_{A\varphi}(\hat{\rho}_{A,0} \otimes |\psi_{\varphi,0}\rangle\langle\psi_{\varphi,0}| \otimes \hat{\rho}_{B,0})\hat{U}^{QST\dagger}_{A\varphi}\hat{U}^{QST\dagger}_{\varphi B}] \tag{6.38}$$

here we have allowed $\hat{U}^{QST}_{\mathrm{A}\phi} = \hat{U}^{QST}_{\mathrm{B}\phi}$, and

$$\mathcal{N}_{A\to B} = \mathrm{Tr}_{AF}[\hat{U}_A\hat{U}_B(\hat{\rho}_{A,0}\otimes|\psi_F\rangle\langle\psi_F|\otimes\hat{\rho}_{B,0})\hat{U}_B^\dagger\hat{U}_A^\dagger] \tag{6.39}$$

with Eq. 6.38 being the channel from Fig. 5.1 and Eq. 6.39 represented by Fig. 6.5.

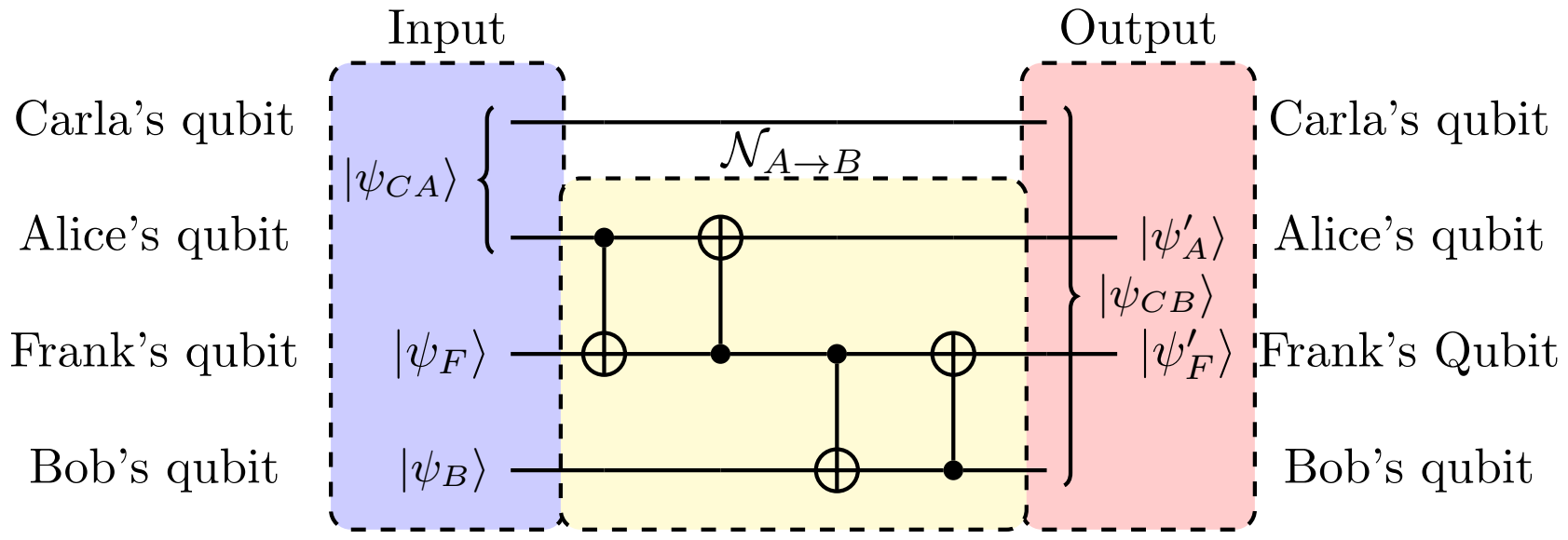

Figure 6.5: Information is transferred from Alice to Bob through Frank as demonstrated with a quantum channel $\mathcal{N}_{\mathcal{A}\to\mathcal{B}}$.

We can compare the diamond norm in Eq. 6.37 to the channel capacity of Eq. 6.38

$$Q(\Xi_{A\to B}) = S(\Xi_{A\to B}) - S(\mathbb{1}\otimes\Xi_{A\to B}) \tag{6.40}$$

where $S$ is the von Neumann entropy (see Fig. 6.6 and Fig. 6.7. The code for this simulation is available in Ref. [1]). Preparation of initial states is different for Eqs. 6.39 and 6.38, which is a result of the different ways that the channels carry out the treatment of the information. However, the SWAP gates do not have such restrictions as the state-transfer is bi-directional.

## 6.4 Universal Quantum Computing

An important criteria of quantum computing, laid out by DiVincenzo [74] is that a system, capable of quantum computing, must have a set of universal quantum gates [97, 98]. Instead of demonstrating that we can achieve one set of universal quantum gates, in this section we demonstrate the ability to extend the UDW logic gate system to its canonical counterpart.

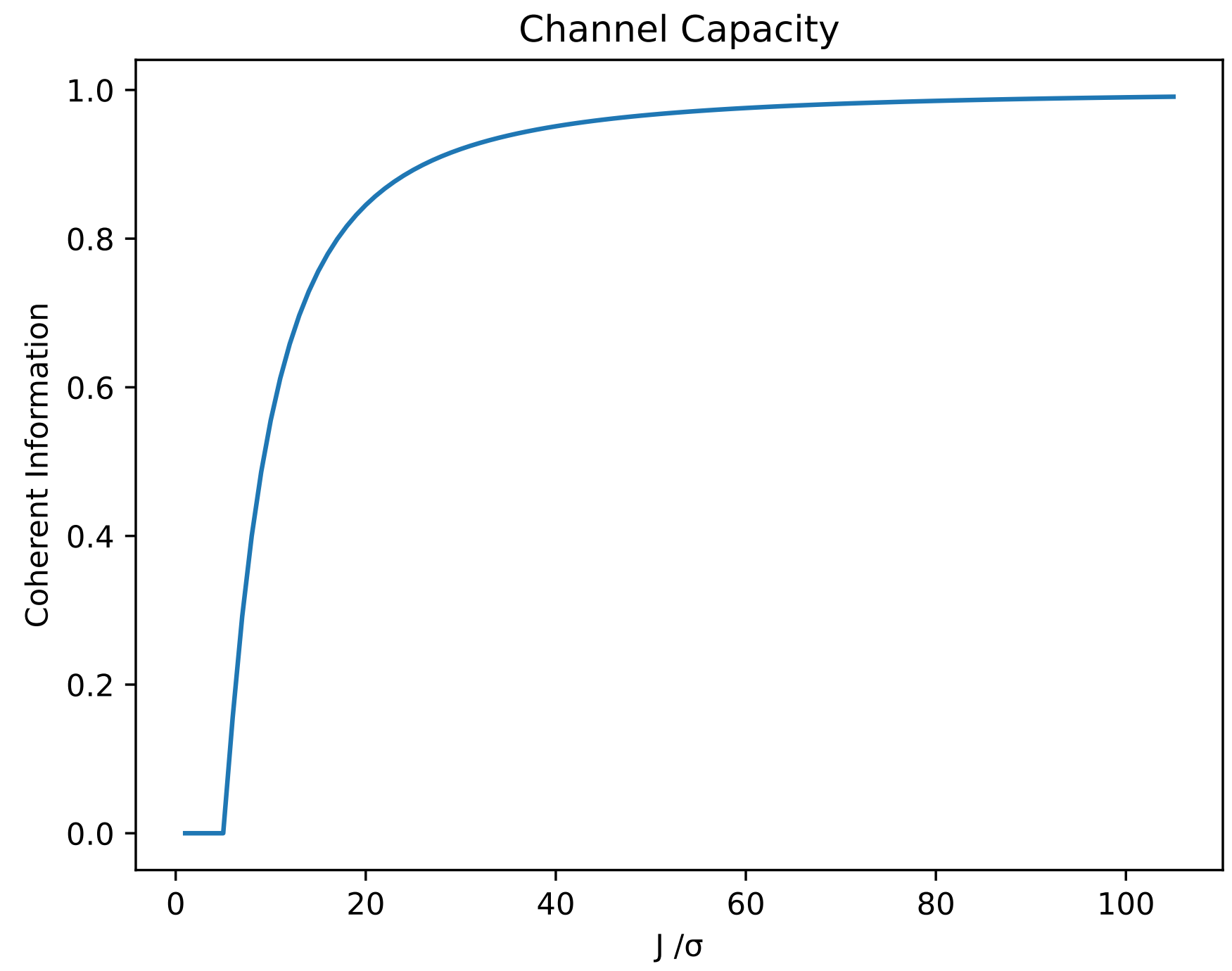

Figure 6.6: A recreation of Fig. (4) from Ref. [2] showing the channel capacity from Eq. 6.38. With the initial state prepared in the $|+_y\rangle$ state.

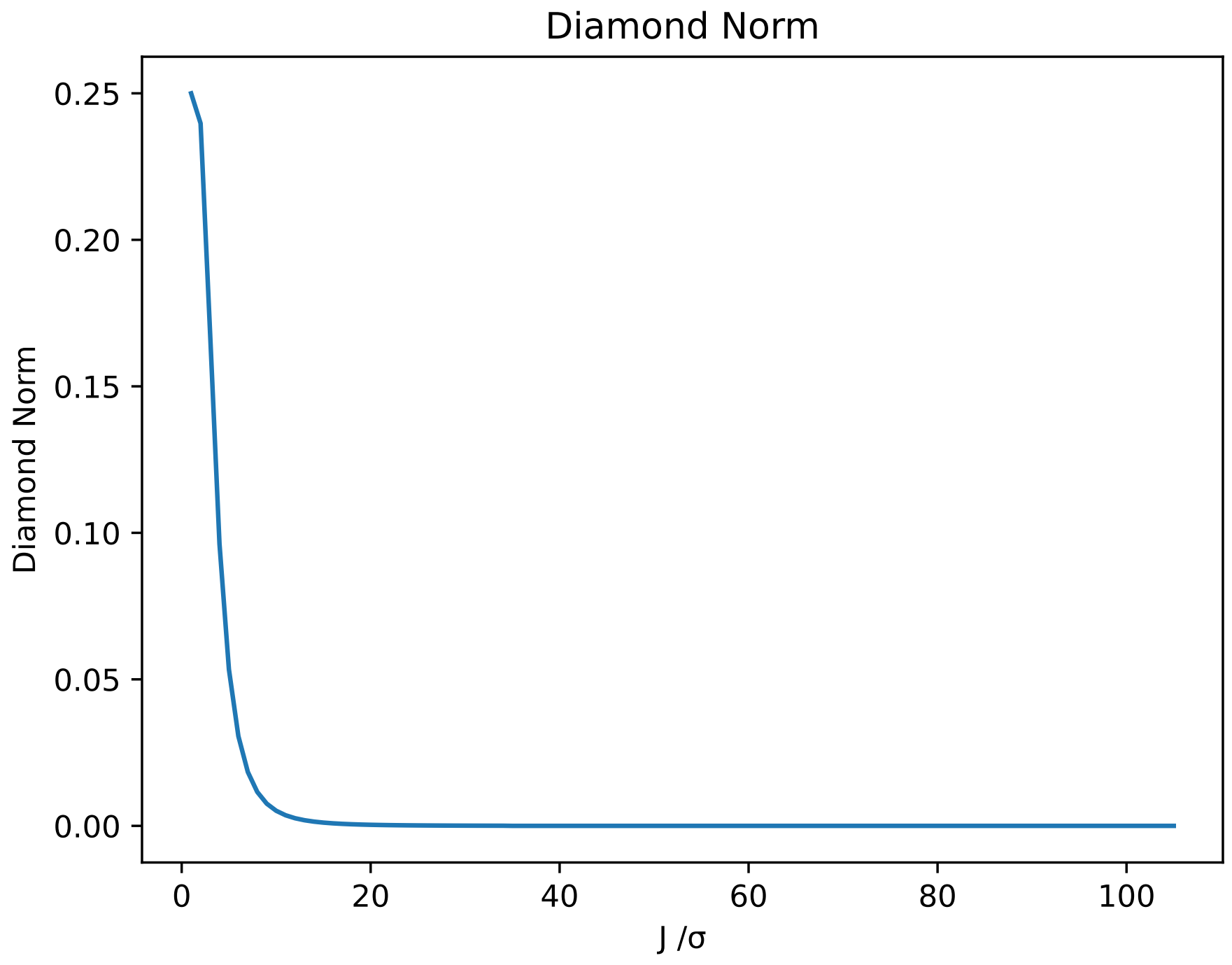

Figure 6.7: The diamond norm is a measure of "idealness" of a channel. This particular graph compares the QST channel both qubit-mediated and field-mediated.

### 6.4.1 Forms of a single CNOT channel

What we showed in this thesis, is that in the limit of strong coupling, two and three CNOT arrangements (QST and SWAP gates) can demonstrate equivalent processes between UDW and canonical formalism. With a little effort we can realize analogies to both qubit-mediated CNOT gates as well as single CNOT gates between two qubits.

Qubit-mediated CNOT channels require firstly acting on two qubits with a single CNOT gate and then, using a QST, transferring the state of the target to a final qubit as shown in Fig. 6.5. To accomplish this with qubit-field gates we recall a single controlled unitary,

$$\hat{U}_{\mathrm{QF}}^{\mathrm{Z}\varphi} = \sum_{\mu\in\pm} \hat{P}_z^{\mu} \otimes e^{i\mu\hat{\varphi}} = (|0\rangle\langle 0| \otimes e^{-i\hat{\varphi}} + |1\rangle\langle 1| \otimes e^{i\hat{\varphi}}). \tag{6.41}$$

which demonstrates an equivalent channel to the qubit-mediated CNOT channel when written out as

$$\Xi_{A\to B} = \mathrm{Tr}_{A\varphi}[\hat{U}_{\varphi B}^{\mathrm{X}\Pi}\hat{U}_{A\varphi}^{\mathrm{Z}\varphi}(\hat{\rho}_{A,0} \otimes |\psi_{\varphi,0}\rangle\langle\psi_{\varphi,0}| \otimes \hat{\rho}_{B,0})\hat{U}_{A\varphi}^{\dagger\mathrm{Z}\varphi}\hat{U}_{\varphi B}^{\dagger\mathrm{X}\Pi}]. \tag{6.42}$$

where $\hat{U}_{\varphi B}^{\mathrm{X}\Pi}$ has taken the other form

$$\hat{U}_{\mathrm{FQ}}^{X\Pi} = \sum_{\mu\in\pm} \hat{P}_x^{\mu} \otimes e^{i\mu\hat{\Pi}} = (|-\rangle\langle -| \otimes e^{-i\hat{\Pi}} + |+\rangle\langle +| \otimes e^{i\hat{\Pi}}). \tag{6.43}$$

When initializing our receiving qubit (of the field mediated setup) in the $|+_y\rangle$ state, the diamond norm rapidly approaches zero as shown in Fig. 6.9.

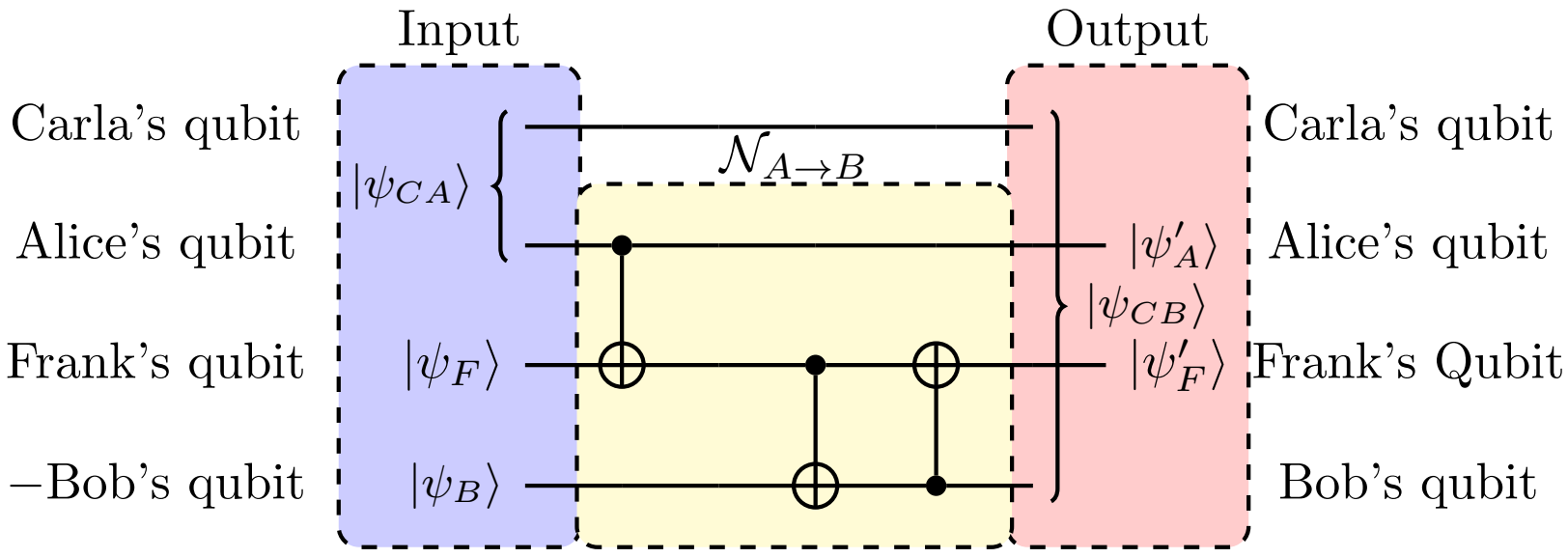

Figure 6.8: Information is transferred from Alice to Bob through Frank as demonstrated with a quantum channel $\mathcal{N}_{\mathcal{A}\to\mathcal{B}}$.

The Hamiltonians representing these unitaries

$$\hat{\mathrm{H}}_{\mathrm{A,int}}(t) = J_{A\varphi}\chi(t)\int_{\mathbb{R}} dx\ F(x)\hat{\mu}_A(t)\otimes\hat{\varphi}(x,t) \tag{6.44}$$

$$\hat{\mathrm{H}}_{\mathrm{B,int}}(t) = J_{B\varphi}\chi(t)\int_{\mathbb{R}} dx\ F(x)\hat{\mu}_B(t)\otimes\hat{\Pi}(x,t) \tag{6.45}$$

$$\tag{6.46}$$

are the electron density and electron current blocks of the Luttinger liquid Hamiltonian coupled with a Kondo impurity via the UDW interaction, a primary result in Ref. [18]. These Hamiltonians indicate physical systems that can realize these gates.

Surprisingly the analogy to the two qubit CNOT gate like that of Eq. 6.4 is bit more of a challenge. Regardless, we can realize this with the introduction of a new unitary

$$\begin{aligned}\hat{U}^{Z\Pi X\varphi}_{\mathrm{QF}} &= e^{-i\hat{\Pi}}\sum_{\mu,\mu'\in\pm}\hat{P}^{\mu}_z\hat{P}^{\mu'}_x\otimes e^{i\mu\hat{\Pi}}e^{i\mu'\hat{\varphi}}\\ &= e^{-i\hat{\Pi}}(|1\rangle\langle 1|\otimes e^{-i\hat{\Pi}} + |0\rangle\langle 0|\otimes e^{i\hat{\Pi}})(|-\rangle\langle -|\otimes e^{-i\hat{\varphi}} + |+\rangle\langle +|\otimes e^{i\hat{\varphi}})\end{aligned} \tag{6.47}$$

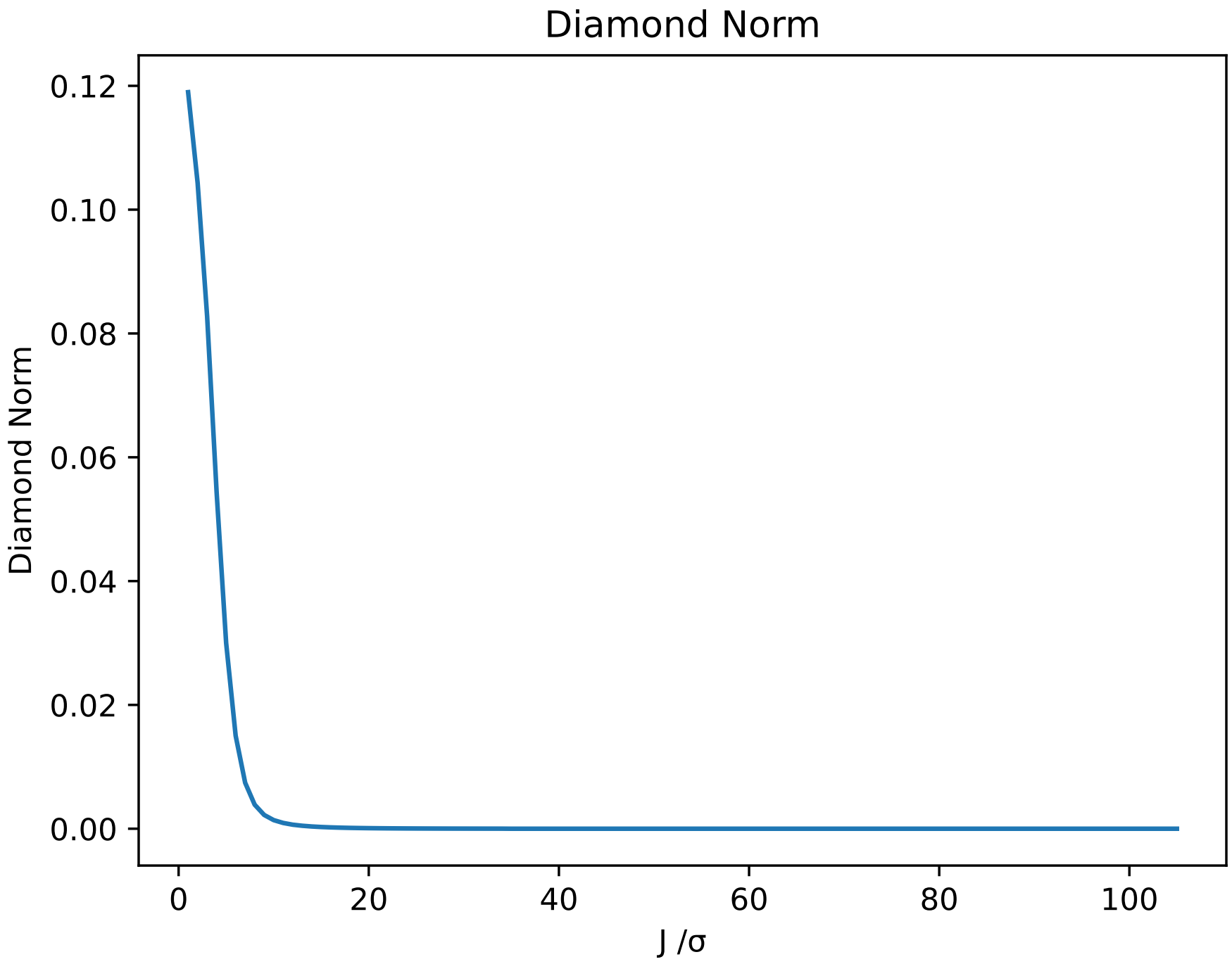

Figure 6.9: The diamond distance of the canonical quantum CNOT gate and our UDW CNOT gate.

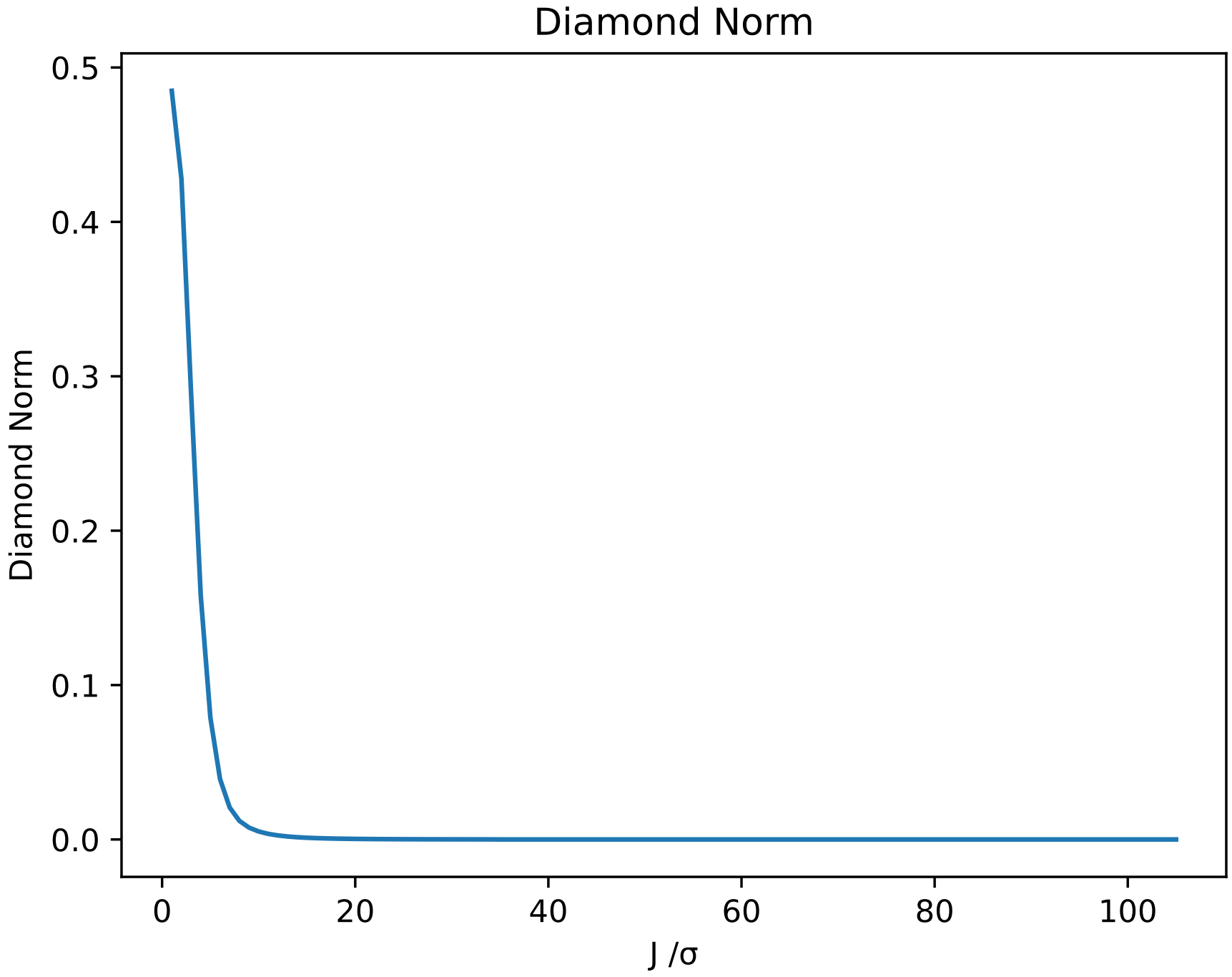

Figure 6.10: The diamond distance of a CNOT between qubits and our qubit-field-qubit gate.

where the $e^{-i\hat{\Pi}}$ exists to absorb a left over phase. The Channel therefore is expressed as

$$\Xi_{A\to B} = \mathrm{Tr}_{A\varphi}[\hat{U}^{QST}_{\varphi B}\hat{U}^{Z\Pi X\varphi}_{A\varphi}(\hat{\rho}_{A,0}\otimes|\psi_{\varphi,0}\rangle\langle\psi_{\varphi,0}|\otimes\hat{\rho}_{B,0})\hat{U}^{Z\Pi X\varphi\dagger}_{A\varphi}\hat{U}^{QST\dagger}_{\varphi B}] \tag{6.48}$$

and can be compared to a single CNOT gate with the diamond distance as shown in Fig.6.10. The same process follows for Tofolli gates as well.

A large subset of the universal quantum gate arrangement is defined by single qubit operations. UDW qubits describe the class of spin-qubits, well-known to be viable for all single qubit operations. However, we can demonstrate how one my construct an equivalent single qubit operator out of the controlled unitaries for our qubit-field interaction. Let's take the Hadamard operation with the following truth table which can be mimicked by defining a new unitary qubit-field operator

Hadamard Gate

| Input | Output |
|---|---|
| $\lvert 0\rangle$ | $\lvert 0\rangle + \lvert 1\rangle$ |
| $\lvert 1\rangle$ | $\lvert 0\rangle - \lvert 1\rangle$ |

$$\hat{U}^{\mathrm{H}}_{\mathrm{FQ}} = \sum_{\mu\in\pm}\hat{P}^{\mu}_{x}\otimes e^{-i\mu\hat{\varphi}} \tag{6.49}$$

combined with the unitary from Eq. 6.41 in the quantum channel

$$\Xi^{\mathrm{H}}_{A\to A} = \mathrm{Tr}_{\varphi}[\hat{U}^{\mathrm{H}}_{\varphi A}\hat{U}^{\mathrm{Z}\varphi}_{A\varphi}(\hat{\rho}_{A,0}\otimes|\psi_{\varphi,0}\rangle\langle\psi_{\varphi,0}|)\hat{U}^{\mathrm{Z}\varphi\dagger}_{\varphi A}\hat{U}^{\mathrm{H}\dagger}_{\varphi A}] \tag{6.50}$$

demonstrates the same computational effects of a single qubit Hadamard gate. Using the same logic we can show that there exist similar analogies to single qubit

S- and T-gates

$$\hat{U}^{\mathrm{S}}_{\mathrm{QF}} = \sum_{\mu\in\pm} \hat{P}^{\mu}_{z} \otimes e^{(-\mu 1+1)i\hat{\Pi}} e^{i\hat{\varphi}}) \tag{6.51}$$

$$\hat{U}^{\mathrm{T}}_{\mathrm{QF}} = \sum_{\mu\in\pm} \hat{P}^{\mu}_{z} \otimes e^{\frac{-\mu 1+1}{2} i\hat{\Pi}} e^{i\hat{\varphi}}). \tag{6.52}$$

which again can be assembled from blocks of the Luttinger liquid Hamiltonian. With theses UDW gates we demonstrate, using the metric of diamond distance, these gates are able to produce the same set of results as a universal set of logic gates.

# Chapter 7

# Future Work

## 7.1 summary of results

In this thesis, we have shown that a combination of the UDW detector formalism with that of the abelian bosonization of a HLL has provided us with a toy-model quantum computer that utilizes qubit-field interactions to realize quantum transduction in quantum materials. We have demonstrated, in earnest, a series of gates that are built from the qubit-field Hamiltonian and replicate quantum logic gates. With these quantum logic gates, tests of entanglement generation and propagation are realizable. Tests of quantum communication through metrics provided by quantum Shannon theory will provide a bias-free, bench-marking platform for comparing the processes to qubit-qubit interactions.

While we don't necessarily expect the UDWQC or other forms of QCs that utilize qubit-field interactions to out-preform current quantum computing technology, yet there are other benefits to gain. From a theoretical framework, we've demonstrated the importance of our UDWQC for progressing quantum Shannon theory in RQI, but these values can be extended into regimes of interest for other fields of physics. Topics such as thermalization of quantum materials will benefit from the perspective of quantum Shannon theory as the purpose of the theory is to track entanglement of systems (and even environments).

We have shown how UDWQCs can fit nicely into a dephasing-like channel such that we can utilize well-established theory of dephasing to extend our UDWQC model to include interactions with the environment with little additional cost computationally, as we can absorb the changes into the dephasing parameter instead of correlators of the fields. These models appear to still be in a type of "toy-model" phase, but will continue to develop into theories that accurately resemble laboratory experiments.

Experiments that can utilize the theory outlined in this thesis are currently under development with experiments under construction and manuscripts being drafted. The goal of this project has been to elevate the theory behind RQI to match that of experimental prospect. We have done this with the two systems focused throughout this thesis. A quantum bus of QdotS coupled to HLLs has been utilized to carefully outline the underlying structure of the theory, and provide a plethora of "low-hanging fruit" research in the area of gate structures to be modelled from the full Tomonaga-Luttinger liquid Hamiltonian. The second system, doped TMDs, have provided us with a low-cost effecient way to realize first-principle studies of the theory presented in this thesis.

## 7.2 Future Work and Open Problems

While attempting to connect the theory of RQI to condensed matter systems, we have unlocked a large amount of open problems, ranging from the low-hanging fruit problems mentioned above, to projects that could last many years. This last section of the thesis aims to recognize these open problems both as shortcomings of the current work and also promising work to be completed.

With the slew of qubit technologies available, such as superconducting qubits, trapped ions, photonics, etc. a metric of quality becomes desirable. QI scientists turn to quantum resource theory to understand appropriate quantitative values to measure a quantum computer's abilities [99], a measure of its "quantumness". Among the resources available, entanglement and "magic" are the two most widely used. For this thesis we primarily stick with entanglement due to the commonality and effectivness of entanglement as a resource [100], but future work into magic would prove to be fruitful.

## 7.3 Entanglement and the DiVincenzo Criteria

Entanglement as a resource provides us with a metric to explore the famous criteria of DiVincenzo [74]. For ease of clarity, a brief list and description of each of the criteria is as follows:

1. Can the system scale to a larger size, with well characterized qubits?
2. Can you prepare your system in a pure fixed state that is the basis of comparison?
3. Does the system have long decoherence times?
4. Does the system have a universal set of quantum gates?
5. Can the system read out the qubits in question?
6. Can your system switch between stationary and flying qubits?

7. Do the Flying qubits make it to where you direct them to?

The work presented in this thesis has been aimed at understanding the theoretical limits to new qubit technologies that employ field-mediated processes. One major caveat here is that experimentally we may find significant technological limits. We can demonstrate progress toward realizing the above criteria but implementing these criteria experimentally is a formidable challenge.

For criteria 1 we demonstrated how an idealized system, sketched by Fig. 1.1 is a possible method for utilizing well defined qubits as we scale the system size larger. Criteria 2 and 5 are subject to the manipulation of the Qdot in a laboratory setting. Qdots have been experimentally implemented into laboratory settings in a vast array of settings. It is well believed that Qdots will make good quantum computers [101]. What ultimately limits criteria 3, is the experimental length of the coherence, recent attempts at measuring this length are approximately around 2 µm coherence length but can reach upards of 5 µm in HgTe nanowires [102, 103]. At this length scale, the switching time-scale estimates above would be increase by several orders of magnitude, possibly reaching nanoseconds. At this times scale, electrical gating can be straightforwardly implemented using electrical waveguides and voltage pulses. It was demonstrated in Ref. [18], these switching times can be achieved with current technology.

In regards to criteria 4, while we have demonstrated a set of universal gates for field-mediated quantum logic gates, this will depend heavily on the progression of the technology. Information loss can be simulated within the field-mediated gates yet comparing these simulations to experiments is left for future work. Furthermore, while this thesis demonstrates a possible arena to realize these quantum computing gates, more work is required to demonstrate this to be true and to what efficiency.

The above discussion refers specifically to criteria 1-5 and regard the minimum capabilities of a quantum computer. We however, are also able to comment on criteria 6 and 7. If one is able to realize this system physically as a UDWQC, it would essentially achieves quantum transduction of information from a stationary qubit to a flying qubit via the UDW detector interaction. Moreover, we are exploring this interaction with right- and left-moving fermionic flying qubits. Many of the restrictions with flying qubits are a consequence of being subjected to Huygen's principle, we avoid this as the right- and left-moving fermions are restricted to a one-dimensional "quantum wire". We have demonstrated in Chapter 6 that interactions between stationary and flying qubits is the primary intent of this system. In Chapter 5 we demonstrate calculable simulations of noise but recognize that far more noise sources exist and can be treated with a similar approach.

## 7.4 A Completed Field Thoery

### 7.4.1 The problems with Impurity Scattering.

Our model relied heavily on the free bosonic field scattering off an impurity. While this interaction Hamiltonian maintains necessary symmetries when coupled to the Kondo impurity, it also introduces physical problems. Specifically, we have failed to address the possible spin-flipping, only briefly mentioning this problem in Chapter 3 in the context of a non-chiral Luttinger liquids. Furthermore, The impurity forward-scattering transmission of information that we evaluated in this thesis has values of coupling that necessarily destroy the transmission and indicate pure reflection [104]. These values of coupling would make for an interesting investigation in this framework, potentially impacting the numerical simulations of channel capacity which directly rely on coupling strength. Alternatively, it could be possible that this interaction is realizable in the context of a different system, but in the Qdot+LL system, to be realized physically, one must evaluate other areas of the model (see Table. 3.1).

### 7.4.2 The Running Coupling

Details of the coupling of Kondo-like impurities to a (1+1D) CFT are very important to realize this model in a laboratory setting. Kondo impurites and their variations are well-known to admit a running coupling constant, and to verify strong coupling in the regimes we are working with, is no simple task. In regards to our model, we are dealing with an effective field theory and we can write down what the ideal coupling should look like when coupling a spin defect to a (1+1D) CFT. Writing down our Hamiltonian in Eq. 2.5 is done with the assumption that the combination of the smearing and coupling will lead to some "reasonable" parameter for achieving a quantum capacity with manageable noise. More investigations are needed into verifying that this coupling exists as it is written down in Eq. 2.5. Moreover, one should aim to understand how the running of the coupling constant will effect the system as a whole given each system will physically be different (with different stresses and torsions etc.) no two couplings will be exactly the same.

### 7.4.3 Nonlinear Couplings

Other methods to realize the coupling, may exist in higher orders of the interactions. Table 3.1 demonstrated a library of interactions that have physical consequences when coupling to a qubit. As mentioned in Chapter. 3, the coherent states needed to evaluate the parameters of this system are very nontrivial and will require a fair deal of caution. Nevertheless, it is demonstrated how to begin this problem and what is necessary to realize the quantum capacity of channels that utilize these interactions.

There are other interactions to consider as well, some of which lead to noise and others that may be beneficial. Interactions between the bulk of the LL and the edge states are amongst the most troublesome. Moreover, time-evolution of the fields as they propagate between qubits, necessarily guarantees that the decoding unitary gate will be different then the encoding gate. It is likely that the correlator of the quantum channel will change by only a phase, when introducing this fact, but exploring time-evolution of the field and the more realistic approach of an interacting fermion will offer up scenarios that are more likely to exist in a laboratory setting.

### 7.4.4 Fermionic Coherent States

Another possible solution to the aforementioned coupling concerns is to skip the bosonization process. Bosonic coherent states are far more understood and developed in quantum Shannon theory and as such, it fits nicely into the RQI framework. However, fermionic coherent states will realize identical results and allow for a direct connection to the right- and left-moving fermions of the LL system, but they are also far more challenging to work with. Unlike the fock space of bosonic coherent states, fermions produce a much more intricate space of states [105,106].

This problem also offers a broad scope of development for quantum Shannon theory. A fully functioning theory of fermionic coherent states for fermionic channels would also give a direct approach to quantify entanglement generation and propagation in fermionic systems that cannot be bosonized. In the absence of qubit-field interactions, Bosonic coherent states have many applications in quantum architecture, and bosonic quantum channels are essential for investigating how QI behaves in these bosonic systems. The same could be attainable with fermionic systems if such a program was developed.

## 7.5 Connections to Cosmology

A key component of this project was to place quantum circuits on a manifold to carry out quantum Shannon calculations. The manifold of choice, corresponding to the (1+1D) CFT, was Minkowski, and therefore obeyed the laws of special relativity. The RQI community have been focused on developing ideas beyond the Minkowski manifold to realize quantum Shannon theory in curved space time.

Working through the LL model has given us a look into the quantum circuits model of a CFT. So a natural extension of this program is to find the holographic duality to a higher dimensional space such as AdS. The AdS/CFT correspondence is well-known in the RQI community but most attempts to develop theories of RQI using this correspondence begin in the higher dimensional space of AdS (usually 2+1D in quantum gravity) and are transformed into a (1+1D) CFT. If the AdS/CFT correspondence holds, the reverse process (developing QI with a lower-dimensional CFT and transforming into AdS) should transform in the same way.

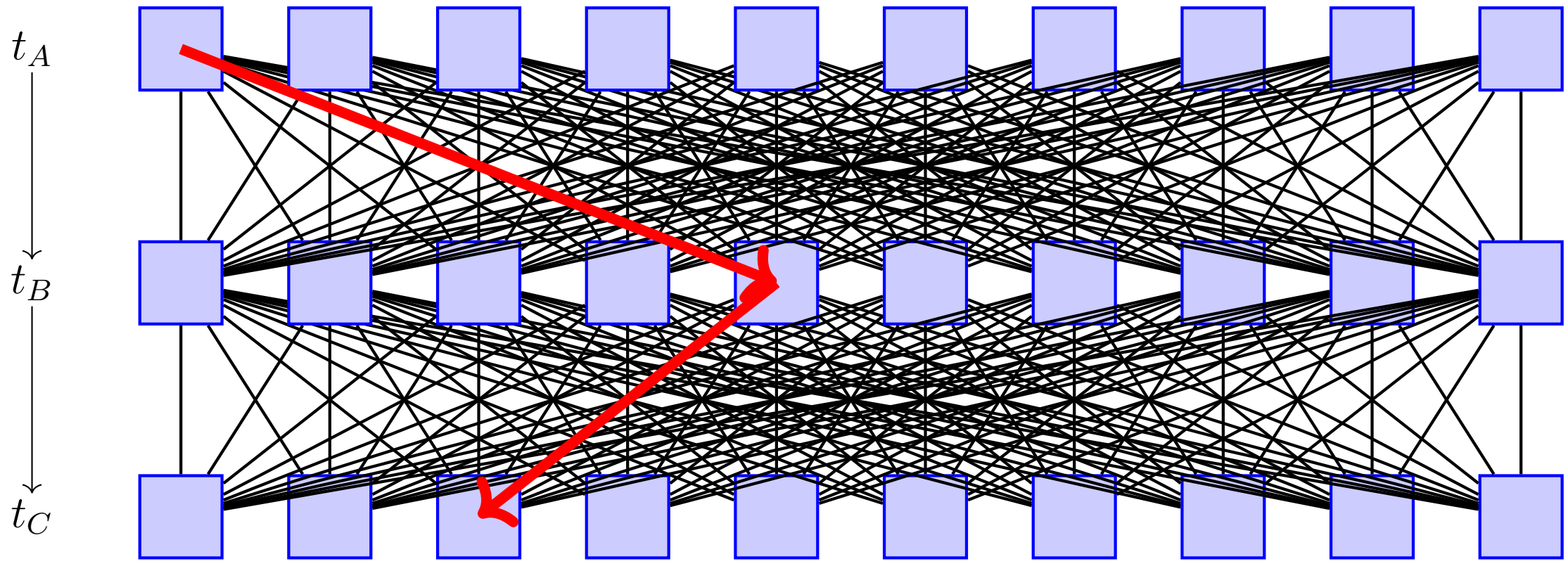

Figure 7.1: All-to-All connectivity means that our Qdots are fully connected. One path here is shown by red arrows. There exists a shortest path from $t_A$ to $t_C$ for information defined on a Finsler Metric for the qubits but also on a 1+1 spacetime between them.

The quantum gravity community has been exploring post-selection as a means to resolve black hole quantum information paradoxes eg. solving Hawking's paradox via final state proposal [107,108] and demonstrating the emergence of the black hole interior from the holographically dual quantum system [109]. Meanwhile in a tangentially related field, post-selection has been utilized in quantum circuits and has maintained the existence of closed time-like curves which are well-known to be non-physical [110, 111]. Placing quantum circuits on a manifold will likely recover the notion that closed time-like curves are non-physical. However, placing quantum circuits of CTCs on a suitable spacetime manifold thus has the potential to find interpretations of the closed timelike curves proposed by Deutsch [112] which is known to cause non-physical quantum computational consequences such as infinite computational power [113, 114].

### 7.5.1 Path Integration and Quantum Shannon Theory

In the quantum channel given in Eq. 5.7, the correlator of vertex operators plays a pivotal role in cancelling out unwanted states. Perturbatively, correlators in QFT are calculated with one of two methods; canonical quantization and path integration. Coincidentally, though we are working in strong coupling, our approach to solve correlators is similar in practice to that of canonical quantization but in the perturbative regime, like that of Ref. [79], extensions to path integration formalism are straightforward.

Recent work has placed an emphasis on understanding the path integral through a quantum computer [115–117]. This path integral formalism has been utilizing Finsler Manifolds in both discrete space as well as continuous space [115, 118, 119] where the Finsler metric was determined to contain the geodesics of quantum information in a quantum circuit. An interesting proposal is one that includes this work when mapping out the correlators such that one can recreate the Finsler metric as we have in the language of quantum Shannon Theory.

The second approach to tackling the path integral formalism also includes the Finsler manifold but allows for an arbitrary approach to the path integral formalism of the fields. The two components of the system are the qubits and the fields, both have associated correlators and it should be possible to carry out separate calculations and then combine them in a process known as Gluing [120]. This allows us to carry out a path integral over a finsler manifold and slice it up as necessary to account for interactions as the fields propagate through the HLL, much like the time-evolution or self-interactions mentioned previously.

## 7.6 Conclusion

In this chapter, we have demonstrated that the UDWQC offers a vast amount of projects to work on. To realize a UDWQC in a lab would shed light on the behavior of entanglement in quantum materials. Moreover, quantum transduction for means of quantum communication is a goal that the QIS community at large aims to achieve, and a laboratory UDWQC would have that same ambition. For these reasons, the theory remains important.

# Appendix A

# Classical Shadow Tomography

To read-out information from a quantum computer requires measurement operators that inherently destroy quantum information. As mentioned in Chapter 1, to know the structure of a quantum state (through the destruction of it) and how it changes requires infinite measurements. Given that this is not realistic we turn to other methods such as Classical Shadows (CS) state tomography to identify, with some degree of error, the structure of a quantum state.

First introduced by Aaronson [121] in 2018, "shadow tomography" offers an approximate classical description of the quantum state, in which $M$ properties of a quantum state can be estimated with error $\epsilon$ by only $O(\frac{\log^4 M}{\epsilon^2})$ copies of the state. We can think of a shadow as an approximation of a quantum state $\rho$ by summing over measurement outcomes $x$, obtained by performing measurements on bases $b$ for a quantum state $x$, i.e.

$$S[\rho] = \sum_{b,x} P(b)\mathcal{P}_{b,x}\rho\mathcal{P}_{b,x}, \tag{A.1}$$

where $\mathcal{P}_{b,x}$ is a projector onto the measurement outcome $x$ on basis $b$, and $P(b)$ is the probability of choosing $b$.

## A.1 Classical Shadows

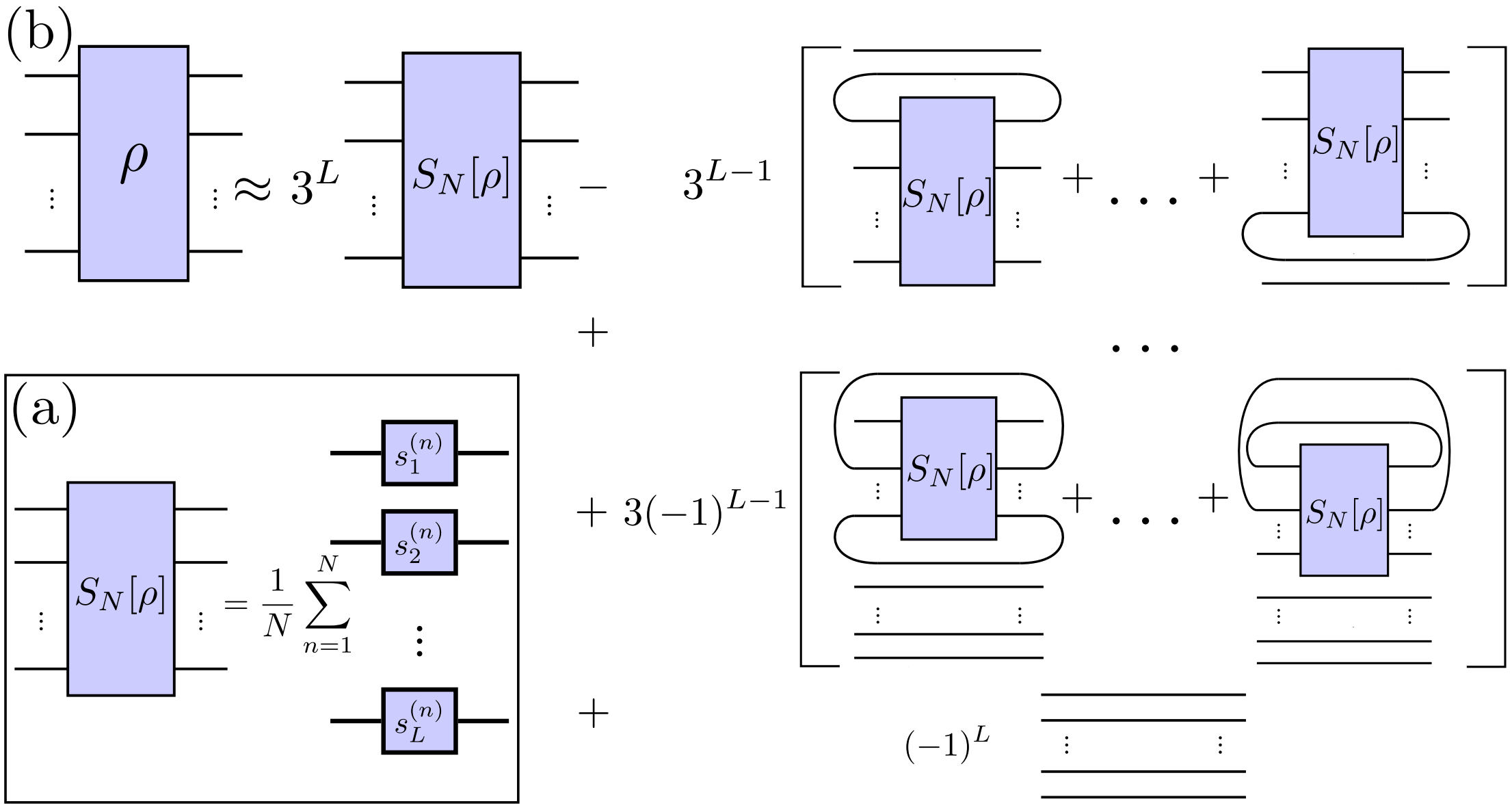

Figure A.1: Diagrammatic description of classical shadows showing a linear relationship between $S_N[\rho]$ and estimator $\sigma_N[\rho] \approx \rho$. (a) The full-density matrix $\rho$ can be approximated by summing over reduced classical shadows with a coefficient that grows exponentially in the number of remaining qubits. (b) A classical shadow $S_N[\rho]$ is obtained by summing over $N$ measurement outcomes on random Pauli bases. For a given $N$, a reduced density matrix, that involves smaller coefficients in the expansion, can be approximated more accurately compared to the full density matrix.

Huang et al. [122–124] developed an algorithm called classical shadows which demonstrated that one can carry out a series of random single-qubit Pauli measurements with a very shallow circuit depth successfully learn properties of many-body systems. The shallow circuit depth meant that these algorithms are suitable for the NISQ-era [125] hardware. After measuring each of the qubits in some random Pauli basis $X$, $Y$ or $Z$ with outcomes $\pm 1$, the post-measurement wavefunction is given by the product state $\left|s^{(n)}\right\rangle = \bigotimes_{l=1}^{L} \left|s_l^{(n)}\right\rangle$. Here, $\left|s_l^{(n)}\right\rangle \in \{|0\rangle, |1\rangle, |+\rangle, |-\rangle, |\mathrm{i}+\rangle, |\mathrm{i}-\rangle\}$ is a Pauli basis state to which the $l^{th}$ qubit has collapsed. The classical shadow $S_N[\rho]$ is obtained by summing over $N$ such randomized measurement outcomes as follows (also see Fig. A.1(b))

$$S_N[\rho] = \frac{1}{N}\sum_{n=1}^{N} |s^{(n)}\rangle\langle s^{(n)}| \tag{A.2}$$

$$= \frac{1}{N}\sum_{n=1}^{N} |s_1^{(n)}\rangle\langle s_1^{(n)}| \otimes \cdots \otimes |s_L^{(n)}\rangle\langle s_L^{(n)}|. \tag{A.3}$$

One can then approximate $\rho$ by adding the reduced classical shadows (see Fig. A.1(a)). This sum simplifies to the following expression from Ref. [122–124]

$$\rho \approx \sigma_N(\rho) = \frac{1}{N}\sum_{n=1}^{N} \sigma_1^{(n)} \otimes \cdots \otimes \sigma_L^{(n)}, \tag{A.4}$$

where

$$\sigma_l^{(n)} = 3|s_l^{(n)}\rangle\langle s_l^{(n)}| - \mathbb{I}. \tag{A.5}$$

The definition of $S_N(\rho)$ presented above is different from Refs. [122,123] which defines it to be the dataset of shots itself and not the density matrix obtained from these shots. But both definitions are complete for Fig. A.1. The density matrix $S_N(\rho)$ defined above is linearly related to the estimator $\sigma_N(\rho)$ of the quantum state $\rho$ obtained by Refs. [122–124]. Hence, the two definitions are informationally equivalent.

One common problem with relevant application in state tomography is studying the phases via ground state preparation, which is a QMA-complete problem [126–130] and cannot be carried out in a reasonable time, even with quantum resources. However, performing dynamics on a quantum state is known to be a BQP-hard problem [131, 132], possible within polynomial time. Likewise, it has been shown that shadow tomography [121] methods such as CS are effective at predicting properties using very few measurements. Thus, if we could combine dynamics simulations and classical shadows, we would have an efficient algorithm to simulate condensed matter systems.

In classical statistical mechanics, ergodicity provides a link between time averages and statistical averages [133]. In quantum mechanics, unitary time evolution retains memory of the initial state, so the link is different yet still enabled by equilibration of macroscopic observables [134]. However, the equilibration time, which may be exponential, depends on a number of factors including the spectral distribution of the initial state, spectral gaps, as well as spectral degeneracies and resonances [135–140].

In Ref. [42] using a one-Dimenstion Transverse Field Ising Model (1DTFIM), we presented an algorithm for identifying phase diagrams and phase transitions of strongly correlated systems (See Fig. A.3)*motivated by how physical quantum systems operate.* It consists of i) identifying an initial state, ii) generating state trajectories by evolving this state in time, iii) using shadow tomography to convert the quantum state to classical data, and iv) applying unsupervised machine learning methods to discover phases of matter and their phase transitions [141–150]. A schematic overview of our approach is shown in Fig. A.2.

Generalizing CS to time-averaged CS (TACS), a shadow tomographic [121] representation of the time-averaged density matrix [134], we then show, on a 20 qubit 1DTFIM, diffusion maps also identify the quantum critical region and cross-overs along a path in the microcanonical phase diagram from a total of 90,000 shots on state trajectories as demonstrated in Fig. A.4. Diffusion maps do so efficiently by learning features from TACS that appear to be susceptibility

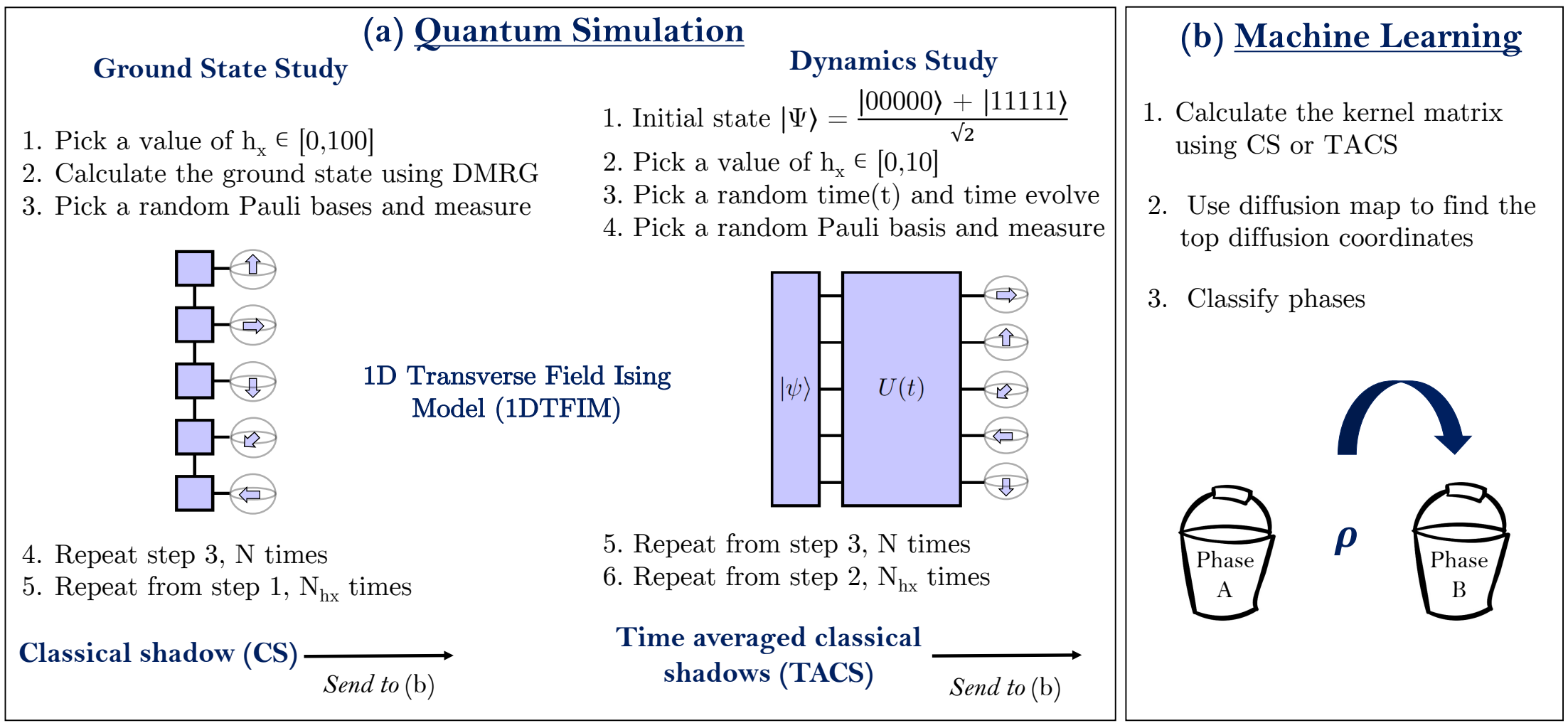

Figure A.2: Schematic overview of our study. (a) Classical shadows(CS) of ground states and time-averaged classical shadows(TACS) from dynamics of a time-reversal invariant GHZ state are generated using quantum simulation. (b) The classical data from quantum simulation is then fed into diffusion maps, an unsupervised machine learning algorithm to learn the phases. Credit to Gaurav Gyawali

and entropy as represented in Fig. A.5. Hence, we can efficiently study the phases and phase transitions of strongly correlated electrons by quantum-simulating state trajectories.

This appendix shows the strength of CS as an algorithmic approach to learning many-body systems through quantum state tomography.

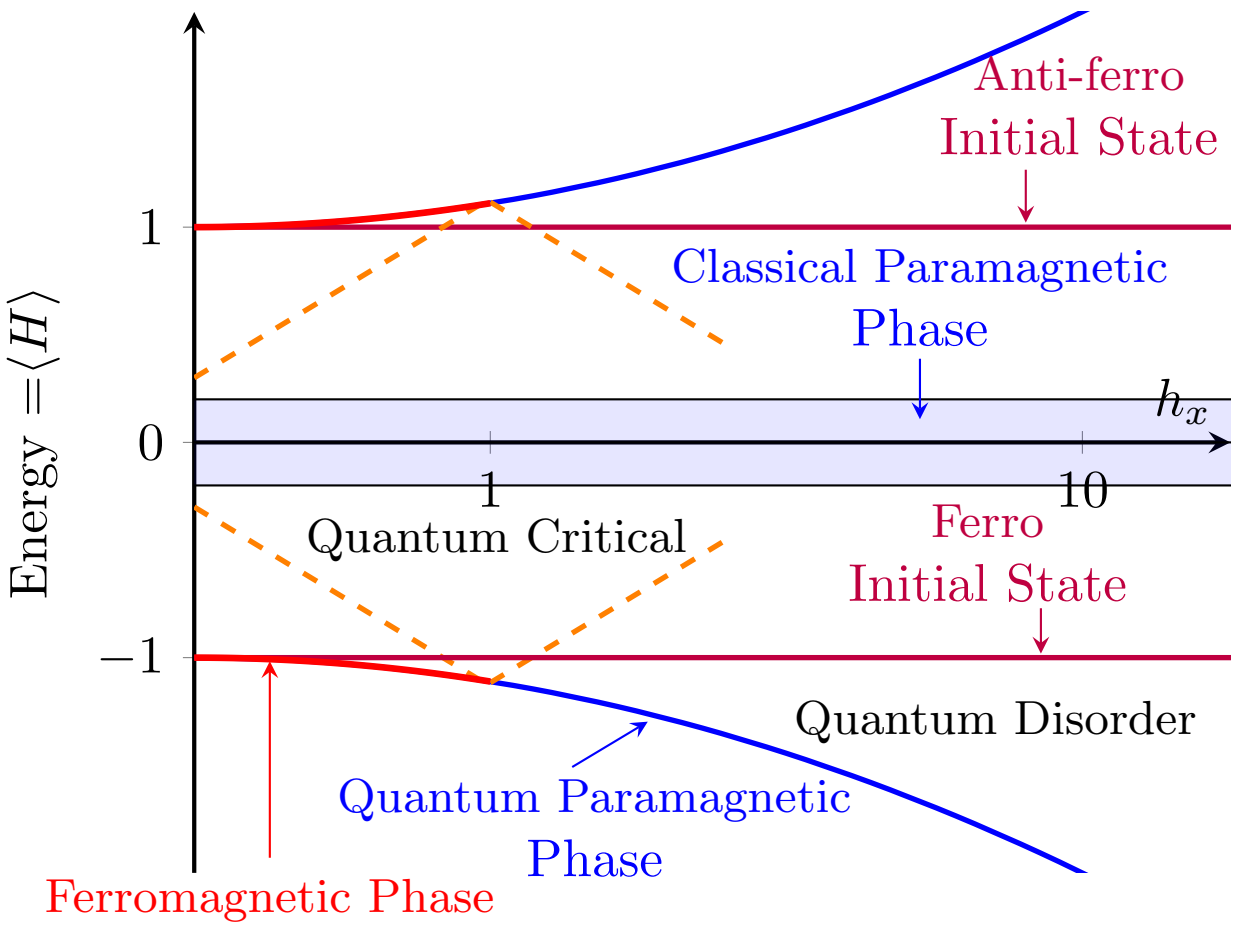

Figure A.3: A sketch of the expected 1DTFIM phase diagram at finite $T$ and finite $N$ as a function of internal energy $E$ and transverse magnetic field $h_x$. This diagram is a modification of the canonical ensemble representation of the phase diagram in Ref. [3], adapted to the microcanonical ensemble. The spectrum is mirror symmetric about $E = 0$ due to the chiral symmetry $\mathcal{C} = ZYZYZY...$

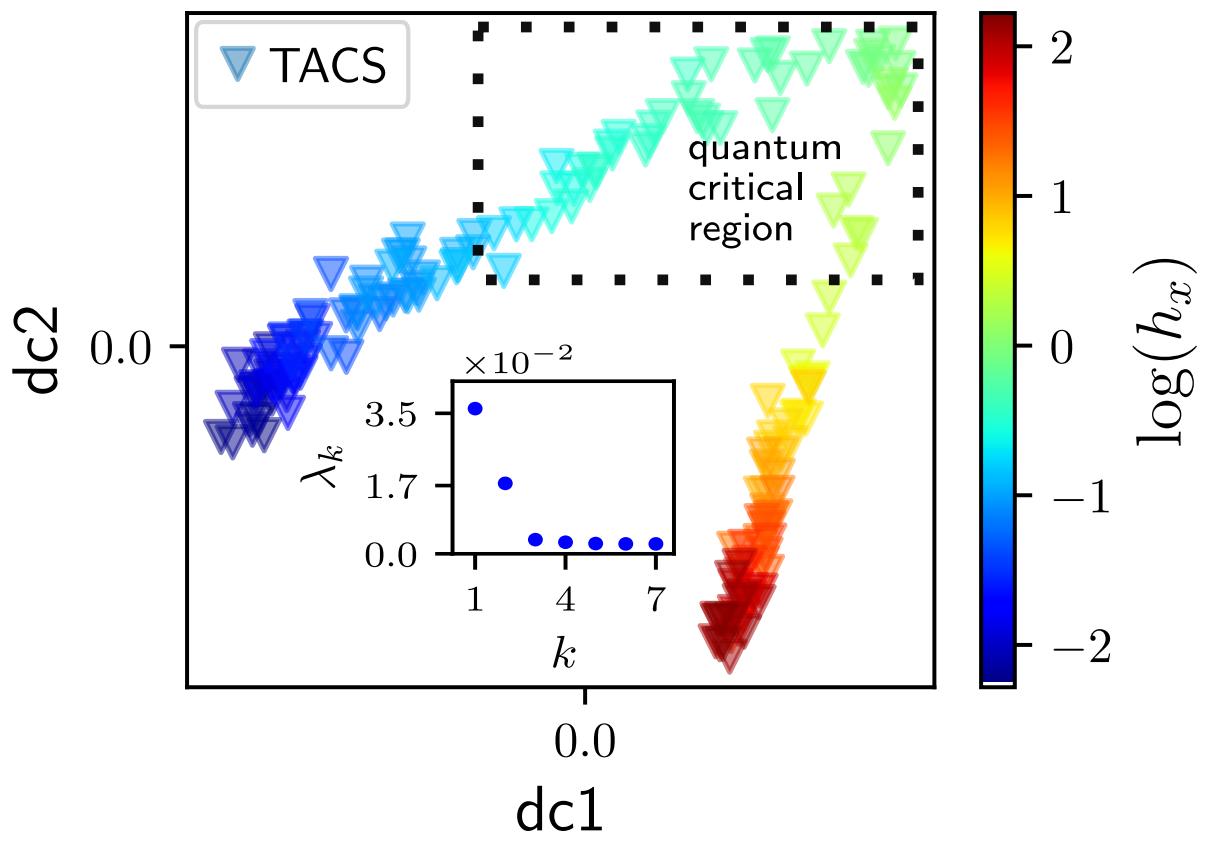

Figure A.4: Phase identification from dynamics data. The eigenvalue spectrum (inset) obtained from diffusion maps shows the two largest eigenvalues corresponding to the two dominant diffusion coordinates dc1 and dc2 (the trivial point $k = 0$, is not shown). The TACS data points largely fall on a two-stranded curve parameterized by $h_x$ in this 2D reduced diffusion space. The quantum critical region (in green) coincides with the inflection point neighborhood of the curve, with points on the left strand belonging mostly to the ordered phase while points on the right strand belong to the disordered phase. Credit to Mabrur Ahmed.

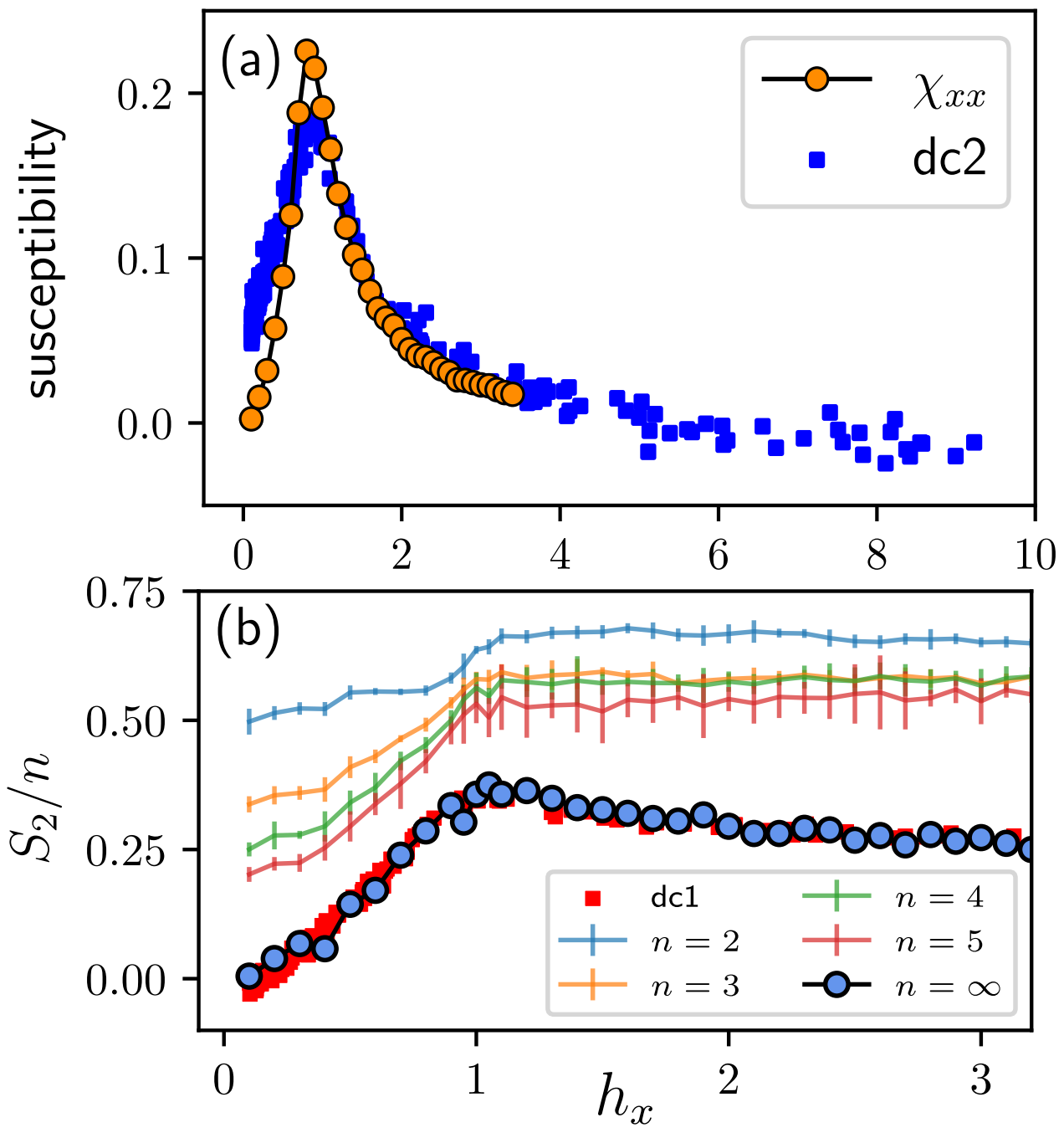

Figure A.5: Interpretation of dc1 and dc2 for microcanonical dynamics of 1DTFIM. The divergent behavior of dc2 qualitatively matches the xx component of the susceptibility, computed using 100k shot TACS data for a 10-site 1DTFIM, denoted by orange circles (b) dc1 matches the Bayesian inference estimate for the second Renyi entropy per site ($S_2/n$) in the thermodynamic limit ($n = \infty$). Bayesian inference is performed on $n$-body entropies for $n = 1 - 5$, also computed using the 10-site dataset. Credit to Mabrur Ahmed.

# Appendix B

# Point-Splitting for a New Hamiltonian

In this appendix we show the procedure for point splitting and explicitly show the results of our Hamiltonian. For these appendices we assume that the fields are normal ordered and therefore can proceed with calculating several of the bosonized terms present in Table 3.1.

## B.1 The $\hat{\psi}^\dagger \hat{\psi}$ coupling

To understand our point splitting of $\hat{\psi}^\dagger_- \hat{\psi}_-$ is we examine the bosonized form of the fields (see Eq. 3.10 as written out,

$$\lim_{\epsilon \to 0} [\hat{\psi}^\dagger_+(z+\epsilon)\hat{\psi}_+(z) = \frac{1}{2\pi} e^{i\sqrt{4\pi}\hat{\phi}(z+\epsilon)} e^{-i\sqrt{4\pi}\hat{\phi}(z)}] \tag{B.1}$$

Which we can rewrite as,

$$\begin{aligned}\hat{\psi}^\dagger_+(z)\hat{\psi}_+(z) &= \frac{1}{2\pi}\lim_{\epsilon\to 0}[e^{i\sqrt{4\pi}\hat{\phi}(z+\epsilon)-i\sqrt{4\pi}\hat{\phi}(z)}\frac{1}{\epsilon}] \\ &= \frac{1}{2\pi}\lim_{\epsilon\to 0}[e^{i\sqrt{4\pi}\epsilon\hat{\phi}(z)}\frac{1}{\epsilon}] \\ &= \frac{1}{2\pi}\lim_{\epsilon\to 0}\frac{1}{\epsilon}[1+i\epsilon\sqrt{4\pi}\partial_z\hat{\phi}(z)+\mathcal{O}(\epsilon^2)] \\ &= \frac{i}{\sqrt{\pi}}\partial_z\hat{\phi}(z)\end{aligned} \tag{B.2}$$

Likewise,

$$\hat{\psi}^\dagger_-(\bar{z})\hat{\psi}_-(\bar{z}) = -\frac{i}{\sqrt{\pi}}\partial_{\bar{z}}\hat{\bar{\phi}}(\bar{z}) \tag{B.3}$$

revisiting our density matrix in Eq. 2.5 after suppressing the $e^{ik_F x}$ terms our Hamiltonian in Eq. 3.2 becomes,

$$\hat{H}_{\text{int}}(t) = J\hat{\mu}(t)(\frac{i}{\sqrt{\pi}}(\frac{1}{v}\partial_t(\hat{\phi}(t,x)+\hat{\bar{\phi}}(t,x)) - \partial_x(\hat{\phi}(t,x)-\hat{\bar{\phi}}(t,x))) \tag{B.4}$$

$$= J\hat{\mu}(t)(\frac{i}{\sqrt{\pi}}(\frac{1}{v}\partial_t\hat{\varphi}(t,x) - \partial_x\hat{\vartheta}(t,x)) \tag{B.5}$$

take note here that $\partial_x\hat{\vartheta} = -\hat{\Pi}$ as well as $\frac{1}{v}\partial_t\hat{\phi}$. this leads to our interaction Hamiltonian,

$$\hat{H}_{\text{int}} = J\hat{\mu}(t)(\frac{i}{\sqrt{\pi}}(2\hat{\Pi})) \tag{B.6}$$

## B.2 For a $\hat{\psi}^\dagger\partial_z\hat{\psi}$ interaction

Another gate can be formed from considering the Dirac Hamiltonian represented by Eq. 5.13. To find a bosonized form of the hamiltonian we wish to evaluate the point splitting of this interaction starting with,

$$\lim_{\epsilon\to 0}[\hat{\psi}_+^\dagger(z+\epsilon)\partial_z\hat{\psi}_-(z) = \partial_z\hat{\phi}(z)\frac{-i}{\sqrt{\pi}}(e^{i\sqrt{4\pi}\hat{\phi}(z+\epsilon)}e^{-i\sqrt{4\pi}\hat{\phi}(z)})] \tag{B.7}$$

and now we rewrite this as,

$$\begin{aligned}
\hat{\psi}_+^\dagger(z)\partial_z\hat{\psi}_+(z) &= \partial_z\hat{\phi}(z)\left(\frac{-i}{\sqrt{\pi}}\lim_{\epsilon\to 0}[e^{i\sqrt{4\pi}\hat{\phi}(z+\epsilon)-i\sqrt{4\pi}\hat{\phi}(z)}\frac{1}{\epsilon}]\right)\\
&= \partial_z\hat{\phi}(z)\left(\frac{-i}{\sqrt{\pi}}\lim_{\epsilon\to 0}[e^{i\sqrt{4\pi}\epsilon\hat{\phi}(z)}\frac{1}{\epsilon}]\right)\\
&= \partial_z\hat{\phi}(z)\left(\frac{-i}{\sqrt{\pi}}\lim_{\epsilon\to 0}\frac{1}{\epsilon}[1+i\epsilon\sqrt{4\pi}\partial_z\hat{\phi}(z)+\mathcal{O}(\epsilon^2)]\right)\\
&= 2(\partial_z\hat{\phi}(z))^2
\end{aligned} \tag{B.8}$$

following the same procedure we find,

$$\hat{\psi}_-^\dagger(\bar{z})\partial_{\bar{z}}\hat{\psi}_-(\bar{z}) = -2(\partial_{\bar{z}}\hat{\bar{\phi}}(\bar{z}))^2 \tag{B.9}$$

Plugging these values into Eq. 5.13 we get,

$$\hat{H}_{\text{D,int}}(t) = J\chi(t)\int_{\mathbb{R}} dy\; p(x(t),y)\hat{\mu}(t)2((\partial_z\hat{\phi}(z))^2+(\partial_{\bar{z}}\hat{\bar{\phi}}(\bar{z}))^2) \tag{B.10}$$

where we simplify by evaluating,

$$\begin{aligned}
(\partial_z\hat{\phi})^2 &= \left(\frac{-i}{2}\left(\frac{1}{v}\partial_t\hat{\varphi}-\partial_x\hat{\varphi}\right)\right)^2\\
&= \frac{1}{4}\left(\hat{\Pi}^2-\frac{2}{v}\partial_t\hat{\varphi}\partial_x\hat{\varphi}+\partial_x\hat{\varphi}^2\right)
\end{aligned} \tag{B.11}$$

and,

$$
\begin{aligned}
(\partial_{\bar{z}}\hat{\bar{\phi}})^2 &= \left(\frac{-i}{2}\left(\frac{1}{v}\partial_t\hat{\varphi} + \partial_x\hat{\varphi}\right)\right)^2 \\
&= \frac{1}{4}\left(\hat{\Pi}^2 + \frac{2}{v}\partial_t\hat{\varphi}\partial_x\hat{\varphi} + \partial_x\hat{\varphi}^2\right)
\end{aligned}
\tag{B.12}
$$

so our Eq. B.10 becomes the same as Eq. 5.14,

$$
\hat{H}_{\text{D,int}} = J\chi(t)\int_{\mathbb{R}} dy\ p(x(t), y)\hat{\mu}(t)(\hat{\Pi}^2 + (\partial_x\hat{\varphi})^2) \tag{B.13}
$$

# Appendix C

# Constraining the Momentum Displacement Operator

In this appendix, we demonstrate the derivation of Eq. 2.27 and to see this we start with the following relation

$$e^{i\hat{\Pi}}\hat{a}e^{-i\hat{\Pi}} = \hat{a} + \alpha_\varphi \tag{C.1}$$

$$-\alpha_\Pi^* e^{i\hat{\Pi}}\hat{a}e^{-i\hat{\Pi}} = -\alpha_\Pi^*\hat{a} - \alpha_\Pi^*\alpha_\varphi \tag{C.2}$$

$$\alpha_\Pi\hat{a}^\dagger - \alpha_\Pi^* e^{i\hat{\Pi}}\hat{a}e^{-i\hat{\Pi}} = \alpha_\Pi\hat{a}^\dagger - \alpha_\Pi^*\hat{a} - \alpha_\Pi^*\alpha_\varphi \tag{C.3}$$

$$\alpha_\Pi\hat{a}^\dagger e^{i\hat{\Pi}}|0\rangle - \alpha_\Pi^* e^{i\hat{\Pi}}\hat{a}|0\rangle = \hat{\Pi}|{+\alpha}\rangle\,\alpha_\Pi^* - \alpha_\varphi|{+\alpha}\rangle \tag{C.4}$$

$$\alpha_\Pi^*\alpha_\varphi e^{i\hat{\Pi}}|0\rangle + e^{i\hat{\Pi}}\hat{\Pi}|0\rangle = \hat{\Pi}|{+\alpha}\rangle - \alpha_\Pi^*\alpha_\varphi|{+\alpha}\rangle \tag{C.5}$$

$$\hat{\Pi}|{+\alpha}\rangle = \alpha_\Pi^*\alpha_\varphi|{+\alpha}\rangle + \alpha_\Pi^*\alpha_\varphi|{+\alpha}\rangle + e^{i\hat{\Pi}}\hat{\Pi}|0\rangle \tag{C.6}$$

where we have utilized $\alpha_\varphi = -\alpha_\Pi$ and the commutation relation

$$[\hat{a}^\dagger, e^{\pm i\hat{\Pi}}] = \mp\alpha_\Pi^* e^{\pm i\hat{\Pi}} \tag{C.7}$$

and if we let $\gamma = 2\alpha_\Pi^*\alpha_\varphi$ and choose parameters carefully such that $\gamma^2 \gg \langle 0|\hat{\Pi}^2|0\rangle$

# Appendix D